\documentclass[aps,prb,onecolumn,superscriptaddress,floatfix,12pt,nofootinbib]{revtex4-2}
\usepackage[utf8]{inputenc}

\usepackage{mathtools}
\usepackage{setspace}
\usepackage{graphicx}
\usepackage{latexsym}
\usepackage{amssymb}
\usepackage{amsmath}
\usepackage{amsfonts}
\usepackage{bm}
\usepackage{enumitem}
\usepackage[draft=false]{hyperref}
\usepackage{lipsum}
\usepackage{braket}
\usepackage{tikz}
\usepackage{comment}
\usepackage{ragged2e}
\usepackage{multirow}
\usetikzlibrary{arrows}
\usepackage{subcaption}
\usepackage[justification=justified, format=plain, font=small]{caption}
\usepackage[normalem]{ulem} 

\DeclareMathAlphabet\mathbfcal{OMS}{cmsy}{b}{n}

\newcommand{\sfigref}[2]{Fig.\,\hyperref[#1]{\ref{#1}(#2)}}

\def\({\left(}
\def\){\right)}
\def\[{\left[}
\def\]{\right]}
\newcommand{\ie}{\begin{equation}\begin{aligned}}
\newcommand{\fe}{\end{aligned}\end{equation}}
\def\d{\partial}

\hypersetup{
  colorlinks = true,
  urlcolor = blue,
  pdfauthor = {},
  pdftitle = {}
}

\begin{document}

MIT-CTP/5493
\\

\title{Gapless Infinite-component Chern-Simons-Maxwell Theories}
\date{\today}

\author{Xie Chen}
\affiliation{Department of Physics and Institute for Quantum Information and Matter, \mbox{California Institute of Technology, Pasadena, California 91125, USA}}

\author{Ho Tat Lam}
\affiliation{Center for Theoretical Physics, \mbox{Massachusetts Institute of Technology,  Cambridge, MA 02139 USA}}

\author{Xiuqi Ma}
\affiliation{Department of Physics and Institute for Quantum Information and Matter, \mbox{California Institute of Technology, Pasadena, California 91125, USA}}

\begin{abstract} 
\begin{spacing}{1.1}
The infinite-component Chern-Simons-Maxwell (iCSM) theory is a $3+1$D generalization of the $2+1$D Chern-Simons-Maxwell theory by including an infinite number of coupled gauge fields. It can be used to describe interesting $3+1$D systems. In Phys.\ Rev.\ B 105, 195124 (2022), it was used to construct gapped fracton models both within and beyond the foliation framework. In this paper, we study the nontrivial features of gapless iCSM theories. In particular, we find that while gapless $2+1$D Maxwell theories are confined and not robust due to monopole effect, gapless iCSM theories are deconfined and robust against all local perturbation and hence represent a robust $3+1$D deconfined gapless order.  The gaplessness of the gapless iCSM theory can be understood as a consequence of the spontaneous breaking of an exotic one-form symmetry. Moreover, for a subclass of the gapless iCSM theories, we find interesting topological features in the correlation and response of the system. Finally, for this subclass of theories, we propose a fully continuous field theory description of the model that captures all these features. 
\end{spacing}
\end{abstract}

\maketitle
\newpage

\tableofcontents

	\section{Introduction and Summary}
	\label{sec:intro}
	
	2+1D abelian topological orders are completely characterized by 2+1D Chern-Simons theories of multiple $U(1)$ gauge groups \cite{PhysRevB.46.2290}. Such a theory is described by the Euclidean Lagrangian
	\ie\label{action_CS}
	\mathcal{L}=\frac{i}{4\pi} K_{IJ}\epsilon^{\mu\nu\rho}{a}_\mu^I\partial_\nu {a}_\rho^J~,
	\fe
	where $a^I_\mu$ are $U(1)$ gauge fields labeled by $I=1,\ldots,N$ and $K$ is an $N\times N$ non-degenerate (no zero eigenvalue), symmetric integer matrix. We omitted the sum over repeated indices. All the topological features of the theory are encoded in the $K$ matrix, including the anyons species, the fractional braiding statistics, the ground state degeneracy, etc.
	
	We can add a Maxwell term (first term) to the Lagrangian \eqref{action_CS}:	
	\ie\label{action}
	\mathcal{L}=\frac{1}{4g^2}{f}^I_{\mu\nu}{f}^{I,\mu\nu}+\frac{i}{4\pi} K_{IJ}\epsilon^{\mu\nu\rho}{a}_\mu^I\partial_\nu {a}_\rho^J~,
	\fe
	where $f_{\mu\nu}^I=\partial_\mu a_\nu^I-\d_\nu a_\mu^I$. The theory is no longer topological but it can accommodate a more general, degenerate $K$ matrix. If the $K$ matrix is degenerate, there are gapless degrees of freedom that do not participate in the Chern-Simons term.
	
	An interesting class of theories of the form \eqref{action} is given by those with a translation invariant $K$ matrix:\footnote{More generally, one can consider a $K$ matrix with periodicity $L$ that obeys $K_{IJ}=K_{I+L,J+L}$. The results in this paper generalize straightforwardly to those cases.}
	\ie\label{Kmatrix}
	K_{I,I\pm k\text{ mod }N}=c_k\in\mathbb{Z}~.
	\fe
	Here, we imposed periodic boundary condition on the index $I$. For such a theory, we can take an alternative point of view and think of $I$ as a discrete coordinate that parameterizes a 1D periodic lattice. It is then natural to take the thermodynamic limit of the hypothetical lattice, which is equivalent to taking the $N\rightarrow\infty$ limit of the theory. This defines an infinite-component Chern-Simons-Maxwell (iCSM) theory \cite{ma2020fractonic}. While the gauge fields are still 2+1D, the infinite components of the gauge fields give rise to an emergent dimension parameterized by $I$.\footnote{The idea of constructing higher-dimensional theories using infinite copies of low-dimensional theories is reminiscent of dimension deconstruction using quiver theories \cite{Arkani-Hamed:2001kyx, Hill:2000mu,Arkani-Hamed:2001wsh}. The connections between quiver theories and fracton topological orders were explored recently in \cite{Razamat:2021jkx,Geng:2021cmq,Franco:2022ziy}. } We will refer to the emergent dimension as the $z$ direction. To avoid long-range interactions, we require that $c_k$ vanish when $k$ is greater than some correlation length $\xi$.
	
	The iCSM theory was considered in the context of 3+1D quantum Hall system consisting of infinitely many stacked 2+1D electron gases in a strong perpendicular magnetic field \cite{PhysRevB.40.11943,PhysRevB.42.1339,Naud_2000,Naud:2000xa}. It was noted that the iCSM theory exhibits various unusual properties, such as irrational braiding statistics and edges states that behaves as ``chiral semi-metals''. 
	
	Recently, the iCSM theory has been revisited as a simple yet rich example of fracton topological order. A fracton topological order \cite{Chamon:2004lew,Haah:2011drr,Vijay:2016phm} is characterized by gapped excitations with restricted mobility, e.g.\ particles (fractons) that cannot move, and sub-extensive ground state degeneracy that depends sensitively on the system size. Many of these peculiar properties can be explained by the underlying exotic global symmetries \cite{Seiberg:2019vrp,Seiberg:2020bhn,Seiberg:2020wsg,Seiberg:2020cxy,Gorantla:2022eem}. See \cite{Nandkishore:2018sel,Pretko:2020cko} for reviews on the fracton topological order. 
	
	The simplest iCSM theory is the one with a diagonal $K$ matrix, which is simply a stack of decoupled 2+1D abelian topological orders. It shares many similarities with the foliated fracton order \cite{Shirley:2017suz,Shirley:2018nhn,Shirley:2018hkm,Shirley:2018vtc,Shirley:2019uou}: the ground state degeneracy grows exponentially with the number of layers and enlarging the system size is achieved by adding decoupled 2+1D topological orders. There is, however, one difference: the gapped excitations (gapped gauge charges) in the iCSM theory are not immobile fractons but planons that move only in 2D planes of constant $z$.

    Interestingly, the iCSM theory realizes more phases than the foliated fracton order. In \cite{ma2020fractonic}, it was used to construct gapped 3+1D fracton models both within and beyond the foliation framework. It was also pointed out that some iCSM theories are gapless without exploring further their properties, and this is the main topic of this paper. Some examples of gapless iCSM theories, in particular the one with $(c_0,c_1)=(-2,1)$, were explored in \cite{Sullivan:2021rbk}.
    
    Below, we summarize the main results of this paper. The iCSM theory has plane waves with dispersion relation
    \ie
    \omega^2=k_x^2+k_y^2+m(q)^2~,\quad m(q)=\frac{g^2}{2\pi}\lambda(q)~,
    \fe
    where $q\sim q+2\pi$ is the momentum in the $z$ direction and $\lambda(q)$ is the eigenvalue of the $K$ matrix for the eigenvector $v_I=e^{iqI}$. The theory is gapless if $\lambda(q)$ vanishes at some $q=q_j$, which we refer to as the \emph{gapless momenta}. Here $j=1,\ldots,\ell$ labels the gapless momenta and $\ell$ is the total number of them. Two pieces of information related to the gapless momenta are crucial. One is whether the gapless momenta $q_j$ are \emph{commensurate} (rational $\frac{q_j}{2\pi}$) or \emph{incommensurate}. The other is the expansion of $\lambda(q)$ around the gapless momenta $q_j$: 
    \ie
    \lambda(q)\approx \lambda_j(q-q_j)^{\gamma_j}~,
    \fe
    which controls the low-energy dispersion relation around the gapless momenta $q_j$. We denote the maximum $\gamma_j$ in a theory by $\Gamma\equiv\text{max}(\gamma_j)$.

    A generic gapless iCSM theory has only incommensurate gapless momenta. In some rare cases, it can have commensurate gapless momenta. If the theory has only incommensurate gapless momenta, it can never be gapless when $N$ is finite. In other words, the gaplessness in such a theory emerges only in the $N\rightarrow\infty$ limit. This is a common phenomenon in lattice models, that the spectral gap vanishes only in the thermodynamic limit.
    
    We emphasize that the $N\rightarrow \infty$ limit and the $g^2\rightarrow\infty$ limit do not commute in a gapless iCSM theory. Since $g^2$ is the only scale in the theory, it means that the $N\rightarrow\infty$ limit also does not commute with the low-energy or long-distance limit. Throughout the paper, when we discuss the iCSM theory, we always work in the strict $N\rightarrow \infty$ limit before considering various low-energy or long-distance limits. When $N$ is finite but large, most of our results are still valid at the energy scale $E$ that obeys $1\gg E/g^2\gg 1/N$ or the length scale $r$ that obeys $N\gg g^2 r\gg 1$.

As a 3+1D system, the gapless iCSM theory has an  \emph{exotic continuous one-form global symmetry} generated by the current
\ie
    J_{\mu\nu}(z)=\frac{i}{g^2}f_{\mu\nu}^z~,\quad J_{\mu z}(z)=\frac{1}{2\pi}\epsilon^{\mu\nu\rho}f^{z}_{\nu\rho}~,
\fe
where the indices $\mu,\nu,\rho$ are restricted to $\tau,x,y$.
The current obeys a conservation equation
\ie\label{eq:intro_conserv}
&\partial_\tau J_{\tau z}+\partial_x J_{x z}+\partial_y J_{y z}=0~,
\\
&\partial_\tau J_{\tau x}+\partial_y J_{y x}-\Delta J_{z x}=0~,
\\
&\partial_\tau J_{\tau y}+\partial_x J_{x y}-\Delta J_{z y}=0~,
\\
&\partial_x J_{x \tau}+\partial_y J_{y \tau}-\Delta J_{z \tau}=0~.
\fe
The first equation is tautologically true because of the Bianchi identity. The other three equations follow from the equations of motion. The operator $\Delta$ is defined as
\ie
\Delta J(z)\equiv \sum_k c_k J(z+k)~,
\fe
where $c_k$ are the entries of the infinite-dimensional $K$ matrix \eqref{Kmatrix}. If the operator $\Delta$ is replaced by $-\partial_z$, the current conservation equation \eqref{eq:intro_conserv} coincides with the current conservation equation of a standard one-form global symmetry  in 3+1D \cite{Gaiotto_2015}. Because of this relation, we refer to the symmetry in the gapless iCSM theory as an exotic one-form global symmetry. The symmetry acts on both the Wilson lines and the extended gauge-invariant monopole operators. In the gapless iCSM theory, the gauge-invariant monopole operators are not local operators but string-like operators that extend in the $z$ direction. Because the monopole operators are extended operators, a gapless iCSM theory is robust against monopole effect in contrast to the 2+1D Maxwell theory. In fact, by examining the correlation functions of the charged operators, we conclude that the exotic continuous one-form symmetry is spontaneously broken, i.e.\ the charged operators obey perimeter law. Also, the gaplessness of a gapless iCSM theory can be understood as a consequence of this spontaneous symmetry breaking. Unlike an ordinary global symmetry, the spontaneous breaking of this exotic continuous one-form global symmetry can lead to multiple Goldstone modes that are gapless at non-zero momenta in the $z$ direction. We conclude that the gapless iCSM theory furnishes a robust 3+1d deconfined gapless order with a spontaneously broken exotic one-form global symmetry.

If we view the gapless iCSM theory as a 2+1D system with infinitely many gauge field components, the 3+1D exotic one-form symmetry reduces to a 2+1D $U(1)^{\ell_c}$ magnetic zero-form symmetry that acts on the gauge-invariant monopole operators and a 2+1D $U(1)^{\ell_c}\times \mathbb{R}^{\ell-\ell_c}$ electric one-form symmetry that acts on the Wilson lines, where $\ell$ is the number of gapless momenta and $\ell_c$ is the number of commensurate gapless momenta. Note that the gauge-invariant monopole operators are now local operators in 2+1D and that the number of independent gauge-invariant monopole operators is given by $\ell_c$.

A gapless iCSM theory also has discrete one-form symmetries. When the $x,y$ direction are compactified on a torus, they give rise to an exact degeneracy at every energy level. In a forthcoming paper \cite{appear}, we will report the calculation of this degeneracy.

Let us briefly summarize the phenomenology of the gapless iCSM theory. See also Table~\ref{tab:compare} for a summary and a comparison with the $U(1)$ gauge theory and the gapped iCSM theory. 

        \renewcommand{\arraystretch}{1.2}
\begin{table}
    \centering
    \begin{tabular}{|c|c|c|c|c|c|}
       \hline 
      {theory}& \multicolumn{2}{|c}{$U(1)$ gauge theory}& \multicolumn{3}{|c|}{iCSM theory}  \\
       \hline
       \multirow{2}{*}{spectrum} & \multirow{2}{*}{gapped} &\multirow{2}{*}{gapless} & \multirow{2}{*}{gapped} &\multicolumn{2}{|c|}{gapless}
       \\
       \cline{5-6}
       &&& & $\Gamma=1$& $\Gamma>1$
        \\
        \hline
        monopole? &no & yes & no & \multicolumn{2}{|c|}{$\ell_c$ number of monopoles}
        \\
        \hline
       robust? & yes & no & yes & \multicolumn{2}{|c|}{yes}
       \\
       \hline
       confine? & no & yes & no & \multicolumn{2}{|c|}{no}
       \\
       \hline
       \vspace*{-4pt}
       \multirow{2}{*}{braiding phase}  & \multirow{2}{*}{finite} & \multirow{2}{*}{zero}& finite& finite & \multirow{2}{*}{divergent}
       \\
        & & & (decay in $z$) & (oscillate in $z$) &
       \\
       \hline
       \vspace*{-4pt}
       longitudinal 
       & \multirow{2}{*}{zero}& \multirow{2}{*}{divergent}& \multirow{2}{*}{zero}& \multirow{2}{*}{finite} &  \multirow{2}{*}{divergent}
       \\
       conductivity &&&&&
       \\
       \hline
       \multirow{2}{*}{Hall conductivity} & \multirow{2}{*}{finite} & \multirow{2}{*}{zero} & finite & \vspace*{-4pt}finite & \multirow{2}{*}{divergent}
       \\
       & & & (decay in $z$) & (oscillate in $z$) &
       \\
       \hline
    \end{tabular}
    \captionsetup{justification=Justified}
    \caption{Properties of the 2+1D $U(1)$ gauge theory (with/without a Chern-Simons term) and the iCSM theory.  A theory is robust if it has no relevant local operator. A divergent braiding phase means that the braiding phase diverges when the distance between two gauge charges approaches infinity. 
    $\Gamma$ is the maximum dynamical exponent of a gapless iCSM theory and $\ell_c$ is the number of commensurate gapless momenta. }
    \label{tab:compare}
\end{table}

A generic gapless iCSM theory has $\Gamma=1$. Such a theory has only low-energy dispersion relations that are linear in $q-q_j$ around gapless momenta $q_j$. It always has an even number of gapless momenta. For each gapless momentum $q_j$, there exists another gapless momentum $q_{j^*}=-q_j$ and $\lambda_{j^*}=-\lambda_j$. 
    
We can couple the $\Gamma=1$ gapless iCSM theory to electrically charged matter and study their interactions. At long distance $g^2r\gg 1$, $|z|\gg 1$, two gauge charges of unit charge separated by $z$ layers and a distance $r$ on the $xy$ plane interact via a deconfined electric potential:
\ie
V(r,z)=\frac{g^2}{4\pi}\sum_{j=1}^\ell\frac{\cos\left(q_j z\right)}{\sqrt{m_j^2r^2+z^2}}~,
\fe
where $m_j=\frac{g^2}{2\pi}\lambda_j$ and the sum is over the gapless momenta $q_j$, $j=1,\ldots,\ell$. When $g^2r\gtrsim|z|\gg 1$, taking one gauge charge around the other along a circle of radius $r$ on the $xy$ plane picks up a braiding phase
\ie\label{eq:intro_braiding}
\exp\left[\sum_{j=1}^{\ell}\frac{g^2 e^{i q_j |z|}}{2m_j}\left(1-\frac{ |z|}{\sqrt{m_j^2r^2+z^2}}\right)\right]~.
\fe
If we further take the limit $g^2r\gg |z|$, the braiding phase approaches its asymptotic value
\ie\label{eq:intro_braiding_long_distance}
\exp\left[\sum_{j=1}^{\ell}\frac{g^2 e^{i q_j |z|}}{2m_j}\right]~,
\fe
which defines the braiding statistics of the gauge charges. It turns out in this limit $g^2r\gg |z|$, the braiding becomes topological in the sense that it does not depend on the details but only the topology of the braiding trajectory as long as the trajectory is sufficiently large. Here, we observe that the braiding statistics is non-local in the $z$ direction i.e.\ that the braiding statistics does not decay with respect to $|z|$. In the picture of flux attachment, the braiding phase \eqref{eq:intro_braiding} is the Aharonov-Bohm phase that measures the fluxes, sourced by the static gauge charges, enclosed by the braiding trajectory. \eqref{eq:intro_braiding} then implies that a flux string that extends in the $z$ direction is attached to a gauge charge. The net flux of the flux string on each layer does not decay with $|z|$ but as $|z|$ increases the flux distribution becomes more and more spread out.

    The correlation functions of Wilson lines in the gapless iCSM theory with $\Gamma=1$ have a scale symmetry (up to modulation) at long distance. At the length scale $r$ that obeys $g^2r\gg 1$, $g^2r\gg|z|$, the correlation function takes the form
    \ie
    \langle W^{I+z}(C_1)W^{I}(C_2)\rangle= \exp\left[\sum_{j=1}^\ell e^{i q_j z} F_j(C_1,C_2)\right] \exp\left[- i\,\text{link}(C_1,C_2)\times \text{PV}\int_{-\pi}^{\pi}dq\, \frac{e^{iq z}}{\lambda(q)}\right]~,
    \fe
    where $F_j(C_1,C_2)$ are scale invariant functions that do not depend on the overall size of the curves $C_1,C_2$, $\text{link}(C_1,C_2)$ is the linking number of the curves $C_1,C_2$ and PV stands for the Cauchy principal value of the integral. 
    Without the Cauchy principal value prescription, the integral is ambiguous because its integrand diverges at the gapless momenta.
    Although the correlation function is not fully topological, its phase is still topological in the sense that it depends only on the linking number of the curves. If we further take the limit $g^2r\gg |z|\gg 1$, the phase of the correlation function reproduces the braiding statistics \eqref{eq:intro_braiding_long_distance}.
    
    We can couple the theory to external electromagnetic gauge fields $A^I_\mu$ localized on the $I$th layer via the coupling $\mathcal{L}\supset\frac{i}{2\pi}\epsilon^{\mu\nu\rho}A_\mu^I\partial_\nu a_\rho^I$ and measure the response current on another layer that is $z$ layers apart. This defines the DC conductivity tensor $\sigma_{ij}(z)$. A gapless iCSM theory with $\Gamma=1$ has a finite DC longitudinal conductivity and a finite DC Hall conductivity:
    \begin{equation}
    \begin{aligned}
    &\sigma_{xx}(z)=\frac{1}{4\pi}\sum_{j=1}^\ell\frac{e^{iq_j z}}{|\lambda_j|}~,
    \\
    &\sigma_{xy}(z)=\text{PV}\int_{-\pi}^\pi\frac{dq}{2\pi}\frac{1}{2\pi}\frac{e^{iq z}}{\lambda(q)}\xrightarrow{|z|\gg 1} \frac{i}{4\pi }\sum_{j=1}^{\ell} \frac{e^{iq_j|z|}}{ \lambda_j}~,
    \end{aligned}
    \end{equation}
    where $\text{PV}$ stands for the Cauchy principal value of the integral. Both the DC longitudinal conductivity and the DC Hall conductivity do not decay with $|z|$.

    We can take the continuum limit in the $z$ direction and derive a fully continuous field theory for the gapless iCSM theory with $\Gamma=1$. Let us introduce a layer spacing $\mathbf{a}$ and define a set of continuum variables
    \ie
    \hat z\equiv z\mathbf{a}~,\quad  \hat g^2 \equiv g^2\mathbf{a}~.
    \fe
    In the continuum limit, we send $\mathbf{a}\rightarrow 0$ while holding the continuum coupling $\hat g$ fixed. This leads to the continuum action
    \ie\label{eq:intro_action_continuum}
    S =\sum_{j=1}^{\ell}\int d^3xd\hat z \left[ \frac{1}{2\hat g^2} \hat f^\dagger_{j,\mu\nu}\hat f^{\mu\nu}_{j}+\frac{\lambda_j}{4\pi}\epsilon^{\mu\nu\rho} \partial_{\hat z}\hat a_{j,\mu}\partial_\nu \hat a^\dagger_{j,\rho}\right]~,
    \fe
    where $\hat a_{j,\mu}$ is a complex continuum gauge field associated to the gapless momenta $q_j$ and $\mu=\tau,x,y$. Recall that in a gapless iCSM theory, each gapless momentum $q_j$ has another paired-up gapless momentum $q_{j^*}=-q_j$. The continuum gauge fields $\hat a_{j,\mu}$ obey $\hat a_{j^*,\mu}=\hat a_{j,\mu}^\dagger$. 
    The continuum theory \eqref{eq:intro_action_continuum} is scale invariant and depends only on the dimensionless couplings $\hat g$ and $\lambda_j$. The continuum gauge fields $\hat a_{j}$ and the microscopic gauge fields $a^z$ are related by
    \ie
    a^z= \sum_{j=1}^{\ell} e^{i q_j z}\hat a_j(z\mathbf{a})~.
    \fe    
    Using this map, the continuum theory reproduces all the physical observables considered in this paper, including the electric potential, the braiding statistics, the correlation function of Wilson lines and the electric conductivity, at long distance or low energy.  

    So far, we have summarized various properties of the gapless iCSM theory with $\Gamma=1$. The gapless iCSM theories with $\Gamma>1$ are rare in the space of gapless iCSM theory. In some sense, these theories behave more singular compared to the ones with $\Gamma=1$. This is related to the softness of the low-energy dispersion relations in the gapless iCSM theory with $\Gamma>1$.
    In a gapless iCSM theory with $\Gamma>1$, the braiding phase between gauge charges diverges as the radius of the braiding trajectory increases, and the DC longitudinal conductivity and the DC Hall conductivity both diverge.
 
The rest of the paper is organized as follows. In Section~\ref{sec:Spectrum}, we study the plane wave spectrum of an iCSM theory and introduce a map from an iCSM theory to a Laurent polynomial. In Section~\ref{sec:global_sym}, we analyze the global symmetry and the robustness of the iCSM theory. In Section~\ref{sec:electric_charged}, we couple the iCSM theory to electrically charged matters and study the electric potential and the braiding statistics between gauge charges. In Section~\ref{sec:Wilson_correlation}, we study the correlation functions of Wilson lines in the iCSM theory. In Section~\ref{sec:Hall}, we compute the longitudinal and the Hall conductivity of the iCSM theory. In Section~\ref{sec:effective_theory}, we take the continuum limit in the $z$ direction and derive a fully continuous field theory for the gapless iCSM theory with $\Gamma=1$. Appendix~\ref{sec:Polynomial} gives an alternative presentation of the action for the iCSM theory using its Laurent polynomial. Appendix~\ref{app:U(1)} reviews the calculations of various observables in 2+1D $U(1)$ gauge theory. Appendix~\ref{app:null_vector} provides more details to a statement in the main text concerning the integer null vectors of an infinite-dimensional $K$ matrix. In Appendix \ref{app:CDW}, we demonstrate the method used in Section~\ref{sec:effective_theory} in a simpler model. Specifically, we discuss a 1+1D lattice model that spontaneous breaks the translation symmetry and derive an effective continuum field theory for it. 

\section{Plane Wave Spectrum}
\label{sec:Spectrum}

The plane wave spectrum of an iCSM theory has been computed in \cite{ma2020fractonic}. We will review it below. The equation of motion of the Lagrangian \eqref{action} is
\ie\label{eq:EOM}
\frac{1}{g^2}\partial_\nu f^{I,\mu\nu}+\frac{i}{2\pi}K_{IJ} \epsilon^{\mu\nu\rho} \d_\nu a_\rho^J=0~.
\fe
If the $K$ matrix is translation invariant, the equation can be solved by plane wave solutions 
\ie\label{planewave}
a_\mu^I(k_{\tau}, k_x,k_y,q)=C_\mu e^{ik_\tau \tau+i k_x x+i k_y y+iq I}~,
\fe
where $q\sim q+2\pi$ parametrizes the eigenvectors $v_I=e^{i q I}$ of the $K$ matrix. When we view the theory as a 3+1D system, $q$ can be interpreted as the momentum in the $z$ direction and the identification $q\sim q+2\pi$ defines a Brillouin zone. When $N$ is finite, $q$ is quantized to be an integer multiple of ${2\pi}/{N}$ due to the periodic boundary condition. In the $N\rightarrow\infty$ limit, the quantization condition is relaxed and $q$ can take any value within the Brillouin zone. 

Solving the equation of motion and Wick rotating from the Euclidean signature to the Lorentzian signature, we obtain the dispersion relation
\ie\label{dispersion}
\omega^2=k_x^2+k_y^2+m(q)^2~,\quad m(q)=\frac{g^2}{2\pi}\lambda(q)~,
\fe
where $\omega =ik_\tau$ and $\lambda(q)$ is the eigenvalue of the $K$ matrix for the eigenvector $v_I=e^{iqI}$. From the dispersion relation, it is clear that the theory is gapless if and only if its $K$ matrix has zero eigenvalues. We will refer to the momenta of these zero eigenvalues as the \textit{gapless momenta} and denote them by $q_j$, where $j=1,\ldots,\ell$ labels the different gapless momenta and $\ell$ is the total number of gapless momenta.

\subsection{Tridiagonal \texorpdfstring{$K$}{} Matrices}

As an example, consider a class of theories with a tridiagonal $K$ matrix
\ie\label{tridiagonal}
K=\left(\begin{array}{cccccc}
	c_0 & c_1 & & &c_1\\
	c_1 & c_0 & c_1 & &\\
	& \ddots & \ddots & \ddots & &\\
	&  & c_1 & c_0 & c_1\\
	c_1&&&c_1 &c_0
\end{array}\right)~,
\fe
where $c_0$ and $c_1$ are integers. 
The eigenvalues of the $K$ matrix are
\ie\label{tridiagonal+eigen}
\lambda(q)=c_1\left(R+2\cos(q)\right)~,
\fe
where we define the ratio $R={c_0}/{c_1}$.

When $N$ is finite, because of the quantization condition, the momentum $q$ must be commensurate, meaning that $\frac{q}{2\pi}$ is a rational number. As $c_0,c_1$ are integers, $\cos(q)=-R/2$ must also be rational at the gapless points. Therefore, the conditions for the theory to be gapless are highly restrictive. According to Niven's theorem \cite{Niven}, $\cos(q)$ and $\frac{q}{2\pi}\in[-\pi,\pi]$ are both rational only when 
\begin{alignat}{5}
\left(\frac{q}{2\pi},\cos(q)\right)= \left(0,1\right),\left(\pm\frac{\pi}{3},\frac{1}{2}\right),\left(\pm\frac{\pi}{2},0\right), \left(\pm\frac{2\pi}{3},-\frac{1}{2}\right),\left(\pm\pi,-1\right)~.
\end{alignat}
It implies that only theories with $R=0,\pm1,\pm2$ can be gapless at finite $N$. Moreover, these theories are typically gapless only on a subsequence of $N$. For example, when $R=1$, the theory is gapless only when $N$ is divisible by 3.

In the $N\rightarrow \infty$ limit, the momentum $q$ can be both commensurate and  incommensurate. And, the theory is gapless if $-2\leq R\leq2$. The gapless momenta are at $q=\pm\arccos\left(-{R}/{2}\right)$. They are generally incommensurate, so most of the gapless iCSM theories can never be gapless at finite $N$. However, in these theories, the gap in the spectrum tends to decrease as $N$ increases and eventually vanishes in the $N\rightarrow\infty$ limit. This phenomenon is common when we take the thermodynamic limit of a lattice model or many-body system.

\subsection{Commensurate vs. Incommensurate}

What we have observed in the tridiagonal $K$ matrix example are very general. That is, not every gapless iCSM theory can be gapless when $N$ is finite and, in fact, most of them are gapless only in the $N\rightarrow\infty$ limit. This is because the gapless momenta are generically incommensurate and the momentum $q$ must be commensurate when $N$ is finite.

As we will discuss later, commensurate and incommensurate gapless momenta lead to different behaviors of the gapless iCSM theories. For example, the number of gauge invariant monopole operators in a gapless iCSM theory is the same as the number of commensurate gapless momenta rather than the total number of gapless momenta. Also, in a gapless iCSM theory that has incommensurate gapless momenta, the braiding statistics between gauge charges do not converge in the $N\rightarrow\infty$ limit, while in a gapless iCSM theory that has only commensurate gapless momenta, the braiding statistics can converge in the $N\rightarrow\infty$ limit if we follow some particular equally-spaced subsequences (Section~\ref{sec:braiding}).

\subsection{Dynamical Exponent}
\label{sec:exponent}

Given a gapless iCSM theory, it is natural to zoom in to the low-energy states. At low energy, the plane wave states split into multiple sectors, each of which centers around a gapless momentum.  For each gapless momentum $q_j$,  we can expand the dispersion relation around it and obtain:
\ie\label{eta_def}
\omega^2=k_x^2+k_y^2+m_j^2(q-q_j)^{2\gamma_j}~,\quad m_j=\frac{g^2}{2\pi}\lambda_j~.
\fe
We define the integer $\gamma_j$ to be the \emph{dynamical exponent} of the gapless momentum $q_j$. 
The low-energy dispersion relation \eqref{eta_def} obeys a Lifshitz scale symmetry:
\ie
\omega\rightarrow\Lambda  \omega,\quad k_x\rightarrow\Lambda k_x,\quad k_y\rightarrow \Lambda k_y ,\quad (q-q_j) \rightarrow \Lambda^{1/\gamma_j}(q-q_j)~.
\fe
When $\gamma_j=1$, the low-energy dispersion relation becomes linear and the scale symmetry becomes the standard isotropic scale symmetry.

We denote the maximal dynamical exponent of a gapless iCSM theory by $\Gamma=\text{max}(\gamma_j)$. A generic gapless iCSM theory has $\Gamma=1$ but in some particular cases we can have $\Gamma>1$. For example, theories with a tridiagonal $K$ matrix all have $\Gamma=1$ except when $R=\pm2$, in which case $\Gamma=2$. 

As we will discuss later, in some sense the $\Gamma=1$ theories are more regular compared to the $\Gamma>1$ theories. Many observables, such as the braiding statistics and the DC conductivity, diverge in the $\Gamma>1$ theories but are finite in the $\Gamma=1$ theories. This is because the low-energy dispersion relations in the $\Gamma>1$ theories are softer compared to the linear low-energy dispersion relations in the $\Gamma=1$ theories. Another consequence of the linear low-energy dispersion relations in the $\Gamma=1$ theories is that such theories develop an isotropic scale symmetry at low energy and their low-energy/long-distance effective field theory descriptions are expected to contain only dimensionless couplings. We will discuss these low-energy/long-distance effective field theories in Section~\ref{sec:effective_theory}. In Table~\ref{tab:compare}, we summarize and compare various properties of the $\Gamma=1$ theories and the $\Gamma>1$ theories. We will discuss them in detail in Sections~\ref{sec:electric_charged}, \ref{sec:Wilson_correlation} and \ref{sec:Hall}.

\subsection{Laurent Polynomial}\label{sec:Laurent}

We now introduce a compact and useful way to encode the translation invariant $K$ matrix \eqref{Kmatrix} using a \emph{Laurent polynomial} $p(u)$. The Laurent polynomial is defined as
\ie\label{polynomial}
p(u)=c_0+\sum^{\xi}_{k=1}c_k(u^k+u^{-k})~,
\fe
where $\xi$ is the maximum $k$ such that $c_k$ is non-zero.
The Laurent polynomial and the eigenvalues of the $K$ matrix are related by $\lambda(q)=p(e^{iq})$.

Given a Laurent polynomial, we can factorize it into
\ie\label{eq:poly_root}
p(u)= c_\xi u^{-\xi}\prod_{j=1}^{L}(u-u_j)^{\gamma_j}~.
\fe
Here, $u_j$ are the distinct roots of the Laurent polynomial. They are labeled by $j=1,\ldots, L$ where $L$ is the number of the distinct roots. $\gamma_j$ is the multiplicity of the root $u_j$ and $\sum_{j=1}^L \gamma_j=2\xi$. Recall that the iCSM theory is gapless if and only if its $K$ matrix has zero eigenvalues, which are  $\lambda(q)=p(e^{iq})$. Therefore, the iCSM theory is gapless if and only if its Laurent polynomial has roots on the unit circle, i.e.\ there are roots that take the form $u_j=e^{iq_j}$ with $q_j$ the gapless momenta. The number of gapless momenta or distinct roots on the unit circle is denoted by $\ell$.

Since $p(u)=p(u^{-1})$, if $u_j$ is a root, $u_j^{-1}$ should also be a root. This implies that the eigenvalues $\lambda(q)=p(e^{iq})$ are symmetric in $q$ and that if $q_j$ is a gapless momentum, $-q_j$ should also be a gapless momentum.

The dynamical exponent of a gapless momentum $q_j$ is given by the multiplicity of the corresponding root $u_j=e^{iq_j}$. We can see this explicitly by expanding the eigenvalue $\lambda(q)=p(e^{iq})$ around $q_j$. This gives
\ie
\lambda(q)= \hat{p}(e^{iq}) (e^{iq}-e^{i q_j})^{\gamma_j}=\lambda_j(q-q_j)^{\gamma_j}+\cdots~,
\fe
where $\lambda_j=\hat p(e^{i q_j})e^{i\gamma_j (q_j+\pi/2)}$. In the first equality, we split $p(u)$ into $(u-u_j)^{\gamma_j}$ and $\hat p(u)$, a polynomial that does not vanish at $u=u_j$.

The roots generally take different values so most of the gapless iCSM theories have $\Gamma=1$. However, in some special cases, there can be repeated roots and we have $\Gamma>1$. For example, consider the Laurent polynomial
\ie
p(u)=(u+1+u^{-1})^\gamma=u^{-\gamma}\left(u-e^{2\pi i/3}\right)^\gamma\left(u-e^{-2\pi i/3}\right)^\gamma~.
\fe
It has roots at $u=e^{\pm 2\pi i/3}$ with a multiplicity $\gamma$ and the corresponding iCSM theory has $\Gamma=\gamma$.  Note that even though these cases are sporadic, fine tuning is not necessary to reach them because the $K$ matrix is a discrete integer matrix.

\section{Global Symmetries and Robustness}
\label{sec:global_sym}

In this section, we analyze the global symmetry and the robustness of the gapless iCSM theory. We focus primarily on the continuous symmetry. We will first review the story in 2+1d $U(1)$ gauge theory, and then generalize it to theory with a finite-dimensional $K$ matrix and finally to iCSM theories. For iCSM theory, we will discuss it from two complementary perspectives: one treats the theory as a 3+1D system and the other treats the theory as a 2+1D system with an infinite number of gauge fields.

When viewing the gapless iCSM theory as a 3+1D system, it has an \textit{exotic continuous one-form global symmetry}, which acts on both the Wilson lines and the string-like gauge invariant monopole operators that extend in the $z$ direction. The exotic global symmetry has a 't Hooft anomaly that obstructs the gauging of the global symmetry. It is spontaneously broken in the sense that the charged operators obey perimeter law. The spontaneous breaking of this exotic continuous one-form global symmetry can lead to multiple Goldstone modes that are gapless at non-zero momenta in the $z$ direction.

When viewing the gapless iCSM theory as a 2+1D system, the exotic continuous one-form global symmetry reduces to a $U(1)^{\ell_c}$ magnetic zero-form symmetry and a $U(1)^{\ell_c}\times \mathbb{R}^{\ell-\ell_c}$ electric one-form symmetry in 2+1D, where $\ell_c$ is the number of commensurate gapless momenta and $\ell-\ell_c$ is the number of incommensurate gapless momenta. The monopole operators are now local operators charged under the magnetic zero-form symmetry.

After analyzing the global symmetries, we argue that the gapless iCSM theory is robust. A theory is robust if it has no relevant local operators (See \cite{Seiberg:2020bhn}, for example, for a review on robustness in high energy physics and in condensed matter physics). Such a theory can emerge as an effective IR theory of some UV systems without fine-tuning because small deformations in the UV cannot trigger any nontrivial renormalization group flow away from the IR theory due to the lack of relevant operators. There is also a weaker notion of robustness enriched by global symmetry. A theory is robust when a global symmetry $G$ is imposed if it has no $G$-symmetric relevant local operators. Such a theory can emerge as an effective IR theory without fine-tuning when the $G$ symmetry is imposed in the UV. 

\subsection{Review of \texorpdfstring{$U(1)$}{} Gauge Theory}

\subsubsection{Maxwell Theory}\label{sec:U(1)_no_CS}

We begin by reviewing 2+1D Maxwell theory. The theory has a $U(1)$ electric one-form symmetry that acts on the Wilson lines \cite{Gaiotto_2015}. The symmetry is not spontaneously broken because the potential between two gauge charges is logarithmically confined
\ie
V(r)=g^2\int\frac{d^2 \vec{k}}{(2\pi)^2} \frac{e^{i\vec{k}\cdot \vec{r}}}{\vec{k}^2}
=
-\frac{g^2}{2\pi}\log(r)+\text{const}~,
\fe
and hence the expectation values of the Wilson loops decay faster than any perimeter law. We review this calculation in Appendix \ref{app:U(1)}. The theory also has a $U(1)$ magnetic zero-form symmetry generated by the conserved current $J_\mu = \frac{1}{2\pi}\epsilon_{\mu\nu\rho}\partial^\nu a^\rho$. The current is trivially conserved $\partial_\mu J^\mu=0$ due to the Bianchi identity $\epsilon^{\mu\nu\rho}\partial_\rho f_{\mu\nu}=0$. The charged operators are the magnetic monopole operators. A minimally charged monopole operator is defined by removing a point in the spacetime and then inserting one unit of flux of the $U(1)$ gauge field on the sphere surrounding that point \cite{Borokhov:2002ib}. 

It is easier to think about the monopole operators in a dual description of the $U(1)$ gauge theory. To dualize the theory, we treat $f_{\mu\nu}$ as an independent two-form field, rather than the field strength of a $U(1)$ gauge field $a_\mu$, and include a Lagrange multiplier in the Lagrangian to impose the Bianchi identity:
\ie\label{Lagrange_mul}
\mathcal{L}\supset\frac{1}{2}\frac{i}{2\pi}\epsilon^{\mu\nu\rho} \varphi \partial_\rho f_{\mu\nu} ~.
\fe
The Lagrange multiplier is a compact boson $\varphi\sim \varphi+2\pi$.
The coefficient of  \eqref{Lagrange_mul} is fixed such that integrating out $\varphi$  constrains the flux of $f_{\mu\nu}$ to be integral. 
After integrating out $f_{\mu\nu}$, we obtain the dual Lagrangian of a free compact boson
\ie\label{eq:compact_boson}
\mathcal{L}=\frac{1}2\frac{g^2}{4\pi^2}\partial_\mu\varphi\partial^\mu\varphi~.
\fe
In the dual description, the $U(1)$ magnetic symmetry acts as $\varphi\rightarrow \varphi +c$ and the basic monopole operator is mapped to $\exp(i\varphi)$. The two point function of the monopole operators approaches a non-zero constant at long distance\footnote{Throughout the paper, when we discuss the monopole operator $\exp(i\varphi)$, we always consider the normal-ordered operator $:\exp(i\varphi):$. If the operator were not normal-ordered, the two-point function \eqref{eq:twopoint_monopole} has an additional constant factor $\exp\left(-\pi/(g^2\varepsilon)\right)$ where $\varepsilon$ is a short distance cutoff. With this addition constant factor, the correlation function is always less 1.}
\ie\label{eq:twopoint_monopole}
\langle \exp(i\varphi)\exp(-i\varphi)\rangle=\exp\left(\frac{4\pi^2}{g^2}\int \frac{d ^3k}{(2\pi)^3}\frac{e^{ik x}}{k^2}\right)=\exp\left(\frac{\pi}{g^2 r}\right)\xrightarrow{r\rightarrow\infty} 1~.
\fe
It implies that the $U(1)$ magnetic symmetry is spontaneously broken and that $\varphi$ is the corresponding Goldstone boson.

The $U(1)$ electric one-form symmetry and $U(1)$ magnetic zero-form symmetry has a mixed 't Hooft anomaly, which is an obstruction to gauging both symmetries simultaneously. It can be seen by coupling the theory to the background gauge fields $A\sim A+d\alpha$ and $B\sim B+d\beta$ for the zero-form and one-form symmetry respectively. The Lagrangian becomes
\ie
\mathcal{L}=\frac{1}{4g^2}(f_{\mu\nu}-B_{\mu\nu})(f^{\mu\nu}-B^{\mu\nu})+\frac{1}{2}\frac{i}{ 2\pi}\epsilon^{\mu\nu\rho} A_{\mu} f_{\nu\rho} ~. 
\fe
The background gauge symmetry also acts on the dynamical gauge field as $a\sim a+\beta$. Under the background gauge transformation, the Lagrangian is shifted by
\ie\label{eq:anomaly_U(1)}
\mathcal{L}\rightarrow\mathcal{L}+\frac{i}{2\pi}\epsilon^{\mu\nu\rho} A_\mu \partial_\nu \beta_\rho~.
\fe
Here we dropped the terms that are total derivatives.
The anomalous background gauge transformation can be canceled by coupling the theory to a 3+1D bulk via the anomaly inflow mechanism \cite{Callan:1984sa}. The bulk is a symmetry-protected topological (SPT) phase described by the classical Lagrangian 
\ie
\mathcal{L}_{\text{SPT}} =-\frac{1}{2}\frac{i}{2\pi} \epsilon^{\mu\nu\rho\sigma} B_{\mu\nu}\partial_{\rho} A_\sigma~.
\fe
The SPT action is gauge invariant on closed manifolds. But on open manifolds, the gauge transformation shifts it by a boundary term that exactly cancels the anomalous transformation \eqref{eq:anomaly_U(1)} of the boundary theory. Since the SPT action is nontrivial on closed manifolds, the anomaly cannot be canceled by adding a counterterm on the boundary.

We now turn to the issue of robustness. We can deform the Lagrangian by the monopole operator. This creates a relevant $\cos(\varphi)$ potential that gaps out the system. This is the renowned Polyakov mechanism \cite{Polyakov:1976fu} that renders the $U(1)$ gauge theory not robust. We can make the theory robust by imposing the $U(1)$ magnetic symmetry. This excludes the monopole operators in the Lagrangian.

\subsubsection{Adding a Chern-Simons Coupling}

We can add to the $U(1)$ gauge theory a nontrivial Chern-Simons coupling 
\ie
\mathcal{L}_{\text{CS}}=\frac{i}{4\pi}K\epsilon^{\mu\nu\rho}a_\mu\partial_\nu a_\rho~.
\fe
This generates a mass $m=\frac{g^2|K|}{2\pi}$ for the photon and therefore gaps the theory.

Because of the Chern-Simons coupling, the monopole operators are not gauge invariant \cite{Affleck:1989qf,Borokhov:2002ib}. The basic monopole operator carries an electric charge $K$ under the $U(1)$ gauge field. It can be seen as follows. Let us place the theory on a sphere with one unit of magnetic flux inserted. Meanwhile, we also insert a charge $Q$ Wilson line in the time direction as a defect, which modifies the Gauss law given by the equation of motion of $a_0$. Integrating the Gauss law over the sphere, we obtain the constraint that the Hilbert space is gauge invariant only if 
\ie
Q=K\oint_{S^2} \frac{f_{12}}{2\pi}=K~.
\fe
By the state-operator correspondence, a basic monopole operator that sources one unit of magnetic flux carries an electric charge $K$. In order to make it gauge invariant, we need to attach to it a charge $K$ Wilson line. Since the monopole operators are not gauge invariant unless Wilson lines are attached to them, the $U(1)$ magnetic symmetry does not act faithfully on the gauge invariant local operators.

The Chern-Simons coupling also breaks the electric one-form symmetry from $U(1)$ to $\mathbb{Z}_K$.\footnote{For odd $K$, the $U(1)_K$ Chern-Simons theory is a spin TQFT so it has an additional $\mathbb{Z}_2$ one-form symmetry generated by the transparent fermion line.} The one-form symmetry shifts the gauge field $a$ by a flat $\mathbb{Z}_K$ gauge field. The one-form symmetry operators are
\ie
\exp\left[i\oint\left(a-\frac{i\pi}{g^2 K}\star f\right)\right]~,
\fe
where $\star f$ is the Hodge dual of $f$.
The one-form symmetry is spontaneously broken because the electric potential between two gauge charges decays exponentially at long distance
\ie
V(r)=g^2\int\frac{d^2 \vec{k}}{(2\pi)^2} \frac{e^{i\vec{k}\cdot \vec{r}}}{\vec{k}^2+m^2}
=
\frac{g^2}{2\pi} K_0(mr)\xrightarrow{r\rightarrow\infty}g^2 e^{-mr}~,
\fe
and hence the Wilson lines obey a perimeter law.
We review the calculation in Appendix~\ref{app:U(1)}.

We now discuss the robustness of the theory in the presence of a nontrivial Chern-Simons coupling. Since the monopole operators are not gauge invariant, they cannot be added to the Lagrangian. We can make them gauge invariant by attaching  Wilson lines to them, but the resulting operators are extended operators, which again cannot be added to the Lagrangian. Hence, in contrast to the theory without a Chern-Simons coupling, the theory with a nontrivial Chern-Simons coupling is robust.

\subsection{Theory with a Finite-dimensional \texorpdfstring{$K$}{} Matrix}
\label{sec:CSN_monopole}

\subsubsection{Global Symmetry and the Smith Normal Form}\label{sec:SNF}

We now generalize the discussions to theories with finite dimensional $K$ matrices. In order to determine the global symmetry of such a theory, it is useful to study the Smith normal form of the $K$ matrix. Recall that any integer matrix $K$ can be put into the Smith normal form using two $\text{GL}(N,\mathbb{Z})$ matrices $W$ and $V$:
\ie\label{eq:SNF_def}
D=VKW=\left(\begin{array}{ccccccc}
	{0}_{\ell\times \ell} &\\
	 & d_1 &   \\
	&  & \ddots &   \\
	&  &  &  d_{N-\ell} 
\end{array}\right)~,
\fe
where $d_j$ are integers and $d_j$ divides $d_{j+1}$ for all $1\leq j<N-\ell$.\footnote{Strictly speaking, the diagonal $0$'s should be at the lower corner of the Smith Normal form. Here, we do a permutation to bring the $0$'s to the upper corner for later convenience.} These $d_j$ are called the invariant factors of the $K$ matrix. Here, $\ell$ is the number of null vectors of $K$.
As $\text{GL}(N,\mathbb{Z})$ matrices, $W$ and $V$ are integer $N\times N$ matrices that are invertible over the integers. The Smith normal form gives a complete solution to the following equation
\ie
\sum_{J} K_{IJ}w_J = 0 \mod 1~.
\fe
The solution takes the form 
\ie\label{eq:Smith_int_sol}
w_I = \sum_{j=1}^\ell p_j W_{I,j}+\sum_{j=1}^{N-\ell}\frac{p_{j+\ell}}{d_j}W_{I,j+\ell}~,
\fe
where $p_j$ are real numbers for $j=1,\ldots,\ell$ and integers for $j=\ell+1,\ldots,N$.

With the above preparation, we are ready to discuss the global symmetry. Let us first consider the electric one-form symmetry $a^I\rightarrow a^I+\beta^I$. For the Maxwell term to be invariant, $\beta^I$ has to be a flat gauge field. It is further constrained by the Chern-Simons coupling
\ie
\mathcal{L}_{\text{CS}}=\frac{i}{4\pi} K_{IJ}\epsilon^{\mu\nu\rho}a^I_\mu\partial_\nu a^J_\rho~,
\fe
which transforms as
\ie
\mathcal{L}_{\text{CS}}\rightarrow\mathcal{L}_{\text{CS}}+\frac{i}{2\pi} K_{IJ}\epsilon^{\mu\nu\rho}\beta^I_\mu\partial_\nu a^J_\rho+\frac{i}{4\pi} K_{IJ}\epsilon^{\mu\nu\rho}\beta^I_\mu\partial_\nu \beta^J_\rho~.
\fe
The shift in the Chern-Simons coupling is trivial, i.e.\ it integrates to an integer multiple of $2\pi i$, if $ \sum_{J}K_{IJ}\beta^J$ has only $2\pi\mathbb{Z}$-valued holonomies. Using \eqref{eq:Smith_int_sol}, such $\beta^I$ take the form 
\ie
\beta^I = \sum_{j=1}^\ell \kappa_j W_{I,j}+\sum_{j=1}^{N-\ell}\kappa_{j+\ell} W_{I,j+\ell}~,
\fe
where $\kappa_j$ is a flat $U(1)$ gauge field for $j=1,\ldots,\ell$ and a flat $\mathbb{Z}_{d_i}$ gauge field for $j=\ell+1,\ldots,N$. It implies that the electric one-form symmetry is a $U(1)^\ell \times \prod_{j=1}^{N-\ell}\mathbb{Z}_{d_j}$ one-form symmetry. The relation between the electric one-form symmetry and the Smith normal form of the $K$ matrix was also discussed in \cite{Pace:2022cnh}.\footnote{The Smith normal form was also featured in a recent construction of fracton models on graphs \cite{Gorantla:2022mrp,Gorantla:2022pii}.}

We now discuss the magnetic zero-form symmetry. Naively, the symmetry is a $U(1)^N$ symmetry generated by the conserved current $J^I_\mu=\frac{1}{2\pi}\epsilon_{\mu\nu\rho}\partial^\nu a^{I,\rho}$. However, this symmetry does not act faithfully on the local operators, specifically the magnetic monopole operators. We can label the monopole operators by their integer-valued magnetic charge vector $M_I=\oint dx dy\,  J_\tau^I$. Because of the Chern-Simons coupling, the monopole operators carry electric charges $Q_I=\sum_J K_{IJ}M_J$ under the gauge field $a_I$. They are not gauge invariant unless their magnetic charge vector $M_I$ is a null vector of the $K$ matrix. Such an integer-valued null vector takes the form 
\ie
M_I=\sum_{j=1}^\ell m_j W_{I,j}~,
\fe
where $m_j$ are integers and $W$ is the $\text{GL}(N,\mathbb{Z})$ matrix in the Smith normal form \eqref{eq:SNF_def}. This means that all the gauge invariant monopole operators can be generated by the $\ell$ basic ones with magnetic charge vector $M_I=W_{I,j}$ where $j=1,\ldots ,\ell$. Therefore, the faithful magnetic zero-form symmetry is the $U(1)^\ell$ subgroup of the naive $U(1)^N$ symmetry.

In summary, the theory has a $U(1)^\ell\times \prod_{j=1}^{N-\ell} \mathbb{Z}_{d_j}$ electric one-form symmetry and a $U(1)^\ell$ magnetic zero-form symmetry. Note that the number of $U(1)$ factors in the electric one-form symmetry or the magnetic zero-form symmetry is identical to the number of gapless modes. We will explain this equality below.

\subsubsection{A Convenient Change of Basis}

We can perform a change of basis  $a\mapsto W a$ using the $\text{GL}(N,\mathbb{Z})$ matrix $W$ of the Smith normal form \eqref{eq:SNF_def}. This brings the $K$ matrix into the form
\ie\label{eq:change_of_basis}
K\mapsto W^{T}KW =\left(\begin{array}{ccccccc}
	{0}_{\ell\times \ell} & \\
	& \hat K\\
\end{array}\right)~,
\fe
where $\hat K$ is a non-degenerate integer symmetric matrix of dimension $N-\ell$. Let us explain why the matrix $W^{T}KW$ takes the above form. Since $W^T KW= W^TV^{-1} D$ and $D$ is the diagonal matrix of the Smith normal form \eqref{eq:SNF_def}, $W^T KW$ should be an upper triangular block matrix
\ie
W^T KW= \left(\begin{array}{ccccccc}
	{0}_{\ell\times \ell} & \check{K}_{\ell\times (N-\ell)}\\
	 & \hat K  \\
\end{array}\right)~.
\fe
The off-diagonal matrix $\check{K}$ vanishes because $W^TKW$ is a symmetric matrix. Therefore, $W^TKW$ takes the form \eqref{eq:change_of_basis}. Here we only consider a $\text{GL}(N,\mathbb{Z})$ transformation because it preserves the independent compactness of the $U(1)$ gauge fields. In other words, the fluxes of the gauge fields remain integer-valued after the basis transformation.

In the new basis, the Lagrangian is
\ie\label{action_new_basis}
\mathcal{L}\mapsto\mathcal{L}=\frac{1}{4g^2}\sum_{I,J=1}^N Z_{IJ}{f}^I_{\mu\nu}{f}^{J,\mu\nu}+\frac{i}{4\pi} \sum_{I,J=1}^{N-\ell}\hat K_{IJ}\epsilon^{\mu\nu\rho}{a}_\mu^{I+\ell}\partial_\nu {a}_\rho^{J+\ell}~,
\fe
where $Z=W^TW$. The first $\ell$ gauge fields $a^1,a^2,\ldots,a^\ell$ do not participate in the Chern-Simons coupling. Quantizing their fluctuations reproduces the $\ell$ gapless modes. The new basis makes the $U(1)^\ell$ subgroup of the electric one-form symmetry and the $U(1)^\ell$ magnetic zero-form symmetry manifest.  In the new basis, the $U(1)^\ell$ one-form symmetry acts only on the first $\ell$ gauge fields and shifts them by flat $U(1)$ gauge fields. It is not spontaneously broken as in the Maxwell theory. The gauge-invariant monopoles in the new basis are those that source only fluxes of the first $\ell$ gauge fields and so the faithful $U(1)^\ell$ magnetic zero-form symmetry is generated by the first $\ell$ conserved currents $J^I_\mu=\frac{1}{2\pi}\epsilon_{\mu\nu\rho}\partial^\nu a^{I,\rho}$ where $I=1,\ldots,\ell$. We will postpone the discussions on the spontaneous symmetry breaking of the $U(1)^\ell$ magnetic zero-form symmetry to the next subsection. Following the same analysis as in Section~\ref{sec:U(1)_no_CS}, we can couple the $U(1)^\ell$ zero-form and one-form symmetry to background gauge fields and show that there is a mixed anomaly between them. The discrete one-form global symmetry is given by the invariant factors of $\hat K$. When we place the theory on a torus, the ground state degeneracy is $|\det(\hat K)|$. Similar result was also noticed in \cite{Sullivan:2021rbk} through the coupled wire construction. We will present a formula for this ground state degeneracy for a translation invariant $K$ matrix in a forthcoming paper \cite{appear}.

We emphasize that although a translation invariant $K$ matrix can be diagonalized using the Fourier transform, the transformation is typically not a $\text{GL}(N,\mathbb{Z})$ transformation. As a result, the gauge fields after the Fourier transform typically have correlated fractional fluxes. As an example, consider a $2\times2$ $K$ matrix that describes a gapless bilayer quantum Hall state (see for example \cite{PhysRevB.47.2265}):
\ie\label{eq:bilayer}
K=\left(\begin{array}{ccccccc}
	1\ & \, 1\\
	1\ & \, 1\\
\end{array}\right)~.
\fe
The matrix is diagonalized by a Fourier transform:
\ie
W=\frac{1}{\sqrt{2}}\left(\begin{array}{ccccccc}
	+1\ & \,1\\
	-1\ & \,1\\
\end{array}\right),\quad W^TKW=\left(\begin{array}{ccccccc}
	0\ &\\
	& \, 2 \\
\end{array}\right)~.
\fe
It is clearly not a $\text{GL}(2,\mathbb{Z})$ transformation because $W$ is not an integer matrix. It is tempting to replace $W$ by $\sqrt{2}W$ so that the transformation matrix is an integer matrix, but again $\sqrt{2}W$ is not a $\text{GL}(2,\mathbb{Z})$ matrix because its inverse is not an integer matrix. The correct $\text{GL}(2,\mathbb{Z})$ transformation that realizes \eqref{eq:change_of_basis} is
\ie
W=\left(\begin{array}{ccccccc}
	+1\ & \,0\\
	-1\ & \,1\\
\end{array}\right),\quad W^TKW=\left(\begin{array}{ccccccc}
	0\ &\\
	& \,1\\
\end{array}\right)~.
\fe
The first column vector of $W$ is the null vector of $K$. The $\text{GL}(2,\mathbb{Z})$ transformation correctly predicts that after gapping out the gapless mode, the system goes into the integer quantum Hall state with $K=1$.

\subsubsection{Duality and Goldston Bosons}\label{sec:duality}

It is illuminating to dualize the first $\ell$ gauge fields, which do not participate in the Chern-Simons coupling, to compact bosons. We will do this explicitly for the case where $\ell=1$. To dualize the gauge field $a_\mu^1$, we treat the field strength $f^1_{\mu\nu}$ as an independent two-form field and introduce a compact scalar field $\varphi\sim\varphi+2\pi$ as a Lagrange multiplier that imposes the Bianchi identiy for $f^1_{\mu\nu}$. This adds to the Lagrangian the term
\ie
\mathcal{L}\supset\frac{1}{2}\frac{i}{2\pi}\epsilon^{\mu\nu\rho} \varphi\partial_\rho f^1_{\mu\nu}~.
\fe 
Integrating out $f^1_{\mu\nu}$, we obtain the dual Lagrangian
\ie\label{action3}
\mathcal{L}=&\,\frac{1}{2}\frac{g^2}{4\pi^2Z_{11}}\partial_\mu\varphi\partial^\mu\varphi+\frac{1}{2}\frac{i}{2\pi }\sum_{I=2}^N\frac{Z_{1I}}{Z_{11}}\epsilon^{\mu\nu\rho}f_{\mu\nu}^I \partial_\rho\varphi
\\
&+\frac{1}{4 g^2}\sum_{I,J=2}^N\hat Z_{IJ}{f}_{\mu\nu}^I{f}^{J,\mu\nu}+\frac{i}{4\pi} \sum_{I,J=2}^N\hat K_{IJ}\epsilon^{\mu\nu\rho}{a}_\mu^I\partial_\nu {a}_\rho^J~,
\fe
where $\hat Z_{IJ}$ is
\ie
\hat Z_{IJ}=Z_{IJ}-\frac{Z_{1I}Z_{1J}}{Z_{11}}~,\quad 2\leq I,J\leq N~.
\fe
Note that the sums over $I,J$ in \eqref{action3} are restricted to $2,\ldots, N$. The second term in the first line of \eqref{action3} is a topological term similar to the $\theta$-term in 3+1D $U(1)$ gauge theory. It is a total derivative but is still nontrivial because both $\varphi$ and $a_\mu^I$ are compact fields that can have nontrivial transition functions. The topological term does not affect the plane wave spectrum and the correlation functions involving only $\varphi$ and $a_\mu^I$. However, it does affect other observables such as the correlation functions involving the vortex line of $\varphi$. In the original duality frame, the vortex line is mapped to the Wilson line $\exp(i\oint a^1)$ and the coefficient of the topological term $Z_{I1}$ appears in the Maxwell term, which directly affects the Green's function of $a_\mu^1$ and the correlation functions involving $\exp(i\oint a^1)$. 

In the dual description \eqref{action3}, the faithful $U(1)$ magnetic symmetry acts as $\varphi\rightarrow\varphi+c$ and the charged gauge invariant monopole operator is mapped to $\exp(i\varphi)$. Since the topological term does not affect the correlation functions involving $\varphi$,  the two-point function of the monopole operator is the same as in a free compact boson theory
\ie\label{eq:monopole_correlation}
\langle\exp(i\varphi)\exp(-i\varphi)\rangle=\exp\left(\frac{ Z_{11}\pi}{g^2r}\right)\xrightarrow{r\rightarrow\infty}1~.
\fe
It approaches a non-zero constant at long distance implying that the $U(1)$ magnetic symmetry is spontaneously broken and that $\varphi$ is the corresponding Goldstone boson. 

The above discussions generalize straightforwardly to situations with $\ell>1$. There, we can dualize the first $\ell$ gauge fields to $\ell$ compact bosons, and the faithful $U(1)^\ell$ magnetic symmetry is also spontaneously broken with the $\ell$ compact bosons as the Goldstone bosons.

\subsubsection{Robustness}\label{sec:CSN_robust}

We now discuss the robustness of the theory. If the theory is gapped, it has no gauge-invariant monopoles and is therefore robust. For a gapless theory, let us first consider the case with $\ell=1$. We can deform the Lagrangian by the relevant gauge invariant monopole operator $\exp(i\varphi)$. This gaps out $\varphi$ in the IR and the remaining Lagrangian is the second line of \eqref{action3}. Since $\varphi$  couples to the other fields only via a topological term, removing it from the Lagrangian does not change the gapped part of the spectrum but only removes the gapless mode. Similar phenomenon appears in theories with $\ell>1$. Adding all the $\ell$ basic gauge invariant monopoles to the Lagrangian gaps out the $\ell$ gapless modes and leaves the gapped part of the spectrum intact. In conclusion, the gapless theories are not robust due to the monopole operators.

We can forbid the monopole operators in the Lagrangian and  make the theory robust by imposing the $U(1)^\ell$ magnetic symmetry microscopically. For example, in the bilayer quantum Hall system described by the $K$ matrix \eqref{eq:bilayer}, the $U(1)^2$ magnetic symmetry of the low-energy Chern-Simons theory is realized microscopically as the conservation of electron number per layer. Because of this microscopic symmetry, the bilayer quantum Hall system can have a robust gapless phase. We can gap this phase by allowing electrons to tunnel between the two layers  \cite{PhysRevB.47.2265}. This effectively introduces a monopole operator with magnetic charge $(1,-1)$ to the Lagrangian and breaks the $U(1)^2$ symmetry to the diagonal $U(1)$.

\subsection{The iCSM Theory}\label{sec:iCS_global_sym}

\subsubsection{Exotic One-form Global Symmetry}

We now move on to discuss the iCSM theory mainly focusing on their continuous global symmetries.\footnote{We thank Shu-Heng Shao for raising a related question that inspires this subsubsection.} We will first discuss it from the perspective of viewing the theory as a 3+1D system. In this case, the flavor index $I$ behaves as the coordinate for the emergent dimension, which we will denote by a different variable $z$ emphasizing the change of perspective. The theory has an exotic continuous one-form global symmetry. The conserved current is a two-form current $J_{[ab]}$:\footnote{We use the Greak alphabets $\mu,\nu,\rho,\ldots$ for the coordinates $\tau, x, y$ and the English alphabets $a,b,c,\ldots$ for the coordinates $\tau,x,y,z$.}
\ie\label{eq:two-form current}
&J_{\tau x}(z)=\frac{i}{g^2}f_{\tau x}^z~,\quad J_{\tau y}(z)=\frac{i}{g^2}f_{\tau y}^z~,\quad J_{\tau z}(z)=\frac{1}{2\pi}f^{z}_{xy}~,
\\
&J_{xy}(z)=\frac{i}{g^2}f_{xy}^z~,\quad J_{yz}(z)=\frac{1}{2\pi}f^{z}_{\tau x}~,\quad J_{xz}(z)=\frac{1}{2\pi}f^{z}_{y\tau}~.
\fe
Define an operator $\Delta$ as
\ie
\Delta J(z)= \sum_{k} c_k J(z+k)~,
\fe
where $c_k$ are the entries of the infinite-dimensional $K$ matrix \eqref{Kmatrix}.
The action of $\Delta$ on $J(z)$ is the same as thinking of $J(z)$ as an infinite-dimensional vector and acting by the infinite-dimensional $K$ matrix on it.
The current $J_{[ab]}$ obeys a conservation equation
\ie\label{eq:current_conservation}
&\partial_\tau J_{\tau z}+\partial_x J_{x z}+\partial_y J_{y z}=0~,
\\
&\partial_\tau J_{\tau x}+\partial_y J_{y x}-\Delta J_{z x}=0~,
\\
&\partial_\tau J_{\tau y}+\partial_x J_{x y}-\Delta J_{z y}=0~,
\fe
and a difference condition
\ie\label{eq:difference_condition}
\partial_x J_{x \tau}+\partial_y J_{y \tau}-\Delta J_{z \tau}=0~.
\fe
The first conservation equation is due to the Bianchi identity. The last two conservation equations and the difference condition follow from the equation of motion \eqref{eq:EOM}. Note that the current conservation equation \eqref{eq:current_conservation} and the different condition \eqref{eq:difference_condition} are similar to the ones for a 3+1D $U(1)$ one-form symmetry \cite{Gaiotto_2015} if we replace the operator $\Delta$ by $-\partial_z$. This is why we call this global symmetry  an exotic continuous one-form symmetry.

What are the conserved charges of this exotic symmetry? To answer this, let us first study
the space of bounded solutions to the equation $\Delta f(z)=0$. It is equivalent to studying the space of bounded null vectors of the infinite-dimensional $K$ matrix. This vector space is spanned by $\ell$ vectors:\ $\{v_j(z)=e^{iq_j z}\,|\,j=1,\ldots,\ell\}$, where $q_j$ are the gapless momenta and $\ell$ is the number of them. We arrange the $\ell$ gapless momenta such that the first $\ell_c$ of them are commensurate and the remaining are incommensurate. In Appendix \ref{app:null_vector}, we prove that the integer-valued bounded null vectors form an $\ell_c$-dimensional lattice within the \emph{commensurate subspace} spanned by the commensurate null vectors $\{v_j(z)\,|\,j=1,\ldots,\ell_c\}$. The basis vectors of this sublattice, denoted by $\{w_j(z)\,|\,j=1,\ldots,\ell_c\}$, are integer-valued \emph{periodic} vectors obeying $\Delta w_j(z)=0$. These basis vectors are also the basis vectors of the commensurate subspace. For the \emph{incommensurate subspace} spanned by the incommensurate null vectors $\{v_j(z)\,|\,j=\ell_c+1,\ldots,\ell\}$, we can construct a set of real-valued basis vectors denoted by $\{w_j(z)\,|\,j=\ell_c+1,\ldots,\ell\}$. If $v_j(z)$ is an incommensurate null vector, $v_j(z)^*$ is also an incommensurate null vector. We can then take $v_j(z)+v_j^*(z)$ and $i(v_j(z)-v_j^*(z))$ to be our real-valued basis vectors. When $v_j(z)$ is real or imaginary, one of these two vectors vanishes. In this case, we include only the nontrivial one to our real-valued basis vectors. The $\ell$ real-valued vectors $\{w_j\,|\, j=1,\ldots,\ell\}$ form a basis for the whole bounded null vector space.

After the above preparation, we are ready to discuss the conserved charges. We will work with periodic boundary conditions in the $x,y,z$ directions. There are two types of conserved charges. The first type of charges are integer-valued magnetic charges integrated over the $xy$-plane:
\ie\label{eq:magnetic_charges}
M(z) = \int dx dy \, J_{\tau z}(z)~.
\fe 
Because of the difference condition \eqref{eq:difference_condition}, they are subject to a constraint
\ie\label{eq:constraint_magnetic}
\Delta M(z)=\int dx dy \left[\partial_x J_{x\tau}(z)+\partial_y J_{y\tau}(z)\right]=0~.
\fe
The integer-valued bounded solutions to \eqref{eq:constraint_magnetic} take the form
\ie
M(z)=\sum_{j=1}^{\ell_c} M_j w_j(z)~, \quad M_j\in\mathbb{Z}~.
\fe
This gives $\ell_c$ independent magnetic charges. These magnetic charges act on the bounded gauge invariant magnetic monopole operators, hence the name magnetic charges. The gauge invariance condition imposes the same constraint as \eqref{eq:constraint_magnetic} on the magnetic charge vectors $M(z)$ of these monopoles. The basis vectors $w_j(z)$ can then be interpreted as the magnetic charge vectors of the basic monopoles that generate all the bounded gauge invariant monopoles. Note that $w_j(z)$ are periodic functions so these monopoles are extended string-like operators that wrap around the $z$ direction with a constant $\tau,x,y$. We impose the boundedness condition on the monopoles because unbounded monopole operators are unphysical as they create states with infinite energy per layer.

The second type of charges are the electric charges integrated over the $yz$-plane or the $xz$-plane. There are $\ell$ charges of this type on the $yz$-plane:
\ie\label{eq:electric_charges}
Q_{j,x}=
\int dy\,\sum_z w_j(z) J_{\tau x}(z)~.
\fe
They are conserved because
\ie
\partial_\tau Q_{j,x}=
-\int dy\,\sum_z w_j(z) \[\partial_y J_{y x}(z)-\Delta J_{z x}(z)\]=0~.
\fe
Here, we perform a sum by part to move the operator $\Delta$ from acting on the currents to acting on $w_j(z)$ and then use the property $\Delta w_j(z)=0$. The difference condition  \eqref{eq:current_conservation} implies that $Q_{j,x}$ is independent of $x$ so the charges are topological:
\ie
\partial_x Q_{j,x}(x)=
\int dy\,\sum_z w_j(z) \left[\partial_y J_{y \tau}(z)-\Delta J_{z \tau}(z)\right]=0~.
\fe
Similarly, there are $\ell$ electric charges $Q_{j,y}$ on the $xz$-plane and they are topological in the sense that they are independent of $y$. 
All these electric charges shift the gauge field $a(z)\rightarrow a(z) +\sum_{j=1}^\ell w_j(z)\kappa_j$ by some flat gauge fields $\kappa_j$, hence the name electric charges. The charged operators are the Wilson lines. Let us discuss the global form of these electric symmetries. The first $\ell_c$ basis vectors $\{w_j(z)\,|\, j=1,\ldots,\ell_c\}$ are integer-valued so when the corresponding symmetry parameter $\kappa_j$ has $2\pi\mathbb{Z}$-valued holonomies the symmetry transformation acts trivially. Therefore the first $\ell_c$ electric symmetries are $U(1)$ symmetries with integer-valued charges. The last $\ell-\ell_c$ electric symmetries are $\mathbb{R}$ symmetries with real-valued charges because any vector in the incommensurate subspace is not integer-valued. In summary, the first $\ell_c$ electric charges are integer-valued and the remaining $\ell-\ell_c$ electric charges are real-valued.

We now change the perspective and view the theory as a 2+1D system. The 3+1D exotic one-form symmetry then decomposes into a 2+1D $U(1)^{\ell_c}$ magnetic zero-form symmetry and a 2+1D $U(1)^{\ell_c}\times \mathbb{R}^{\ell-\ell_c}$ electric one-form symmetry.  Note that the magnetic monopoles  are now viewed as local operators and hence the magnetic symmetry is a zero-form symmetry.

\subsubsection{'t Hooft Anomaly}

We can couple the exotic one-form symmetry \eqref{eq:current_conservation} to a background gauge field. At the linearized order, it adds to the Lagrangian the coupling
\ie
\mathcal{L}\supset iC_{ab} J^{ab}=i\left(\frac{1}{2}
C_{\mu\nu}J^{\mu\nu}+C_{\mu z} J^{\mu z}\right)~.
\fe
Because of the current conservation equation \eqref{eq:current_conservation} and the difference condition \eqref{eq:difference_condition}, there is a background gauge symmetry:
\ie\label{eq:gauge_symmetry_background}
C_{\mu\nu}\sim C_{\mu\nu}+\partial_\mu \sigma_\nu -\partial_\nu \sigma_\mu~, \quad
C_{\mu z}\sim C_{\mu z}+\partial_\mu \sigma_z-\Delta \sigma_\mu~.
\fe
After including appropriate sub-leading terms, the Lagrangian becomes
\ie\label{eq:coupling}
\mathcal{L}=\frac{1}{4g^2}({f}_{\mu\nu}-C_{\mu\nu})({f}^{\mu\nu}-C^{\mu\nu})+\frac{i}{4\pi} \epsilon^{\mu\nu\rho}\Delta{a}_\mu\partial_\nu {a}_\rho+\frac{i}{2\pi}\epsilon^{\mu\nu\rho}C_{\mu z}\partial_\nu a_\rho~.
\fe
The dynamical fields transform under the background gauge symmetry as
\ie\label{eq:gauge_symmetry_dynamical}
a_\mu \sim a_\mu + \sigma_\mu~.
\fe
Combining \eqref{eq:gauge_symmetry_background} and \eqref{eq:gauge_symmetry_dynamical}, the background gauge symmetry transforms the Lagrangian in an anomalous way as
\ie\label{eq:gauge_transform_L}
\mathcal{L}\rightarrow\mathcal{L}-\frac{i}{4\pi} \epsilon^{\mu\nu\rho}\Delta \sigma_\mu\partial_\nu {\sigma
}_\rho+\frac{i}{2\pi}\epsilon^{\mu\nu\rho}C_{\mu z}\partial_\nu \sigma_\rho~,
\fe
where we omit terms that are total derivatives. 

The anomaly can be canceled by coupling the system to an SPT phase in one higher dimension. The SPT phase is described by the classical Lagrangian
\ie
\mathcal{L}_{\text{SPT}}=-\frac{i}{4\pi}\epsilon^{\mu\nu\rho\lambda}C_{\mu \nu}\partial_\rho C_{\lambda z}-\frac{1}{4}\frac{i}{4\pi}\epsilon^{\mu\nu\rho\lambda}C_{\mu\nu}\Delta C_{\rho\lambda}~.
\fe
The gauge field $C_{ab}=(C_{\mu\nu}, C_{\mu z})$ is extended from the boundary to the bulk by extending the indices $\mu,\nu$ from $\{\tau,x,y\}$ to $\{\tau,x,y,w\}$.\footnote{Such extension is also used for anomaly inflow for subsystem symmetries \cite{Burnell:2021reh}.} The gauge symmetry acts in the same way as \eqref{eq:gauge_symmetry_background} with $\mu,\nu=\{\tau,x,y,w\}$.
Under the gauge transformation, the Lagrangian transforms as
\ie\label{eq:gauge_transform_SPT}
\mathcal{L}_{\text{SPT}}\rightarrow\mathcal{L}_{\text{SPT}}&\,+\frac{1}{2}\frac{i}{4\pi}\epsilon^{\mu\nu\rho\lambda}(C_{\mu\nu} \Delta\partial_\rho\sigma_\lambda-\Delta C_{\mu\nu}\partial_\rho \sigma_\lambda)
\\
&\,
-\frac{i}{2\pi}\epsilon^{\mu\nu\rho\lambda}\partial_\mu \left(C_{\nu z}\partial_\rho\sigma_\lambda  \right)+\frac{i}{4\pi}\epsilon^{\mu\nu\rho\lambda}\partial_\mu \left(\Delta\sigma_\nu \partial_\rho  \sigma_\lambda\right)
~.
\fe 
On a closed manifold, the action is invariant. On an open manifold with periodic boundary condition in the $z$ direction, the gauge transformation of the bulk SPT Lagrangian \eqref{eq:gauge_transform_SPT} generates a boundary term which exactly cancels the anomalous gauge transformation of the boundary Lagrangian \eqref{eq:gauge_transform_L}. Since the classical action is nontrivial on closed manifold, the anomalous gauge transformation \eqref{eq:gauge_transform_L} cannot be removed by adding a local counterterm on the boundary. This means that the anomalous gauge transformation \eqref{eq:gauge_transform_L} is a genuine 't Hooft anomaly.

\subsubsection{Spontaneous Symmetry Breaking} \label{sec:iCS_SSB}

We now discuss the spontaneous breaking of the exotic one-form symmetry. We first approach it from the 3+1D perspective. The charged operators are the Wilson lines and the string-like gauge invariant monopole operators. Following the criterion for one-form symmetries \cite{Gaiotto_2015}, we interpret a perimeter law for the charged  operators as an indicator for the spontaneous breaking of the exotic one-form symmetry. In Section~\ref{sec:potential}, we compute the electric potential between two gauge charges and find that it decays to zero as a power law at long distance. This implies that the Wilson lines obey a perimeter law and hence the exotic one-form symmetry is spontaneously broken. Let us also consider the correlation function of two string-like monopole operators separated by a distance $r$. For simplicity, assume the theory has $\ell_c=1$; the same conclusion holds when $\ell_c>1$. When $N$ is finite, the correlation function is given by \eqref{eq:monopole_correlation}. Note that $Z_{11} = \sum_I W_{I1}^2$ and $W_{I1}$ is a periodic vector so $Z_{11}$ is proportional to $N$. We can interpret $N$ as the length of the string-like monopole operator. Then \eqref{eq:monopole_correlation} implies that the string-like monopole operators obey a perimeter law and hence the exotic one-form symmetry is spontaneously broken. It is consistent with above conclusion. 

Since the exotic one-form symmetry is a continuous symmetry, when it is spontaneously broken, there should be gapless Goldstone modes. These Goldstone modes are the $\ell$ gapless modes in the gapless iCSM theories. Note that the electric charges \eqref{eq:electric_charges} and the magnetic charges \eqref{eq:magnetic_charges} are constructed using $w_j(z)$ rather than a constant vector so they are generally not translation invariant and hence the corresponding Goldstone modes can have non-zero momenta in the $z$ direction.

We now change the perspective and view the theory as a 2+1D system with an infinite number of gauge fields. The exotic one-form symmetry then decomposes into a $U(1)^{\ell_c}$ zero-form symmetry and a $U(1)^{\ell-\ell_c}\times\mathbb{R}^{\ell_c}$ one-form symmetry. The one-form symmetry is spontaneously broken because the charged Wilson lines obey a perimeter law as discussed above. Naively, it seems to contradict the Coleman-Mermin-Wagner theorem for higher-form symmetries that a continuous $p$-form symmetry cannot be spontaneously broken at dimension $D\leq p+2$ \cite{Gaiotto_2015,Lake:2018dqm}. The resolution is that the theory has an infinite number of gauge fields, which allows it to evade the theorem. As discussed in Section \ref{sec:duality}, the zero-form symmetry is spontaneously broken when $N$ is finite. However, in the $N\rightarrow\infty$ limit, the opposite happens and the zero-form symmetry is restored. The two-point function of the monopole operators behaves as $\exp\left[\frac{\mathcal{O}(N)}{g^2r}\right]$. When we take the $N\rightarrow\infty$ limit, it appears to diverge but it is too fast to reach this conclusion. When we take the limit, we should normalize the operators appropriately. One such normalization is to demand the two-point function of the normalized operators to be 1 at a some finite distance $r_0$. Then the two-point function of the normalized monopole operators is $\exp\left[\frac{\mathcal{O}(N)}{g^2}\left(\frac{1}{r}-\frac{1}{r_0}\right)\right]$, which vanishes in the $N\rightarrow\infty$ limit when $r>r_0$. Hence the symmetry is restored in the $N\rightarrow\infty$ limit. In \cite{Sullivan:2021rbk}, the monopoles operators are interpreted as string-like operators rather than local operators and the magnetic zero-form symmetry is said to be ``weakly broken'' in the sense that the order parameters, namely the charged monopole operators, are extended operators.

\subsubsection{Robustness} 

We now discuss the robustness of the gapless iCSM theories. Recall that when the size of the $K$ matrix is finite, all the gapless theories are not robust due to the gauge invariant monopole operators. For this reason, we should pay attention to the monopole operators in the gapless iCSM theories.
First of all, let us consider the gapless iCSM theories with only incommensurate gapless momenta. These theories have no bounded gauge invariant monopole operators so they are robust. Next, we consider more general gapless iCSM theories that have commensurate gapless momenta. These theories have bounded gauge invariant monopole operators. Nevertheless we will argue that they are robust. We will discuss it from both the 3+1D and the 2+1D perspective. When viewing the theories as 3+1D systems, the monopole operators are extended string-like operators. They cannot be included into the Lagrangian because of locality in the $z$ direction. Hence the theories are robust. When viewing the theories as 2+1D systems, the gauge invariant monopole operators are point-like operators and they can be included in the Lagrangian. As discussed in Section~\ref{sec:CSN_robust}, including these monopole operators in the Lagrangian has the effect of gapping out the $\ell_c$ exactly gapless modes with vanishing $m(q)$ at the commensurate gapless momenta and leaving the remaining spectrum invariant. Since there is a continuum spectrum of light modes with arbitrarily small $m(q)$ around each gapless momentum, the effect of removing a finite number of exactly gapless modes is not significant. Hence, we conclude that the theory remains gapless after the deformation and therefore is robust.

In summary, we conclude that the gapless iCSM theories are all robust independent of whether the theories have bounded gauge invariant monopoles or not. This is to be contrasted with the non-robustness when the size of the $K$ matrix is finite.

We end this section with a comment on spontaneous symmetry breaking and robustness. If an internal zero-form symmetry is spontaneously broken, we can gap out the corresponding Goldstone bosons by introducing symmetry-violating operators into the Lagrangian. Therefore theories with a spontaneously broken internal zero-form symmetry are not robust. The robustness of the gapless iCSM theories is consistent with this lore because viewing the theories as 2+1D systems, the $U(1)^\ell$ magnetic zero-form symmetry is not spontaneously broken as discussed in Section~\ref{sec:iCS_SSB}. 

What happens when an internal higher-form symmetry or an exotic one-form symmetry \eqref{eq:current_conservation} is spontaneously broken? Since the symmetry-violating operators are extended operators, they cannot be included into the Lagrangian. Therefore, unlike the cases of zero-form symmetries, theories with a spontaneously broken higher-form symmetry or an exotic one-form symmetry, such as the gapless iCSM theories, are robust. 

What happens when the translation symmetry is spontaneously broken, also known as staging \cite{PhysRevB.37.4792,PhysRevB.40.11943,PhysRevB.42.1339}? It happens for example in the gapless iCSM theories when the gap closes at non-zero gapless momenta. This spontaneously breaks the translation symmetry in the $z$ direction. We can add to the Lagrangian small deformations that explicitly break the translation symmetry such as
\ie
\mathcal{L}\supset\frac{1}{4g^2}\delta Z_{IJ} f^I_{\mu\nu}f^{J,\mu\nu}~,
\fe
where $\delta Z_{IJ}$ is a matrix that does not respect the translation symmetry. Such deformations alter the spectrum but do not gap out the gapless modes. The new spectrum is given by the eigenvalues of the matrix $(1+\delta Z)^{-1}K$, which include the same number of zero eigenvalues as the $K$ matrix. In order to gap out the gapless modes, one would need to deform the $K$ matrix but such deformations cannot be a small deformation because the entries of the $K$ matrix are discrete integers. In conclusion, although the gapless iCSM theories break the translation symmetry, they are still robust.

\section{Electrically Charged Matters}
\label{sec:electric_charged}

In this section, we couple the iCSM theory to electrically charged matters and study their interactions mediated by the gauge fields, such as their electric potentials and their braiding statistics. Denote the current of the charged matters on the $I$th layer by $j^I_\mu$ where $\mu=\tau,x,y$. The current $j^I_\mu$ is coupled to the gauge field $a_\mu^I$ on the same layer via the coupling $\mathcal{L}\supset ia^I_\mu j^{I,\mu}$. Since the current only has components in the $\tau,x,y$ direction, if we interpret the iCSM theory as a 3+1D system, these electrically charged matters are planons that can move only on the $xy$-plane.

The coupling modifies the equation of motion \eqref{eq:EOM} to
\ie
\frac{i}{g^2}\partial_\nu f^{I,\mu\nu}-\frac{1}{2\pi}K_{IJ} \epsilon^{\mu\nu\rho} \d_\nu a^J_\rho=j^I_\mu~.
\fe
Let us study the gauge field sourced by a static gauge charge at the $I=0$ layer. The current is $j_\mu^I=\delta_{\mu,\tau}\delta_{I,0}\delta(\vec x)$. The solution to the equation of motion is
\ie\label{eq:profile_iCS}
&a^I_\tau =-ig^2\int_{-\pi}^\pi\frac{dq}{2\pi}\int \frac{d^2\vec{k}}{(2\pi)^2}\frac{e^{i\vec{k}\cdot\vec{x}+iq I}}{\vec{k}^2+m(q)^2}~,
\\
&a^I_i =-ig^2\int_{-\pi}^\pi\frac{dq}{2\pi}\int \frac{d^2\vec{k}}{(2\pi)^2}\frac{m(q)\epsilon_{ij}k^je^{i\vec{k}\cdot\vec{x}+iq I}}{\vec{k}^2(\vec{k}^2+m(q)^2)}~.
\fe
A quick way to obtain the solution is to work in the momentum basis $a_\mu(q)=\sum_{I}a_\mu^Ie^{iqI}$ where the action \eqref{action} is diagonalized. Then we can use the solution \eqref{eq:guage_source} in the $U(1)$ gauge theory reviewed in Appendix \ref{app:U(1)} and do a Fourier transform to obtain the solution in the iCSM theory.

\subsection{Electric Potential}\label{sec:potential}

We now analyze the electric potential $V(r,z)$ between two gauge charges, separated by $z$ layers and a distance $r$ on the $xy$ plane. We adopt the convention that $z=0$ when the two gauge charges are on the same layer. The question that particularly interests us is whether the potential is confined or deconfined. Recall that in the $U(1)$ gauge theory, the electric potential is logarithimically confined when there is no Chern-Simons term and is deconfined in the presence of a nontrivial Chern-Simons term. Since the existence or absence of a Chern-Simons term also determines whether the theory is gapped or gapless, the confinement in the $U(1)$ gauge theory is tied to the gaplessness of the theory. In contrast, we will show below that in the iCSM theory, the electric potential is always deconfined independent of whether the theory is gapped or gapless, but that the falloff of the electric potential depends on the details of the spectrum.

The electric pontential $V(r,z)$ is given by the solution $ia_\tau^z$ of \eqref{eq:profile_iCS}
\ie\label{eq:potential_iCS}
V(r,z)=g^2\int_{-\pi}^\pi\frac{dq}{2\pi}\int_0^{\infty} \frac{kdk}{2\pi}\int_0^{2\pi}\frac{d\theta}{2\pi}\frac{e^{ikr\cos(\theta)+iq z}}{k^2+m(q)^2}= \frac{g^2}{2\pi}\int_{-\pi}^\pi\frac{dq}{2\pi} e^{iqz}K_0(|m(q)|r)~.
\fe
Here, $K_0(x)$ is a modified Bessel function of the second kind. It behaves like $-\log(x)$ for $0<x\ll 1$ and $x^{-1/2}e^{-x}$ for $x\gg 1$. 
Below, we will discuss three situations.

\bigskip
\hfill
    \emph{Gapped iCSM theory}
\hfill\bigskip

If the theory is gapped, we can first take the long-distance limit $g^2r\gg1$ on the integrand and then do the integration in \eqref{eq:potential_iCS}. This gives a potential that decays exponentially in $r$ and thus is deconfined.

\bigskip
\hfill
    \emph{Gapless iCSM theory with $\Gamma=1$}
\hfill\bigskip   

If the theory is gapless with $\Gamma=1$, the potential at long distance $g^2r\gg1$ is dominated by the light modes near the gapless momenta, so we can approximate the integral \eqref{eq:potential_iCS} by 
    \ie\label{potential_eta=1}
    V(r,z)=\frac{g^2}{2\pi}\sum_{j=1}^\ell\int_{-\infty}^{+\infty}\frac{dq}{2\pi} e^{i(q+q_j) z}K_0\big(|m_j| |q| r\big) 
    =\frac{g^2}{4\pi}\sum_{j=1}^\ell\frac{\cos\left(q_j z\right)}{\sqrt{m_j^2r^2+z^2}}~,
    \fe
    where we used the expansion $m(q)\approx m_j(q-q_j)$ near the gapless momentum $q_j$. In the second equality, we used the fact that the list of gapless momenta is invariant under $q\rightarrow -q$ to replace $e^{iq_j z}$ with $\cos(q_j z)$. The potential decays as $1/r$ and $1/z$ for large $r$ and large $z$ respectively, so it is a deconfined potential.

\bigskip
\hfill
    \emph{Gapless iCSM theory with $\Gamma>1$}
\hfill\bigskip   

If the theory is gapless with $\Gamma>1$, we can also approximate the long-distance potential in the regime $g^2r\gg 1$ by including only the contributions from the light modes near the gapless momenta. 
    
    As an example, consider a theory with only a pair of gapless momenta at $\pm q_1$ and $m(q)\approx m_1 (q-q_1)^{\gamma_1}$ near $q_1$. The long-distance potential is approximated by
    \ie
    V(r,z)&\,=\frac{g^2}{\pi}\int_{-\infty}^{+\infty}\frac{dq}{2\pi} e^{iqz}K_0\big(|m_1| |q|^{\gamma_1} r\big)\cos\left(q_1 z\right)~.
    \fe
    The potential is deconfined for any $\gamma_1>1$. It decays as a power-law in both $r$ and $z$:
    \ie\label{eq:large_rz_potential}
    V(r,z)\sim \begin{dcases}
    \dfrac{\# g^2\cos(q_1 z)}{|z|}~,\quad & |m_1|r\ll |z|^{\gamma_1}
    \\
    \dfrac{\# g^2\cos(q_1 z)}{|m_1 r|^{1/\gamma_1}}~,\quad & |m_1|r\gg |z|^{\gamma_1}
    \end{dcases}~.
    \fe
    Here $\#$ represents the order 1 coefficients that can appear in the falloff of the potential.
    
    When $\gamma_1$ increases, the potential decays slower. Therefore, in theories with multiple gapless momenta, the potential is dominated by the gapless momenta with the largest dynamical exponent and hence it decays as 
    \ie\label{eq:large_rz_potential_Gamma}
    V(r,z)\sim \begin{dcases}
    \dfrac{g^2 f(z)}{|z|}~,\quad &\text{}g^2r\ll |z|^\Gamma
    \\
    \dfrac{g^2 \tilde f(z)}{(g^2r)^{1/\Gamma}}~,\quad &\text{}g^2r\gg |z|^\Gamma
    \end{dcases}~.
    \fe
    Here we only keep track of how the potential decays and we are ignorant about the oscillatory $z$-dependent functions $f(z), \tilde f(z)$ that can appear in the numerator.

\subsection{Braiding Statistics}
\label{sec:braiding}

Next, we analyze the braiding statistics of two gauge charges. Suppose that we fix a gauge charge and move another one, $z$ layers apart, around a circle of radius $r$ on the $xy$-plane with the circle centered at the position of the first gauge charge. This generates an Aharonov-Bohm phase
\ie\label{eq:braiding}
\exp(i\phi(r,z))&=\exp\left(i\oint \vec{a}^z\cdot d\vec{x}\right)
\\
&=\exp\left[-g^2\int_{-\pi}^\pi\frac{dq}{2\pi}\int_0^{\infty} \frac{kdk}{2\pi}\int_0^{2\pi} rd\theta \frac{m(q)k\cos(\theta)e^{ikr\cos(\theta)+iq z}}{k^2(k^2+m(q)^2)}\right]
\\
&=\exp\left[-ig^2\int_{-\pi}^{\pi}\frac{dq}{2\pi}\frac{e^{iq z}}{m(q)}\Big(1-|m(q)|rK_1(|m(q)|r)\Big)\right]~.
\fe 
where $\vec{a}^z=(a_x^z,a_x^z)$ is from the solution \eqref{eq:profile_iCS}. Here, $K_1(x)$ is a modified Bessel function of the second kind. It behaves as $x^{-1}$ for $0<x\ll1$ and $x^{-1/2} e^{-x}$ for $x\gg1$. 
We are interested in the asymptotic limit of the braiding phase $\exp(i\phi_\infty(z))\equiv\lim_{r\rightarrow\infty}\exp(i\phi(r,z))$, which defines the braiding statistics of the gauge charges. Below, we will discuss three situations.

\bigskip
\hfill
    \emph{Gapped iCSM theory}
\hfill\bigskip   

If the theory is gapped, we can first take the long-distance limit $g^2r\gg1$ on the integrand and then do the integration in \eqref{eq:braiding}. The braiding phase converges exponentially fast to
    \ie\label{eq:braiding_gapped}
    \exp(i\phi_\infty(z)) =\exp\left[- 2\pi i\int_{-\pi}^{\pi}\frac{dq}{2\pi}\frac{e^{iq z}}{\lambda(q)}\right]=\exp\left(-2\pi i\left( K^{-1}\right)_{I,I+z}\right)~.
    \fe
    The last expression is independent of $I$ because of the translation symmetry.
    
    Given the Laurent polynomial $p(u)$ of the $K$ matrix, we can evaluate the integral \eqref{eq:braiding_gapped} explicitly. Define the complex variable $u=e^{iq}$. The integral can be expressed as a contour integral along the unit circle on the complex $u$-plane
    \ie\label{eq:contour_gapped}
    \exp(i\phi_\infty(z)) &=\exp\left[- 2\pi i\oint\frac{du}{2\pi i}\frac{u^{z+\xi-1}}{c_\xi}\prod_{j=1}^{L}\left(\frac{1}{u-u_j}\right)^{\gamma_j}\right]~,
    \fe
    where we used $\lambda(q)=p(e^{iq})$ and the factorization  \eqref{eq:poly_root} of $p(u)$. Here, $u_j$ are the roots of $p(u)$.
    Since the theory is gapped, half of the roots are inside the unit circle while the other half are outside the unit circle. Using the residue theorem, we can express the contour integral as a sum over residues of poles inside the unit circle. This calculation is straightforward when $z\ge0$, in which case the poles that contribute are precisely those $u_j$ inside the unit circle. When $z\leq 0$, other than these poles, there is another pole inside the unit circle at $u=0$ if $z\leq-\xi$. We could perform the contour integral again to obtain the result, but instead we observe that the braiding phase is symmetric in $z$ because of the reflection symmetry $z\rightarrow -z$. This can be seen by a change of variable $u\rightarrow 1/u$, which returns the same contour integral with $z\rightarrow -z$. Therefore, the final result is
    \ie\label{eq:gapped_poles}
    \exp(i\phi_\infty(z)) &=\exp\left[-2\pi i\sum_{|u_j|<1}u_j^{|z|-1}\text{Res}_{u_j}\left[\frac{1}{p(u)}\right] \right]~,
    \fe
    where the sum is over distinct roots inside the unit circle and $\text{Res}_{u_j}\left[1/p(u)\right]$ is the residue of $p(u)^{-1}$ at $u_j$.
    
    By \eqref{eq:gapped_poles}, the braiding phase $\exp\left(i\phi_\infty(z)\right)$ decays exponentially in $|z|$ and the decay rate is given by the largest magnitude of the roots inside the unit circle. Since the braiding phase measures the total magnetic flux on $z$th layer, in the picture of flux attachment, the gauge charge is attached with a flux string that extends in the $z$ direction. The flux of the flux string on the $z$th layer decays with respect to $|z|$. 
    
    As an example, let us work out the braiding phase for the tridiagonal $K$ matrix \eqref{tridiagonal} when $|R|>2$. For simplicity, we assume $c_0,c_1>0$. In this case, the roots of $p(u)$ are
    \ie
    u_1&=-\frac{c_0}{2c_1}+\sqrt{\frac{c_0^2}{4c_1^2}-1}~,\\
    u_2&=-\frac{c_0}{2c_1}-\sqrt{\frac{c_0^2}{4c_1^2}-1}~.
    \fe
    $u_1$ is inside the unit circle and $u_2$ is outside the unit circle. Using \eqref{eq:gapped_poles}, we obtain
    \ie\label{braiding_tri_gapped}
    \exp(i\phi_\infty(z))=\exp\left[-2\pi i\frac{u_{1}^{|z|}}{\sqrt{c_0^2-4c_1^2}}\right]~.
    \fe
    
\bigskip    
\hfill
    \emph{Gapless iCSM theory with $\Gamma=1$}
\hfill\bigskip   

If the theory is gapless with $\Gamma=1$, the integral \eqref{eq:braiding_gapped} is not well-defined because $\lambda(q)=0$ at the gapless momenta. We cannot take the large-distance limit $g^2r\gg1$ inside the integral \eqref{eq:braiding}. Instead, we should return to \eqref{eq:braiding} and take the limit more carefully.
    
\begin{figure}
        \centering
        \includegraphics[width=0.5\textwidth]{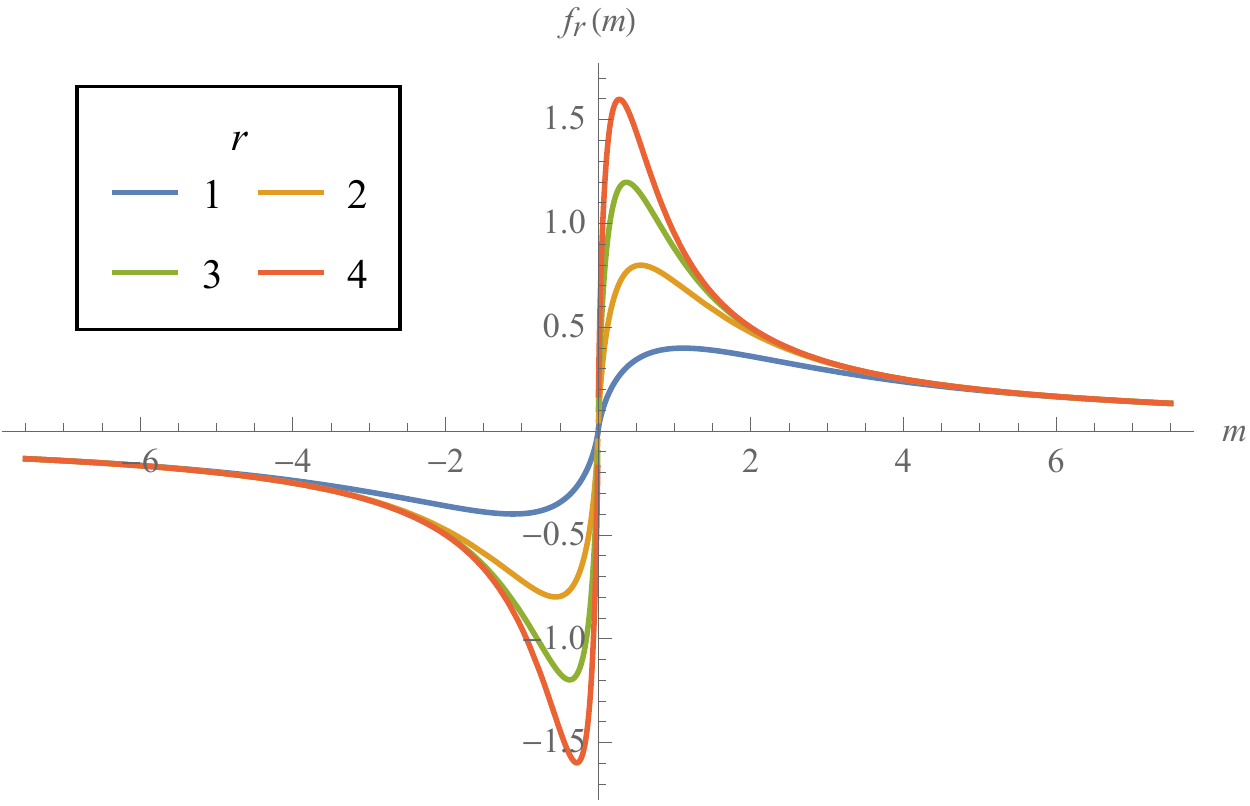}
        \caption{Function $f_r(m)$ in \eqref{f_r(m)} for $r=1,2,3,4$.
        The larger $r$ is, the more $f_r(m)$ looks like $1/m$.}
        \label{fig:PV}
    \end{figure}

    The integrand in \eqref{eq:braiding} involves the function
    \ie\label{f_r(m)}
    f_r(m)=\frac{1}{m}\big(1-|m|rK_1(|m|r)\big)~.
    \fe
    In Figure~\ref{fig:PV}, we plot the function $f_r(m)$ for a range of $r$'s. It is everywhere finite and, in particular, vanishes at $m=0$. As $r$ increases, it becomes a better and better approximation to $1/m$ except when $|m|$ is small. Therefore, the parameter $r$ can be viewed effectively as a cutoff which prevents the divergence of $1/m$ in the integral \eqref{eq:braiding}.
    
    Since $f_r(m)$ is a function symmetric in $m$ and $m(q)$ is linear in $q-q_j$ near the gapless momenta $q_j$, in the $r\rightarrow\infty$ limit, the braiding phase converges to the Cauchy principal value of the integral \eqref{eq:braiding_gapped}:
    \ie\label{eq:braiding_gapless_linear_crossing}
    \exp(i\phi_\infty(z)) =\exp\left[-2\pi i\,\text{PV}\int_{-\pi}^{\pi}\frac{dq}{2\pi}\frac{ e^{iq z}}{\lambda(q)}\right]~.
    \fe
    The Cauchy principal value of an integral with an integrand $g(x)$ that diverges at $x_0$ inside the integration domain $(x_1,x_2)$ is defined as
    \ie\label{PV_def}
    \text{PV}\int_{x_1}^{x_2} dx\, g(x)=\lim_{\epsilon\to 0^+}\left(\int_{x_1}^{x_0-\epsilon}dx\, g(x)+\int_{x_0+\epsilon}^{x_2}dx\, g(x)\right).
    \fe
    In \eqref{eq:braiding}, $1/r$ effectively plays the role of the cutoff $\epsilon$ in the definition of the Cauchy principal value.
    
\begin{figure}
        \centering
        \includegraphics[width=0.6\textwidth]{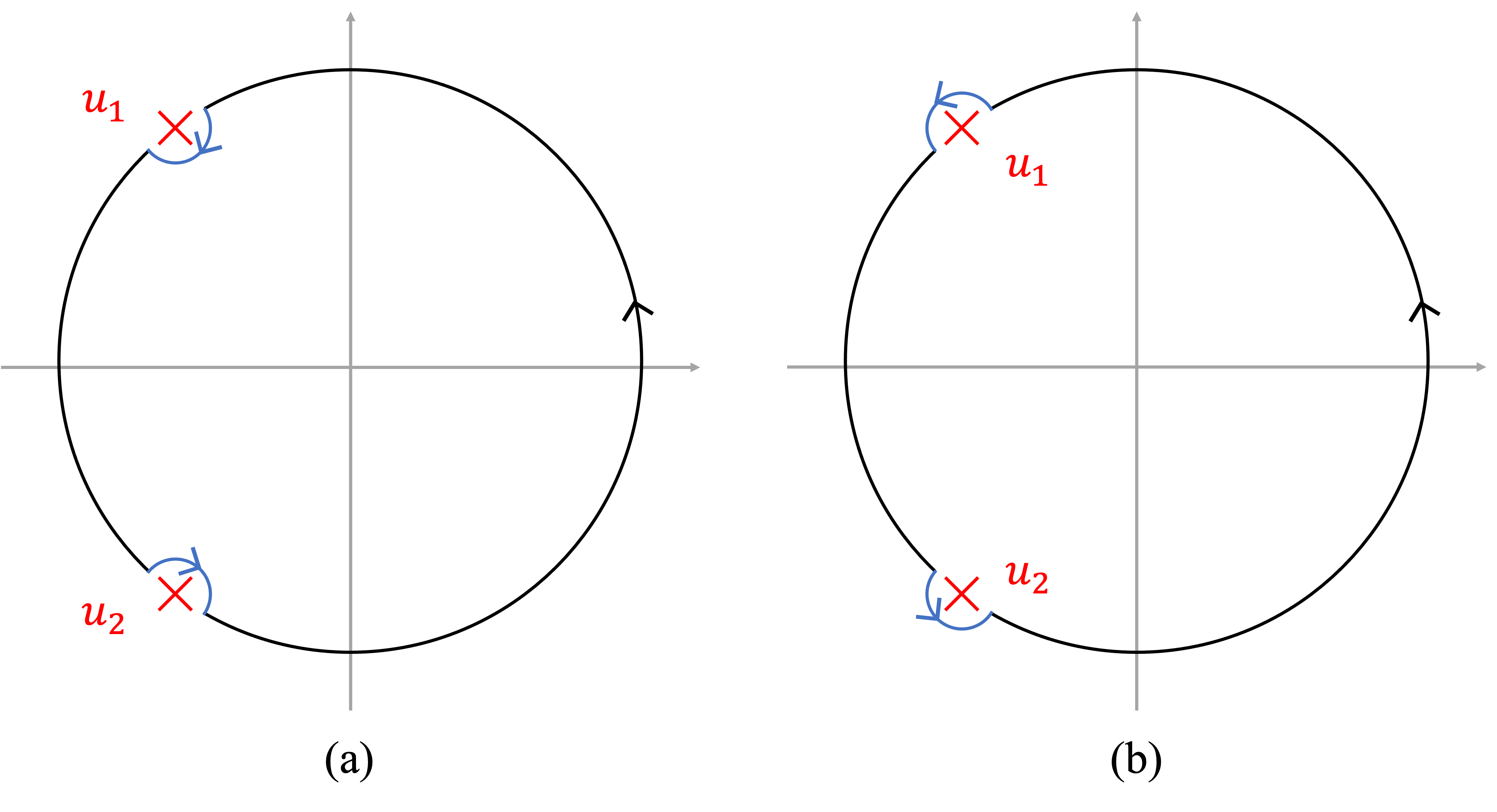}
        \captionsetup{justification=Justified}
        \caption{Evaluation of the Cauchy principal value integral \eqref{eq:contour_gapless} by adding small semicircles (drawn in blue). The contour is cut open in the vicinity of the roots $u_1$ and $u_2$ which, for simplicity, we assume to be the only roots of $p(u)$. In subfigure (a), we add clockwise semicircles inside the unit circle, so that the closed contour picks up no pole. Alternatively, as shown in subfigure (b), we can add counterclockwise semicircles outside the unit circle, so that the closed contour picks up both poles. The two approaches give the same answer after correctly subtracting the contributions of the semicircles.}
        \label{fig:contour}
    \end{figure}

    As in the gapped case, given the polynomial  $p(u)$ of the $K$ matrix, we can express the integral \eqref{eq:braiding_gapless_linear_crossing} as a contour integral along the unit circle on the complex $u$-plane
    \ie\label{eq:contour_gapless}
    \exp(i\phi_\infty(z)) &=\exp\left[- 2\pi i\, \text{PV}\oint\frac{du}{2\pi i}\frac{u^{z+\xi-1}}{c_\xi}\prod_{j=1}^{L}\left(\frac{1}{u-u_j}\right)^{\gamma_j}\right]~.
    \fe
    Since the theory is gapless, the integrand has poles not only inside and outside the unit circle but also on the unit circle. The poles on the unit circle are simple poles because the theory has $\Gamma=1$. Using the residue theorem, the Cauchy principal value of the integral is given by the sum over the residues of poles inside the unit circle and \textit{half} of the residues of poles on the unit circle
    \ie\label{eq:gapless_poles}
    \exp(i\phi_\infty(z)) &=\exp\left[-2\pi i\left(\sum_{|u_j|<1}u_j^{|z|-1}\text{Res}_{u_j}\left[\frac{1}{p(u)}\right]+\frac{1}{2}\sum_{|u_j|=1}u_j^{|z|-1}\text{Res}_{u_j}\left[\frac{1}{p(u)}\right] \right)\right]~.
    \fe
    We now explain why the poles on the unit circle only contribute half of their residues. According to the Cauchy principal value prescription, the contour in \eqref{eq:contour_gapless} is cut open in the vicinity of every root on the unit circle. To evaluate the integral, we can close the contour by including a small semicircle around every root on the unit circle, and then subtracting them. If the semicircles are inside the unit circle as drawn in Figure~\ref{fig:contour}(a), the closed contour does not pick up the residues on the unit circle. Meanwhile, the added clockwise semicircle integral is minus half of the residues on the unit circle because the added pieces are semicircles. Subtracting the latter from the former yields the Cauchy principal value, which includes half of the residues on the unit circle. As a consistency check, we can also choose the semicircle to be outside the unit circle as shown in Figure~\ref{fig:contour}(b). The closed contour picks up the residues on the unit circle. Subtracting the counterclockwise semicircle integrals, we again find that only half of the residues on the unit circle contributes to the Cauchy principal value. These two different ways of closing the contour indeed give the same answer.
    
    As an example, let us work out the braiding phase for the tridiagonal $K$ matrix \eqref{tridiagonal} when $|R|<2$. For simplicity, we assume $c_0,c_1>0$. In this case, the roots of $p(u)$ are
    \ie
    u_1&=-\frac{c_0}{2c_1}+ i\sqrt{1-\frac{c_0^2}{4c_1^2}}~,\\
    u_2&=-\frac{c_0}{2c_1}- i\sqrt{1-\frac{c_0^2}{4c_1^2}}~.
    \fe
    Both roots are on the unit circle and we can express them as $u_{1,2}=e^{i q_{1,2}}$ where $q_2=-q_1$. Using \eqref{eq:gapless_poles}, we obtain
    \ie\label{eq:braiding_phase_tri_gapless}
    \exp(i\phi_\infty(z)) =\exp\left[-2\pi i\frac{\sin( q_1|z|)}{\sqrt{4c_1^2-c_0^2}}\right].
    \fe
    The braiding phase oscillates as a function of $z$. Interestingly, two gauge charges on the same layer braid trivially. Therefore, they should be interpreted as bosons or fermions. Note that this observation does not hold in more general iCSM theories. It is valid if all the roots of the Laurent polynomial are on the unit circle and have multiplicity 1. 
    
    For a more general $K$ matrix, the braiding phase receives contributions from both the poles inside and those on the unit circle. For large $|z|$, the poles $u_j$ inside the unit circle are suppressed by a factor of $|u_j|^{|z|}$, and therefore the braiding phase is dominated by the poles on the unit circle
    \ie\label{eq:braiding_phase_gapless_general}
    \exp(i\phi_\infty(z))=\exp\left[-\pi i\sum_{|u_j|=1}u_j^{|z|-1}\text{Res}_{u_j}\left[\frac{1}{p(u)}\right]\right]=\exp\left[\sum_{j=1}^{\ell} \frac{g^2}{ 2m_j}e^{iq_j|z|}\right]~,
    \fe
    where the sum is over the gapless momenta $q_j$ and $m_j$ is the coefficient in the expansion $m(q)\approx m_j(q-q_j)$ near the gapless momentum $q_j$.
    
    Unlike in the gapped iCSM theories, the braiding phase does not decay but only oscillates in $|z|$. In the picture of flux attachment, it implies that a flux string that extends in the $z$ direction is attached to a gauge charge and that the total flux on each layer oscillates as $|z|$ increases. In Section~\ref{sec:Wilson_correlation}, we will determine the shape of this flux string and show that the flux string is more and more spread out as $|z|$ increases.

\bigskip
\hfill
    \emph{Gapless iCSM theory with $\Gamma>1$}
\hfill\bigskip   

If the theory is gapless with $\Gamma>1$, the braiding phase diverges in the $g^2r\rightarrow \infty$ limit. To explain this divergence, let us consider the example of a tridiagonal $K$ matrix \eqref{tridiagonal} with $c_0=2$ and $c_1=-1$. The theory is gapless at $q=0$, around which we have $\lambda(q)\approx q^2$ and $\gamma=2$. For $g^2r\gg 1$, the braiding phase is dominated by the contributions from small $\lambda(q)$ around $q=0$, so we can approximate
   \eqref{eq:braiding} by
    \ie
    \exp(i\phi(r,z)) 
    =\exp\left[- 2\pi i\int_{-\infty}^{\infty}\frac{dq}{2\pi} \frac{e^{iq z}}{q^2}
    \left(1-\frac{g^2q^2r}{2\pi}K_1\left(\frac{g^2q^2r}{2\pi}\right)\right)\right]~.
    \fe
    For $g^2r\gg z^2$, we find
    \ie
    \exp(i\phi(r,z)) =\exp\left(-\# ig\sqrt{r}\right)~,
    \fe
    where we omit the numerical factor $\#$. The braiding phase diverges in the $g^2r\rightarrow \infty$ limit.\footnote{Here, we showed that the braiding statistics between two elementary gauge charges is not well-defined when the $K$ matrix is a tridiagonal matrix with $c_0=2$ and $c_1=-1$. However, we can also consider dipoles consisting of two elementary gauge charges of opposite charge at nearby layers. The braiding statistics of these dipoles is well-defined and is trivial. We thank Meng Cheng for pointing this out to us.} The divergence is related to the fact that $\lambda(q)$ has the same sign on both sides of $q=0$. This is in contrast to the $\Gamma=1$ case, where $\lambda(q)$ changes sign across the gapless momenta and the divergence cancels exactly, leading to a finite braiding phase in the $g^2r\rightarrow\infty$ limit. 
    For the same reason, any gapless iCSM theory with $\Gamma=2$ has a divergent braiding phase in the $g^2r\rightarrow \infty$ limit.

    Next, we consider the gapless iCSM theories with $\Gamma>2$. Around a gapless momentum $q_j$ with the exponent $\gamma_j$, we have
    \ie\label{mass_higher_order}
    m(q)= m_j(q-q_j)^{\gamma_j}\left[1+\alpha_1(q-q_j)+\alpha_2(q-q_j)^2+\cdots\right]~,
    \fe
    where $\alpha_1,\alpha_2$ are some coefficients which are generically non-zero. Expanding $m(q)^{-1}$ around $q_j$, we find
    \ie
    \frac{1}{m(q)}=\frac{1}{m_j(q-q_j)^{\gamma_j}}\left[1-\alpha_1(q-q_j)+(\alpha_1^2-\alpha_2)(q-q_j)^2+\cdots\right]~.
    \fe
    If $\gamma_j\geq 2$, either the first or the second term diverges as an even power of $(q-q_j)$. It then leads to a divergence in the braiding phase in the $g^2r\rightarrow\infty$ limit. Therefore, we conclude the braiding phase diverges in the $g^2r\rightarrow\infty$ limit in the gapless iCSM theories with $\Gamma>1$. 

\bigskip
\bigskip
To summarize, the long-distance braiding phase is finite in the gapped iCSM theories and the gapless iCSM theories with $\Gamma=1$, and diverges in the gapless iCSM theories with $\Gamma>1$.

We now discuss the relation between the iCSM theory and the theory with a finite dimensional $K$ matrix. Since the iCSM theory is the $N\rightarrow \infty$ limit of the theory with a finite $N$-dimensional $K$ matrix, it is natural to ask whether one can obtain the long-distance braiding phase $\exp(i\phi_\infty)$ in the iCSM theory by taking the $N\rightarrow \infty$ limit of the finite $N$ counterpart. In a finite $N$ theory, the braiding phase is
\ie\label{eq:braiding_CSN}
\exp(i\phi(r,z)) =\exp\left[-\frac{ig^2}{N}\sum_{q}\frac{e^{iq z}}{m(q)}\Big(1-|m(q)|rK_1(|m(q)|r)\Big)\right]~,
\fe 
where the sum is over quantized momenta $q=\frac{2\pi k}{N}$ with $k=1,\ldots, N$. In the $g^2r\rightarrow \infty$ limit, the braiding phase converges to
\ie\label{CSN_braiding_large_r}
\exp(i\phi_\infty(z)) =\exp\left(-\frac{2\pi i}{N}\sum_{ |\lambda(q)|>0}\frac{e^{iq z}}{\lambda(q)}\right)~,
\fe
where the sum is now restricted to quantized, gapped momenta. 

\begin{figure}
     \centering
     \begin{subfigure}[b]{0.32\textwidth}
         \centering
         \includegraphics[width=\textwidth]{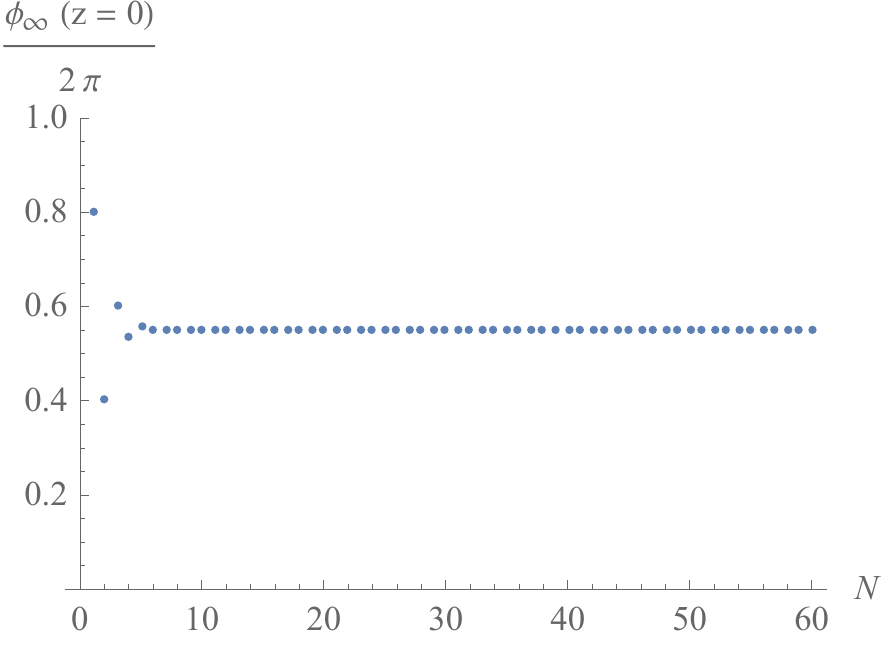}
         \caption{$p(u)=u^{-1}+3+u$.}
         \label{fig:phase_gapped}
     \end{subfigure}
     \hfill
     \begin{subfigure}[b]{0.32\textwidth}
         \centering
         \includegraphics[width=\textwidth]{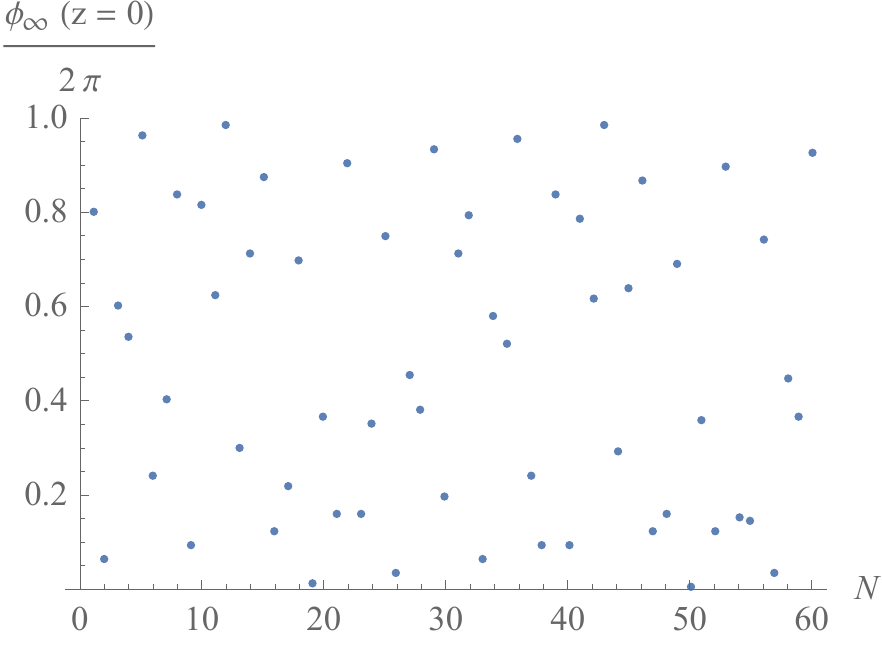}
         \caption{$p(u)=2u^{-1}+1+2u$.}
         \label{fig:phase_incommensurate}
     \end{subfigure}
     \hfill
     \begin{subfigure}[b]{0.32\textwidth}
         \centering
         \includegraphics[width=\textwidth]{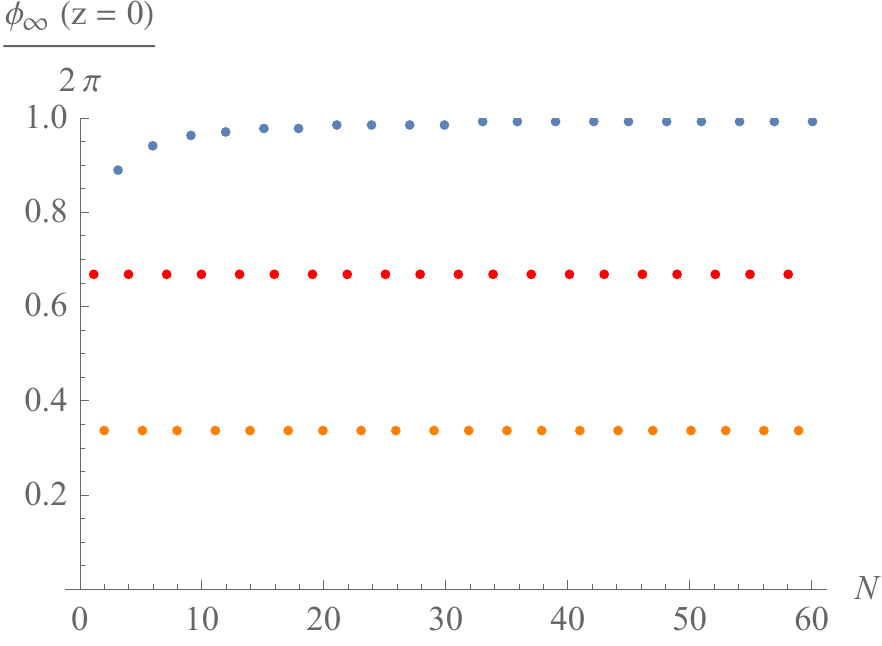}
         \caption{$p(u)=u^{-1}+1+u$.}
         \label{fig:phase_commensurate}
     \end{subfigure}
     \captionsetup{justification=Justified}
        \caption{Long-distance braiding phases $\phi_\infty(z=0)$ for two gauge charges on the same layer in theories with finite $N$-dimensional tridiagonal $K$ matrices, as a function of $N$. Figure~\ref{fig:phase_gapped} is a gapped theory. Its braiding phase converges as $N\rightarrow \infty$. Figure~\ref{fig:phase_incommensurate} is a gapless theory with incommensurate gapless momenta. Its braiding phase does not converge as $N\rightarrow \infty$.
        Figure~\ref{fig:phase_commensurate} is a gapless theory with only commensurate gapless momenta. Its braiding phase does not converge as $N\rightarrow \infty$ but there are converging subsequences: $N=0$ mod 3 (blue), $N=1$ mod 3 (red) and $N=2$ mod 3 (orange). Only the $N=0$ mod 3 branch is gapless, and only the braiding phase on this branch converges to the braiding phase $\exp(i\phi_\infty(z=0))=1$ in the corresponding iCSM theory.}
        \label{fig:phases}
\end{figure}

If the iCSM theory is gapped, the finite $N$ long-distance braiding phase \eqref{CSN_braiding_large_r} indeed converges to the long-distance braiding phase \eqref{eq:braiding_gapped} in the iCSM theory in the $N\rightarrow\infty$ limit (see the example illustrated in Figure~\ref{fig:phase_gapped}). Since $g^2$ is the only scale in the theory, the agreement of the long-distance braiding phase is related to the fact that the $g^2\rightarrow\infty$ limit and the $N\rightarrow \infty$ limit commute if the iCSM theory is gapped. 

What about gapless iCSM theories? We will focus on a theory with $\Gamma=1$, which has a finite long-distance braiding phase.

First, consider an iCSM theory that has incommensurate gapless momenta. In such a theory, the finite $N$ long-distance braiding phase \eqref{CSN_braiding_large_r} does not converge as $N\rightarrow\infty$ (see the example illustrated in Figure \ref{fig:phase_incommensurate}). This is because in the summation of \eqref{CSN_braiding_large_r}, the quantized momenta do not approach the incommensurate gapless momenta at the same rate from the two sides as $N$ increases and therefore the divergences do not cancel as $N\rightarrow \infty$.

Next, consider an iCSM theory that has only commensurate gapless momenta. The finite $N$ long-distance braiding phase \eqref{CSN_braiding_large_r} also does not converge as $N\rightarrow \infty$ (see the example illustrated in Figure~\ref{fig:phase_commensurate}). However, some subsequences may converge. For example, consider the subsequence of $N=0$ mod $M$, where $M$ is the minimal integer such that $ Mq_j=0\text{ mod }2\pi$ for all gapless momenta $q_j$. In the example illustrated in Figure \ref{fig:phase_commensurate}, $M=3$ and the subsequence is the $N=0$ mod 3 branch. On this subsequence, all the theories are gapless and they have the same gapless momenta as the iCSM theory. As $N\rightarrow\infty$, the quantized momenta in the summation of \eqref{CSN_braiding_large_r} approach the gapless momenta at the same rate from the two sides. Therefore, in the $N\rightarrow \infty$ limit, this subsequence converges and moreover, it converges to the long-distance braiding phase \eqref{eq:braiding_gapless_linear_crossing} of the iCSM theory. The Cauchy principal value prescription of \eqref{eq:braiding_gapless_linear_crossing} follows from the fact that for every gapless momentum, the two nearest quantized momenta in the summation of \eqref{CSN_braiding_large_r} are always equally far away on this subsequence. We can also consider other subsequences with $N=P$ mod $M$. They also converge in the $N\rightarrow \infty$ limit but their limits are different from \eqref{eq:braiding_gapless_linear_crossing}. 

In conclusion, starting from the finite $N$, finite-distance braiding phase, if we first take $N\rightarrow\infty$ and then take $r\rightarrow \infty$, we obtain the long-distance braiding phase \eqref{eq:braiding_gapless_linear_crossing} in the iCSM theory. On the other hand, if we first take $r\rightarrow\infty$ and then take $N\rightarrow\infty$, the sequence does not converge if the iCSM theory is gapless. In particular, it means that the two limits do not commute for gapless iCSM theory. Relatedly, the $g^2\rightarrow\infty$ limit and the $N\rightarrow \infty$ limit also do not commute for the gapless iCSM theory. We will discuss this non-commutativity further in Section \ref{sec:effective_theory}.

\section{Correlation Functions of Wilson Lines}
\label{sec:Wilson_correlation}

In this section, we study the correlation functions of the Wilson lines in the iCSM theories, focusing on their long-distance behavior. We will show that the long-distance correlation functions are topological in the gapped iCSM theories, not topological but scale invariant (up to a modulation factor) in the gapless iCSM theories with $\Gamma=1$ and decays more slowly than the perimeter law in the gapless iCSM theories with $\Gamma>1$. Here, ``topological'' means that the correlation function is invariant under smooth deformations of the curves that support the Wilson lines. All the curves are restricted to be on some constant $z$ slices so the smooth deformations are restricted to be only in the $\tau,x,y$ direction and deformations in the $z$ direction are forbidden.  Although the full long-distance correlation functions are not topological in the gapless iCSM theories, their phases are still topological if the theory has $\Gamma=1$. We can interpret the Wilson lines as the worldlines of some gapped gauge charges. Then, when the two Wilson lines link nontrivially, the phases of the correlation functions are the braiding phases of two gauge charges. The fact that the phases of the correlation functions are topological implies that for gapless iCSM theories with $\Gamma=1$, the braiding statistics computed in Section \ref{sec:braiding} is independent of the details of the braiding trajectories as long as the the trajectories are sufficiently large.

In Appendix \ref{app:U(1)}, we review the correlation functions of the Wilson lines in the $U(1)$ gauge theory. They can be computed from the Green's function
\ie
G_{\mu\nu}(m,x)=\frac{1}{g^2}\langle a_\mu(x) a_\nu(0)\rangle
=
\int \frac{d^3k}{(2\pi)^3} \frac{k^2\delta_{\mu\nu}-k_\mu k_\nu-m\epsilon_{\mu\nu\rho}k^\rho}{k^2\left(k^2+m^2\right)} e^{i kx}~.
\fe
The integral is computed in
\eqref{eq:U(1)_green_re} and \eqref{eq:U(1)_green_im}. It gives
\ie\label{eq:U(1)_green}
G_{\mu\nu}(m,x)=&\,-\frac{i}{4\pi r}\frac{1-e^{-|m|r}(1+|m|r)}{mr}\frac{\epsilon_{\mu\nu\rho}x^\rho}{r}
\\
&\,+\frac{e^{-|m|r}}{6\pi r}\delta_{\mu\nu}+\frac{1}{4\pi r}\left[\frac{1}{|m|^2r^2}-\left(\frac{1}{3}+\frac{1}{|m|r}+\frac{1}{|m|^2r^2}\right)e^{-|m|r}\right]Q_{\mu\nu}~,
\fe
where $Q_{\mu\nu}$ is the (traceless) quadrupole tensor
\ie
Q_{\mu\nu}=\frac{3x_\mu x_\nu-r^2\delta_{\mu\nu}}{r^2}~.
\fe

In the iCSM theory, the action \eqref{action} is diagonalized in the momentum basis so the Green's function can be obtained by doing a Fourier transform on the Green's function \eqref{eq:U(1)_green} of the $U(1)$ gauge theory
\ie\label{eq:CSinfty_green}
G_{\mu\nu}(x,z)=\frac{1}{g^2}\langle a_\mu^{I+z}(x) a_\nu^I(0)\rangle=\int_{-\pi}^{\pi}\frac{dq}{2\pi}G_{\mu\nu}\big(m(q),x\big)e^{iq z}~.
\fe
From the Green's function, we can compute the correlation functions of the Wilson lines by
\ie\label{eq:CSinfty_Wilson_correlation}
\langle W^{I}(C_1) W^{I+z}(C_2)\rangle=\exp\left(-g^2\oint_{C_1} dx_1^\mu\oint_{C_2} dx_2^\nu\, G_{\mu\nu}(x_1-x_2,z) \right)~.
\fe
Below, we discuss three situations focusing on the long-distance behavior of the correlation functions.

\bigskip
\hfill
    \emph{Gapped iCSM theory}
\hfill\bigskip   

If the theory is gapped, we can first take the long-distance limit $g^2r\gg1$ inside the integral of \eqref{eq:CSinfty_green}. It gives
\ie
G_{\mu\nu}(x,z)=-\frac{i\epsilon_{\mu\nu\rho}x^\rho}{4\pi r^2}\int_{-\pi}^{\pi}\frac{dq}{2\pi}\frac{e^{iq z}}{m(q)}~.
\fe
Substituting this into \eqref{eq:CSinfty_Wilson_correlation}, we obtain a topological correlation function at long distance
\ie
\langle W^I(C_1)W^{I+z}(C_2)\rangle&=\exp\left(- 2\pi i\int_{-\pi}^{\pi}\frac{dq}{2\pi} \frac{e^{iq z}}{\lambda(q)} \text{link}(C_1,C_2)\right)~,
\fe
which depends only on the linking number $\text{link}(C_1,C_2)$ of the two curves $C_1$ and $C_2$.

\bigskip
\hfill
    \emph{Gapless iCSM theory with $\Gamma=1$}
\hfill\bigskip   

If the theory is gapless with $\Gamma=1$, we cannot take the long-distance limit $g^2r\rightarrow\infty$ inside the integral \eqref{eq:CSinfty_green}  because  at the gapless momenta $m(q)$ vanishes and the integrand diverges. We need to take the limit more carefully. 

Let us first consider the real part of the Green's function $G_{\mu\nu}(x,z)$:
\ie
\text{Re}[G_{\mu\nu}(x,z)]
=\int_{-\pi}^{\pi}\frac{dq}{2\pi}\int \frac{d^3k}{(2\pi)^3} \frac{k^2\delta_{\mu\nu}-k_\mu k_\nu}{k^2\left(k^2+m(q)^2\right)} e^{i kx+i q z}~.
\fe
It determins the magnitude of the correlation function.
At long distance $g^2r\gg 1$, the integral is dominated by light modes with small $m(q)$ around the gapless momenta, so we approximate it by
\ie\label{eq:CSinf_Green_Re_mom}
&\text{Re}[G_{\mu\nu}(x,z)]=\sum_{j=1}^{\ell}\int_{-\infty}^{\infty}\frac{dq}{2\pi}\int \frac{d^3k}{(2\pi)^3} \frac{k^2\delta_{\mu\nu}-k_\mu k_\nu}{k^2\left(k^2+m_j^2 q^2\right)} e^{i kx+i (q+q_j) z}~,
\fe
where the sum is over gapless momenta $q_j$ and $m_j$ is the coefficient in the expansion $m(q)\approx m_j(q-q_j)$ near $q=q_j$. Evaluating the integral, we get
\ie\label{eq:CSinf_Green_Re}
\text{Re}[G_{\mu\nu}(x,z)]=\sum_{j=1}^\ell\frac{e^{iq_j z}}{4\pi^2|m_j|r^2}\left[\frac{2}{3(1+\zeta_j^2)}\delta_{\mu\nu}+\left(\frac{2+3\zeta_j^2}{3(1+\zeta_j^2)}-\zeta_j\,\text{arccot}(\zeta_j)\right)Q_{\mu\nu}\right]~,
\fe
where we define dimensionless variables
\ie\label{eq:define_zeta}
\zeta_j=\frac{z}{|m_j|r}~.
\fe
As a consistency check of the approximation, \eqref{eq:CSinf_Green_Re} decays as $1/(m_j r^2)$ for large $r$, which is slower than the $1/(m^2r^3)$ decay of the integrand \eqref{eq:U(1)_green} away from the gapless momenta. Substituting \eqref{eq:CSinf_Green_Re} into \eqref{eq:CSinfty_Wilson_correlation}, we obtain the magnitude of the correlation function $|\langle W^{I+z}(C_1)W^{I}(C_2)\rangle|$. Note that in the summation of \eqref{eq:CSinf_Green_Re}, if we strip off the $e^{iq_j z}$ factors, the remaining functions transform as $\Lambda^{-2}$ under the scale transformation $(t,x,y,z)\rightarrow (\Lambda t,\Lambda x,\Lambda y,\Lambda z)$. Therefore, the magnitude takes the form
\ie
|\langle W^{I+z}(C_1)W^{I}(C_2)\rangle|= \exp\left[\sum_{j=1}^\ell e^{i q_j z} F_j(C_1,C_2)\right]~,
\fe
where $F_j(C_1,C_2)$ is a scale invariant function. This is related to the scale invariance of the low-energy dispersion relation \eqref{eta_def} of the gapless iCSM theories with $\Gamma=1$.

Next, let us consider the imaginary part of the Green's function $G_{\mu\nu}(x,z)$:
\ie\label{eq:prop_imaginary}
\text{Im}[G_{\mu\nu}(x,z)]
&=\int_{-\pi}^{\pi}\frac{dq}{2\pi}\int \frac{d^3k}{(2\pi)^3} \frac{im(q)\epsilon_{\mu\nu\rho}k^\rho}{k^2\left(k^2+m(q)^2\right)} e^{i kx+iqz}
\\
&=-\int_{-\pi}^\pi\frac{dq}{2\pi}\frac{\epsilon_{\mu\nu\rho}x^\rho }{4\pi r^2}h_r(m(q))e^{iq z}~.
\fe
It determines the phase of the correlation function. Here, 
\ie
h_r(m)=\frac{1-e^{-|m|r}(1+|m|r)}{mr}~.
\fe
As a function of $m$, $h_r(m)$ is similar to $f_r(m)$ defined in \eqref{f_r(m)}. It vanishes at $m=0$ for all $r$. As $r$ increases, it becomes a better and better approximation to the function $1/m$ except when $m$ is close to zero. Since $h_r(m)$ is a function symmetric in $m$ and $m(q)$ is linear in $q-q_j$ near the gapless momenta $q_j$, the integral of \eqref{eq:prop_imaginary} is approximated by the Cauchy principal value of the following integral when $g^2r\gg1$ 
\ie\label{eq:CSinf_Green_Im_long}
\text{Im}[G_{\mu\nu}(x,z)]=-\,\text{PV}\int_{-\pi}^\pi\frac{dq}{2\pi}\frac{e^{iq z}}{m(q)} \frac{\epsilon_{\mu\nu\rho}x^\rho}{4\pi r^3}+\cdots~.
\fe
Here, we took $g^2r$ to be the largest variable in the approximation. In particular, we implicitly assumed $g^2r\gg|z|$. 

In \eqref{eq:CSinf_Green_Im_long}, the ``$\cdots$'' denotes the correction to this approximation. The leading order correction comes from the light modes with small $m(q)$ for which $h_r$ deviates from $1/m$ significantly. These light modes are distributed around the gapless momenta so the leading order correction is given by 
\ie\label{eq:CSinf_Green_Im_mom}
\sum_{j=1}^{\ell}\int_{-\infty}^\infty\frac{dq}{2\pi}\frac{e^{i(q+q_j) z}}{4\pi r}\frac{e^{-|m_j q|r}(1+|m_j q|r)}{m_j qr}\frac{\epsilon_{\mu\nu\rho}x^\rho}{r}
=i\sum_{j=1}^{\ell}\frac{e^{i q_j z}}{ \pi m_j}\left(\frac{\zeta_j}{1+\zeta_j^2}+\text{arctan}(\zeta_j)\right)\frac{\epsilon_{\mu\nu\rho}x^\rho}{4\pi r^3}~,
\fe
where $\zeta_j$ is defined in \eqref{eq:define_zeta}. The correction is significant if $\zeta_j\sim 1$. 

Combining \eqref{eq:CSinf_Green_Im_long} and \eqref{eq:CSinf_Green_Im_mom}, the imaginary part of the Green's function in the regime $g^2 r\gtrsim |z|$, $g^2r \gg 1$  is
\ie\label{eq:CSinf_Green_Im}
\text{Im}[G_{\mu\nu}(x,z)]=\left[-\text{PV}\int_{-\pi}^\pi\frac{dq}{2\pi}\frac{e^{iq z}}{m(q)} +i\sum_{j=1}^{\ell}\frac{e^{i q_j z}}{\pi m_j}\left(\frac{\zeta_j}{1+\zeta_j^2}+\text{arctan}(\zeta_j)\right)\right]\frac{\epsilon_{\mu\nu\rho}x^\rho}{4\pi r^3}~.
\fe
Here, we didn't assume $|z|\gg 1$. If we further assume $|z|\gg 1$, then we are in the regime $g^2 r\sim|z|\gg 1$ and the Green's function is
\ie\label{eq:CSinf_Green_Im_large_z}
\text{Im}[G_{\mu\nu}(x,z)]=i\sum_{j=1}^{\ell}e^{i q_j z}\left[-\frac{\text{sign}(z)}{2m_j}  +\frac{1}{\pi m_j}\left(\frac{\zeta_j}{1+\zeta_j^2}+\text{arctan}(\zeta_j)\right)\right]\frac{\epsilon_{\mu\nu\rho}x^\rho}{4\pi r^3}~,
\fe
where $\text{sign}(z)=+1$ for $z>0$ and $\text{sign}(z)=-1$ for $z<0$.
Here, we approximated the integral of \eqref{eq:CSinf_Green_Im} but kept only the piece that does not decay with $|z|$. It is the same approximation as in \eqref{eq:braiding_phase_gapless_general}.

Substituting \eqref{eq:CSinf_Green_Im} into \eqref{eq:CSinfty_Wilson_correlation}, we obtain the phase of the correlation function $\langle W^{I+z}(C_1)W^{I}(C_2)\rangle$. In the regime where $g^2 r\gg |z|\gg 1$, we have $\zeta_j\ll 1$, $\text{Im}[G_{\mu\nu}(x,z)]$ is given by the $-\text{sign}(z)/2m_j$ part of \eqref{eq:CSinf_Green_Im_large_z}, and the phase of the correlation function is
\ie
\exp\left[\sum_{j=1}^{\ell} \frac{g^2}{ 2m_j}e^{iq_j|z|}\text{link}(C_1,C_2)\right]~.
\fe
The phase depends on $|z|$ instead of $z$ because there is a $\text{sign}(z)$ in \eqref{eq:CSinf_Green_Im_large_z} and the gapless momenta come in pairs, i.e.~for each $j$ there is a $j^*\neq j$ such that $q_{j^*}=-q_j$ and $m_{j^*}=-m_{j}$.
The phase is topological in the sense that it depends only on the linking number $\text{link}(C_1,C_2)$ of the two curves $C_1,C_2$. 

In the regime where $g^2 r\sim|z|$, $g^2r\gg 1$, we have $\zeta_j\sim1$, $\text{Im}[G_{\mu\nu}(x,z)]$ is given by \eqref{eq:CSinf_Green_Im_large_z} and the phase of the correlation function is no longer topological. Let us specialize to the configuration where $C_1$ is a curve extending in the time-direction at the spatial origin and $C_2$ is a circle at a fixed time centered at the origin. The phase of the correlation function is 
\ie\label{eq:phase_gamma=1}
\exp\left[\sum_{j=1}^{\ell}\frac{g^2 e^{i q_j |z|}}{2m_j}\left(1-\frac{ |z|}{\sqrt{m_j^2r^2+z^2}}\right)\right]~.
\fe
The phase of the correlation function \eqref{eq:phase_gamma=1} can be interpreted as the Aharonov-Bohm phase that measures the magnetic fluxes attached to a static gauge charge, at the $z$th layer inside a circle of radius $r$. A flux string that extend in the $z$ direction is attached to an gauge charge. The net flux of the flux string on each layer does not decay with $|z|$ but the distribution becomes more and more spread out as $|z|$ increase. 

\bigskip
\hfill
    \emph{Gapless iCSM theory with $\Gamma>1$}
\hfill\bigskip   

If the theory is gapless with $\Gamma>1$, as above, we can decompose the correlation function into its magnitude and its phase. Let $r$ be the characteristic distance between $C_1$ and $C_2$. Using the long-distance behavior of the electric potential \eqref{eq:large_rz_potential_Gamma}, the magnitude of the correlation function is expected to decay as $\exp\left(-|g^2r|^{1-1/\Gamma}\right)$ at long-distance $g^2r\gg 1$. The decay rate is slower than the perimeter law. This signals the deconfinement in the gapless iCSM theory. What about the phase of the correlation function? The discussions in Section~\ref{sec:braiding} suggest that the phase diverge in the $g^2r\rightarrow \infty$ limit so the correlation function does not have a well-defined long-distance limit.

\section{Electric Conductivity}
\label{sec:Hall}

The iCSM theory has a $U(1)$ symmetry per layer. On the $I$th layer, it is generated by the conserved current $J_\mu^I=\frac{1}{2\pi}\epsilon_{\mu\nu\rho}\partial^\nu a^{I,\rho}$. As we discussed in Section \ref{sec:iCS_global_sym}, these $U(1)$ symmetries do not act faithfully on the local operators. Nevertheless, we can couple the current $J_\mu^I$ to an external electromagnetic gauge field $A_\mu^I$ localized on the $I$th layer via the coupling $\mathcal{L}\supset iA^I_\mu J^{I,\mu}$, and measure the response current on another layer that is $z$ layers apart. At frequency $\omega$, this defines the AC conductivity tensor $\sigma_{ij}(\omega,z)$.

The conductivity can be calculated using the linear response theory. We review the calculation in the $U(1)$ gauge theory in Appendix \ref{app:U(1)} and the conductivity \eqref{single_12_U(1)} is:
\ie\label{single_12_U(1)_maintext}
    \sigma_{xx}(\omega)&=\sigma_{yy}(\omega)=\frac{g^2}{4\pi^2}\frac{-i(\omega+i0^+)}{m^2-(\omega+i0^+)^2}~,\\
    \sigma_{xy}(\omega)&=-\sigma_{yx}(\omega)=\frac{g^2}{4\pi^2}\frac{m}{m^2-(\omega+i0^+)^2}~.
\fe
Here, we include the proper $i0^+$ prescription. In the iCSM theory, the action \eqref{action} is diagonalized in the momentum basis so we can use the results \eqref{single_12_U(1)_maintext} in the $U(1)$ gauge theory and do a Fourier transform  to obtain the AC conductivity in the iCSM theory:
\ie\label{CSi_AC}
    \sigma_{xx}(\omega,z)&=\sigma_{yy}(\omega,z)=\frac{g^2}{4\pi^2}\int_{-\pi}^\pi\frac{dq}{2\pi}\frac{-i(\omega+i0^+)}{m(q)^2-(\omega+i0^+)^2}e^{iq z}~, \\
    \sigma_{xy}(\omega,z)&=-\sigma_{yx}(\omega,z)=\frac{g^2}{4\pi^2}\int_{-\pi}^\pi\frac{dq}{2\pi}\frac{m(q)}{m(q)^2-(\omega+i0^+)^2}e^{iq z}~.
\fe
We are particularly interested in the DC conductivity $\sigma_{ij}(0,z)$, which is obtained by taking the $\omega\rightarrow0$ limit of the AC conductivity $\sigma_{ij}(\omega,z)$. Recall that without the Chern-Simons term, the $U(1)$ gauge theory is gapless and has a divergent DC longitudinal conductivity and a vanishing DC Hall conductivity. On the other hand, with the Chern-Simons term, the $U(1)$ gauge theory is gapped and it has a vanishing DC longitudinal conductivity and a nontrivial finite DC Hall conductivity.  Below, we will discuss three situations.

\bigskip
\hfill
    \emph{Gapped iCSM theory}
\hfill\bigskip 

If the theory is gapped, in the DC limit $\omega\rightarrow 0$, we get
    \ie
        \sigma_{xx}(0,z)&=0, \\
        \sigma_{xy}(0,z)&=\frac{1}{2\pi}\int_{-\pi}^\pi\frac{dq}{2\pi}\frac{e^{iq z}}{\lambda(q)} 
        =\frac{1}{2\pi}\left(K^{-1}\right)_{I,I+z}~.\label{Hall_gapped}
    \fe
    Because of the translation symmetry, the second expression of the DC Hall conductivity does not depend on $I$.
    The DC longitudinal conductivity vanishes because the theory is gapped. The DC Hall conductivity agrees with the long-distance braiding phase \eqref{eq:braiding_gapped} and, in particular, it decays exponentially in $|z|$.
    
    As an example, consider a gapped iCSM theory with a tridiagonal $K$ matrix \eqref{tridiagonal} that has $R>2$ and $c_0,c_1>0$. The DC conductivities are
    \ie
    &\sigma_{xx}(0,z)=0~,
    \\
    &\sigma_{xy}(0,z)=\frac{1}{2\pi}\frac{u_1^{|z|}}{\sqrt{c_0^2-4c_1^2}}~,
    \fe
    where $u_{1}=(-c_0+\sqrt{c_0^2-4c_1^2})/{2c_1}$.
    
    \bigskip
\hfill
    \emph{Gapless iCSM theory with $\Gamma=1$}
\hfill\bigskip 
    
    If the theory is gapless with $\Gamma=1$,  it has a finite DC longitudinal conductivity and a finite DC Hall conductivity.
    
    Let us first consider the longitudinal conductivity. For small $|\omega|\ll g^2$, the longitudinal conductivity $\sigma_{xx}(\omega,z)$ is dominated by light modes with small $m(q)$, so it can be approximated by 
    \ie\label{eq:sigma_11_small_omega}
    \sigma_{xx}(\omega,z)&\approx\frac{g^2}{4\pi^2}\sum_{j=1}^\ell\int_{-\infty}^{\infty}\frac{d q}{2\pi} \frac{-i\omega}{m_j^2q^2-(\omega+i0^+)^2} e^{i (q+q_j)z}\\
    &=\frac{g^2}{4\pi^2}\sum_{j=1}^\ell\frac{e^{iq_j z+i\omega|z|/|m_j|}}{2|m_j|}~,
    \fe
    where the sum sums over the gapless momenta $q_j$. In the DC limit $\omega\rightarrow 0$, we obtain the DC longitudinal conductivity
    \ie
    \sigma_{xx}(0,z)=\frac{1}{4\pi}\sum_{j=1}^\ell\frac{e^{iq_j z}}{|\lambda_j|}~.
    \fe
    This finite DC longitudinal conductivity should be contrasted with the divergent DC longitudinal conductivity of a gapless $U(1)$ gauge theory.
    
    We now consider the Hall conductivity. Taking the DC limit $\omega\rightarrow 0$ in  \eqref{CSi_AC}, we obtain the DC Hall conductivity 
    \ie
    \sigma_{xy}(0,z)=\frac{g^2}{4\pi^2}\int_{-\pi}^\pi\frac{dq}{2\pi}\frac{m(q)}{m(q)^2+0^+}e^{iq z}
    =\frac{1}{2\pi}\text{PV}\int_{-\pi}^\pi\frac{dq}{2\pi}\frac{e^{iq z}}{\lambda(q)}~.
    \fe
    Note that without the $0^+$ prescription, the integral in the first expression is ambiguous because $m(q)$ vanishes at the gapless momenta. The $0^+$ prescription gives a unique way to evaluate the integral that is we should take the Cauchy principal value of the integral as defined in \eqref{PV_def}. If $0^+$ is replaced by a finite positive number, the integrand in the first expression behaves like Figure~\ref{fig:PV} near the gapless momenta, where the divergence of the integrand is regularized is a symmetric way on the two sides of gapless momenta.  As the positive number decreases to $0^+$, the integral converges to its Cauchy principal value with $0^+$ effectively serving as the cutoff $\epsilon$ in \eqref{PV_def}. The DC Hall conductivity agrees with the long-distance braiding phase \eqref{eq:braiding_gapless_linear_crossing}. In particular, when $|z|\gg1$, it becomes
    \ie
    \sigma_{xy}(0,z)=\frac{i}{4\pi }\sum_{j=1}^{\ell} \frac{e^{iq_j|z|}}{ \lambda_j}~,
    \fe
    which does not decay but only oscillates in $|z|$. The finite non-zero DC Hall conductivity should be contrasted with the vanishing DC Hall conductivity of a gapless $U(1)$ gauge theory.
    
    As an example, consider a gapless iCSM theory with a tridiagonal $K$ matrix that has $0<c_0<2c_1$. The DC conductivities are
    \ie
    &\sigma_{xx}(0,z)=\frac{1}{2\pi}\frac{\cos(q_1|z|)}{\sqrt{4c_1^2-c_0^2}}~,
    \\
    &\sigma_{xy}(0,z)=\frac{1}{2\pi}\frac{\sin(q_1|z|)}{\sqrt{4c_1^2-c_0^2}}~,
    \fe
    where $e^{iq_1}=(-c_0+i\sqrt{4c_1^2-c_0^2})/{2c_1}$.
    
    \bigskip
\hfill
    \emph{Gapless iCSM theory with $\Gamma>1$}
\hfill
\bigskip

    If the theory is gapless with $\Gamma>1$, both the DC longitudinal conductivity and the DC Hall conductivity diverge. 
    
    For small $|\omega|\ll g^2$, both $\sigma_{xx}$ and $\sigma_{xy}$ are dominated by the contributions around the gapless momenta. Consider a gapless momentum $q_j$ with an exponent $\gamma_j>1$. Around it, we have $m(q)\approx m_j(q-q_j)^{\gamma_j}$ and we can approximate the integral of $\sigma_{xx}$ and $\sigma_{xy}$ by
    \ie
    \sigma_{xx}(\omega,z)&=\frac{g^2}{4\pi^2}\int_{-\infty}^{\infty}\frac{d q}{2\pi}\frac{-i\omega}{m_j^2q^{2\gamma_j}-\omega^2}e^{i (q+q_j) z} + \cdots~,
    \\
    \sigma_{xy}(\omega,z)&=\frac{g^2}{4\pi^2}\int_{-\infty}^\infty\frac{dq}{2\pi}\frac{m_jq^{\gamma_j}}{m_j^2q^{2\gamma_j}-\omega^2}e^{i(q+q_j) z} + \cdots~.
    \fe
    where $\cdots$ denotes the integral away from $q=q_j$.
    In the DC limit $\omega\rightarrow 0$, both integrals diverge as $\omega^{-1+\gamma_j^{-1}}$. The larger $\gamma_j$ is, the faster the integral diverges. Therefore, including the contributions from all gapless momenta, both the DC longitudinal conductivity and the DC Hall conductivity diverge as $\omega^{-1+\Gamma^{-1}}$.  The divergent DC Hall conductivity is consistent with the divergent long-distance braiding phase.

\bigskip
\bigskip

In summary, the DC longitudinal conductivity vanishes for the gapped iCSM theories, is finite for the gapless iCSM theories with $\Gamma=1$, and diverges for the gapless iCSM theories with $\Gamma>1$; the DC Hall conductivity is finite for the gapped iCSM theories and the gapless iCSM theories with $\Gamma=1$ and diverges for the gapless iCSM theories with $\Gamma>1$.

Let us comment on whether the DC limit $\omega\rightarrow 0$ and the $N\rightarrow \infty$ limit commute. The discussions on the Hall conductivity is similar to the discussions in Section \ref{sec:braiding} that concern whether the $r\rightarrow \infty$ limit and the $N\rightarrow\infty$ limit of the braiding phase commute. The finite $N$, AC conductivity is
\ie\label{CSN_AC}
\sigma_{11}(\omega,z)=\frac{1}{N}\sum_{q}\frac{g^2}{4\pi^2}\frac{-i\omega}{m(q)^2-\omega^2}e^{i q z}~,\\
    \sigma_{12}(\omega,z)=\frac{1}{N}\sum_{q}\frac{g^2}{4\pi^2}\frac{m(q)}{m(q)^2-\omega^2}e^{i q z}~,
\fe
where the sum is over quantized momenta $q= \frac{2\pi k}{N}$ with $k=1,\ldots,N$.  Starting from the AC conductivity with finite $N$, if we first take the $N\rightarrow\infty$ limit and then take the DC limit $\omega\rightarrow0$, we obtain the DC conductivity in the iCSM theory. On the other hand, if we first take the DC limit, we obtain
\ie\label{eq:conductivity_CSN}
    \sigma_{11}(\omega,z)&=\frac{g^2}{4\pi^2}\frac{i}{ \omega}\frac{1}{N}\sum_{|\lambda(q)|=0} e^{i q z}+\mathcal{O}(\omega)~,\\
    \sigma_{12}(0,z)&=\frac{1}{2\pi }\frac{1}{N}\sum_{|\lambda(q)|>0} \frac{e^{i q z}}{\lambda(q)}~,
\fe
where the sum in the first line sums over gapless, quantized momenta $q$ and the sum in the second line is over gapped, quantized momenta. If we further take the $N\rightarrow\infty$ limit, we recover the DC conductivity of an iCSM theory if it is gapped. However, for a gapless iCSM theory, the answer depends on whether the gapless momenta are commensurate or incommensurate:
\begin{itemize}[leftmargin=*]
\item
For a theory with only incommensurate gapless momenta, we get a vanishing DC longitudinal conductivity because such a theory cannot be gapless at finite $N$, and an ambiguous answer for the DC Hall conductivity because the limit does not exist.
\item
For a theory with only commensurate gapless momenta, the limit does not exist for both the DC longitudinal conductivity and the DC Hall conductivity. Such a theory can be gapless on certain subsequences of $N$, on which the DC longitudinal conductivity diverges, and gapped on other subsequences of $N$, on which the DC longitudinal conductivity vanishes. Therefore, the $N\rightarrow\infty$ limit of the DC longitudinal conductivity does not exist. Regarding the DC Hall conductivity, although the limit does not exist, there are different converging subsequences. If we follow a particular subsequence on which the theories have the same gapless momenta as the iCSM theory, the $N\rightarrow\infty$ limit of the DC Hall conductivity does exist and it converges to the DC conductivity of the iCSM theory.
\item
For a theory that has both commensurate and incommensurate gapless momenta, the $N\rightarrow\infty$ limit again does not exist for both the DC longitudinal conductivity and the DC Hall conductivity. 
\end{itemize}
To conclude, the DC limit $\omega\rightarrow 0$ and the $N\rightarrow \infty$ limit do not commute in the gapless iCSM theory. Since $g^2$ is the only scale in the theory, this non-commutativity is related to the fact that the $g\rightarrow\infty$ limit and $N\rightarrow \infty$ limit do not commute in the gapless iCSM theory. We will discuss this non-commutativity further in Section \ref{sec:effective_theory}.

\section{Effective Field Theory}
\label{sec:effective_theory}

The iCSM theory is defined with a continuous $\tau,x,y$ coordinate but a discrete $z$ coordinate. In this section, we will take the continuum limit in the $z$ direction and derive a fully continuous effective field theory for the iCSM theory. We will focus on a gapless iCSM theory with $\Gamma=1$. Such a theory has linear, scale invariant dispersion relations at low energy, so its continuum field theory is expected to be also scale invariant (up to some modulations). The method used in this section is similar to the one used in Appendix~\ref{app:CDW} for deriving an effective continuum field theory for a 1+1d lattice model that spontaneously breaks the translation symmetry.

Consider a gapless iCSM theory with $\Gamma=1$. The gapless momenta are $q_j$, $j=1,\ldots,\ell$ and the dispersion relation near each $q_j$ is
\ie\label{eq:low_energy_spec_tridiag}
\omega^2=k_x^2+k_y^2+m_j^2(q-q_j)^2~,\quad m_j=\frac{g^2}{2\pi}\lambda_j~.
\fe
For each gapless momentum $q_j$, there exists a different $q_{j^*}$ such that $q_{j^*}=-q_j$.

Let us define a set of continuum variables
\ie\label{eq:lattice_to_continuum}
\hat z\equiv z\mathbf{a}~,\quad  \hat g^2 \equiv g^2\mathbf{a}~,
\fe
where $\mathbf{a}$ is the layer spacing. In the continuum limit, we send $\mathbf{a}\rightarrow 0$ while holding the dimensionless continuum coupling $\hat g$ fixed. Effectively, this zooms in to the low-energy spectrum \eqref{eq:low_energy_spec_tridiag} near the gapless momenta. This motivates us to define a slowly varying, \textit{complex} continuum gauge field $\hat a_j(\hat z)$ for each $q_j$. They are related to the original gauge field $a^z$ by
\ie\label{eq:lattice_continuum_field_CS}
a^z= \sum_{j=1}^{\ell} e^{i q_j z}\hat a_j(z\mathbf{a})+\cdots~,
\fe
where we omit the other heavy fields with momenta away from the gapless momenta. Since $a^z$ is a real gauge field, we have the relation $\hat a_{j^*}(\hat z)=\hat a_j^\dagger(\hat z)$.
Substituting \eqref{eq:lattice_continuum_field_CS} into the action \eqref{action} and then taking the continuum limit, we obtain the continuum effective  action
\ie\label{eq:effective_action_3+1d}
S =\sum_{j=1}^{\ell}\int d^3xd\hat z \left[ \frac{1}{2\hat g^2} \hat f^\dagger_{j,\mu\nu}\hat f^{\mu\nu}_{j}+\frac{\lambda_j}{4\pi}\epsilon^{\mu\nu\rho} \partial_{\hat z}\hat a_{j,\mu}\partial_\nu \hat a^\dagger_{j,\rho}\right]~,
\fe
where  $\mu,\nu=\tau,x,y$. Here, we replaced $\mathbf{a}\sum_{z}\rightarrow \int d\hat z$ and used the fact that $\hat a_{j}(\hat z)$ are slowly varying fields and thus obey the relation
\ie
\sum_{z} \Big(e^{i q_{i} z}\hat a_{ i}( z\mathbf{a})\Big)\Big(e^{-i q_{j} z}\hat a^\dagger_{j}( z\mathbf{a})\Big)=\sum_{z} \delta_{ij} \hat a_{ j}( z\mathbf{a})\hat a^\dagger_{ j}( z\mathbf{a})~.
\fe
The effective continuum field theory \eqref{eq:effective_action_3+1d} develops an isotropic scale symmetry
\ie
(t,x,y,\hat z)\rightarrow (\Lambda t,\Lambda x,\Lambda y,\Lambda \hat z)~.
\fe
In particular, all the couplings in the action are dimensionless.

We now check that this continuum field theory \eqref{eq:effective_action_3+1d} indeed reproduces the low-energy/long-distance behavior of the iCSM theories. Specifically, we will look at the low-energy dispersion relation and the long-distance Green's function. Reproducing this information guarantees that all the physical observables discussed in Sections~\ref{sec:electric_charged}, \ref{sec:Wilson_correlation} and \ref{sec:Hall} can be reproduced by the continuum field theory.

The continuum field theory has plane wave states
\ie
\hat a_{j,\mu}(k_{\tau}, k_x,k_y,q)=\hat C_{j,\mu} e^{i\omega t+i k_x x+i k_y y+i\hat q \hat z}
\fe
with the dispersion relation
\ie
\omega=k_x^2+k_y^2+(\hat m_j \hat q)^2~,\quad \hat m_j=\frac{\hat g^2}{2\pi}\lambda_j~.
\fe
This reproduces the low-energy spectrum \eqref{eq:low_energy_spec_tridiag} if we relate the continuum momentum $\hat q$ and the lattice momentum $q$ by $\hat q = (q\mp q_j)/\mathbf{a}$ near the gapless momentum $\pm q_j$. The number of degrees of freedom also matches. In the iCSM theory, the plane waves have real coefficients for each gapless momentum $q_j$ while in the continuum theory, the plane waves have complex coefficients $\hat C_{j,\mu}$ for each pair of gapless momenta $q_j$ and $q_{j^*}$.

The continuum field theory \eqref{eq:effective_action_3+1d} has Green's function
\ie
\frac{1}{\hat g^2}\langle  \hat a_{i,\mu}(x,\hat z) \hat a_{j,\nu}^\dagger(0,0)\rangle=\delta_{ij}\int\frac{d^3 k d \hat q}{(2\pi)^4}\frac{k^2\delta_{\mu\nu}-k_\mu k_\nu-\hat m_j\hat q\epsilon_{\mu\nu\rho}k^\rho}{k^2\left(k^2+\hat m_j^2\hat q^2\right)}e^{i(kx+\hat q \hat z)}~.
\fe
Using the relation \eqref{eq:lattice_continuum_field_CS}, the long-distance Green's function of the gauge field $a^I$ in the iCSM theory is expected to be reproduced by
\ie\label{eq:low_energy_prop}
G_{\mu\nu}(x,z)=\frac{1}{g^2}\langle  a^{I+z}_\mu(x)a^I_\nu(0)\rangle
=\frac{1}{g^2}\sum_{j=1}^{\ell}e^{i q_j z}\langle  \hat a_{j,\mu}(x,z\mathbf{a})\hat a_{j,\nu}^\dagger(0,0) \rangle ~.
\fe
Let us check if this is true. The real part of \eqref{eq:low_energy_prop} is
\ie
\text{Re}[G_{\mu\nu}(x,z)]=\sum_{j=1}^{\ell} \int\frac{d^3 k d (\hat q\mathbf{a})}{(2\pi)^4}\frac{k^2\delta_{\mu\nu}-k_\mu k_\nu}{k^2\left(k^2+\hat m_j^2\hat q^2\right)}e^{i(kx+\hat q \hat z+q_jz)} ~.
\fe
This reproduces the real part of the long-distance Green's function \eqref{eq:CSinf_Green_Re_mom} if we use the relation \eqref{eq:lattice_to_continuum} and rewrite the integral using $q=\hat q\mathbf{a}$.  The imaginary part of \eqref{eq:low_energy_prop} is
\ie
\text{Im}[G_{\mu\nu}(x,z)]=i\sum_{j=1}^{\ell}\int\frac{d^3 k d (\hat q\mathbf{a})}{(2\pi)^4}\frac{\hat m_j \hat q \epsilon_{\mu\nu\rho} k^\rho}{k^2\left(k^2+\hat m_j^2\hat q^2\right)}e^{i(kx+\hat q \hat z+q_jz)} ~.
\fe
After doing the integration on $k$, using the relation \eqref{eq:lattice_to_continuum} and rewriting the integral in terms of $q=\hat q \mathbf{a}$, we obtain
\begin{align}
&\text{Im}[G_{\mu\nu}(x,z)]=-\sum_{j=1}^{\ell}\int_{-\infty}^\infty\frac{dq}{2\pi}\frac{e^{i(q+q_j) z}}{4\pi r}\frac{1-e^{-|m_j q|r}(1+|m_j q|r)}{m_j qr}\frac{\epsilon_{\mu\nu\rho}x^\rho}{r} \nonumber
\\
&=\sum_{j=1}^{\ell}\left[-\,\text{PV}\int_{-\infty}^\infty\frac{dq}{2\pi}\frac{e^{i(q+q_j) z}}{m_j q} +\text{PV}\int_{-\infty}^\infty\frac{dq}{2\pi}\frac{e^{i(q+q_j) z-|m_j q|r}(1+|m_j q|r)}{m_j q}\right]\frac{\epsilon_{\mu\nu\rho}x^\rho}{4\pi r^3} \nonumber
\\
&=i\sum_{j=1}^{\ell}e^{i q_j z}\left[-\frac{\text{sign}(z)}{2m_j}+\frac{1}{\pi m_j}\left(\frac{\zeta_j}{1+\zeta_j^2}+\text{arctan}(\zeta_j)\right)\right]\frac{\epsilon_{\mu\nu\rho}x^\rho}{4\pi r^3}~,\label{eq:eff_imaginary_2}
\end{align}
where $\zeta_j=z/|m_j|r$. The integrand in the first line is regular everywhere, but in the second line we split it into two functions that have opposite singularities at $q=0$. To accommodate the singularities such that the result agrees with the first line, we use the Cauchy principal value prescription. The first Cauchy principal value integral can be performed exactly using
\ie
\text{PV}\int_{-\infty}^\infty\frac{dq}{2\pi}\frac{e^{iq z}}{q}=\frac{i}{2}\text{sign}(z)~.
\fe
The second Cauchy principal value integral is precisely \eqref{eq:CSinf_Green_Im_mom}. This exactly reproduces the imaginary part of the long-distance Green's function \eqref{eq:CSinf_Green_Im_large_z} in the regime $g^2 r\sim|z|\gg 1$. Note that the first Cauchy principal value integral in the second line of \eqref{eq:eff_imaginary_2} is not the same integral that appears in \eqref{eq:CSinf_Green_Im_long}. The two integrals agree only in the regime $g^2 r\sim|z|\gg 1$. In this regime, the integral of \eqref{eq:CSinf_Green_Im_long} is dominated by the residues on the unit circle.

In summary, the fully continuous effective field theory \eqref{eq:effective_action_3+1d} reproduces both the low-energy dispersion relations and the long-distance Green's function of a gapless iCSM theory with $\Gamma=1$, and therefore also reproduces all the physical observables considered in Sections~\ref{sec:electric_charged}, \ref{sec:Wilson_correlation} and \ref{sec:Hall} at the low-energy/long-distance limit.

The continuum effective field theory \eqref{eq:effective_action_3+1d} is derived by taking the continuum limit of the miscroscopic theory \eqref{action} with finite $N$ number of gauge fields. It amounts to dialing the miscroscopic coupling constant $g^2$ while we increase $N$ such that $g^2/N$ is held fixed. A more natural setting is to fix the microscopic coupling $g^2$ and view the continuum effective field theory as a low-energy/long-distance description of the system. This means that we would like to consider the system at a scale $g^2r,|z|\gg 1$. However notice that if we consider a finite $N$ theory at such scale $g^2r\gg1$, most of the Maxwell terms are washed away and the action is dominated by the Chern-Simons terms and some decoupled Maxwell terms that solely describe the exactly gapless degrees of freedom at the commensurate gapless momenta. It means that if we take the long-distance limit before taking the $N\rightarrow\infty$ limit, we get a rather boring theory that is very different from the gapless iCSM theory. On the other hand, throughout this paper, we always take the $N\rightarrow\infty$ limit first and then take the long-distance limit. In such an order of limit, the theory already develops a continuum gapless spectrum before we take the $N\rightarrow\infty$ limit and hence we cannot decouple the gapless degrees of freedom from the gapped ones and the whole system is described by the effective field theory \eqref{action}.
In practice, we don't have a system with infinite $N$ and in such a system, the effective field theory \eqref{action} is a good approximate description in the regime $N\gg g^2r,|z|\gg 1$ or equivalently the energy scale $E$ that obeys $1\gg E/g^2 \gg1/N$.

We end this section with a comment on the global aspect of the continuum theory. The continuum theory is not uniquely specified by the continuum action \eqref{eq:effective_action_3+1d}. It should be supplemented with more information such as which bundles are summed over in the path integral and what large gauge symmetries are allowed. This information is crucial because it determines what the gauge invariant observables are and whether $\lambda_j$ should be quantized. Using the map \eqref{eq:lattice_to_continuum}, we should at least demand that $\exp(i\oint a^z)=\exp\left(\sum_{j=1}^{\ell} e^{i q_j z}\hat a_j(z\mathbf{a})\right)$ be a gauge invariant operator under the large gauge transformation. Understanding the global aspects of the continuum theory is beyond the scope of this paper and we will leave it for future work.

\begin{acknowledgments}
We are indebted to discussion with Jason Alicea, Steve Girvin, Pranay Gorantla, Biao Lian, Nathan Seiberg, Shu-Heng Shao and Samson Shatashvili. X.M. and X.C. are supported by the National Science Foundation under award number DMR-1654340, the Simons collaboration on ``Ultra-Quantum Matter'' (grant number 651440), the Simons Investigator Award (award ID 828078) and the Institute for Quantum Information and Matter at Caltech. X.C. is also supported by the Walter Burke Institute for Theoretical Physics at Caltech.  H.T.L. is supported in part by a Croucher fellowship from the Croucher Foundation, the Packard
Foundation and the Center for Theoretical Physics at MIT. H.T.L. thanks the Simons Center for Geometry and Physics for their hospitality
during the ``Geometrical Aspects of Topological Phases of Matter: Spatial Symmetries, Fractons and Beyond" program. The authors of this paper were ordered
alphabetically. 

\end{acknowledgments}
\appendix

\section{A Representation of the iCSM Lagrangian Using Laurent Polynomial}
\label{sec:Polynomial}

In this appendix, we give a representation of the iCSM Lagrangian \eqref{action} using the Laurent polynomial \eqref{polynomial}. Define 
\ie\label{eq:def_au}
a_\mu(u)=\sum_{I} a_\mu^I u^I~.
\fe
The iCSM Lagrangian \eqref{action} can be written as
\ie\label{integral_Lagrangian}
\mathcal{L}&=\oint \frac{du}{2\pi i }\frac{1}{u}\left[\frac{1}{4g^2}f_{\mu\nu}(u)f^{\mu\nu}(u^{-1})+\frac{i}{4\pi}p(u)\epsilon^{\mu\nu\rho} a_\mu(u) \partial_\nu a_\rho(u^{-1})\right]~.
\fe
where the complex counter integral is along the unit circle with $|u|=1$.\footnote{We thank Samson Shatashvili for helpful discussions.} To recover the original Lagrangian \eqref{action}, we can substitute \eqref{eq:def_au} and \eqref{polynomial} into \eqref{integral_Lagrangian} and then use the residue theorem to pick up the residues at $u=0$. Substituting $u=e^{iq}$ into \eqref{eq:def_au} and \eqref{integral_Lagrangian}, one recognizes that $a_\mu(e^{iq})$ is the Fourier transform of $a_\mu^I$ and that the complex contour integral \eqref{integral_Lagrangian} becomes the Lagrangian in the momentum basis.

\section{Review of \texorpdfstring{2+1D $U(1)$}{} Gauge Theory}\label{app:U(1)}

In this appendix, we review the calculations of various observables in 2+1D $U(1)$ gauge theory:
\ie\label{U(1)_action}
\mathcal{L} = \frac{1}{4g^2}f^{\mu\nu} f_{\mu\nu}+\frac{iK}{4\pi}\epsilon^{\mu\nu\rho}a_\mu\partial_\nu a_\rho~.
\fe
They include the Green's function, the correlation functions of the Wilson lines, the electric potential and braiding statistics between gauge charges, and the electric conductivity. 

\subsection{Green's Function}

The Euclidean Green's function
\ie\label{eq:green}
G_{\mu\nu}(x)=\frac{1}{g^2}\langle a_\mu(x) a_\nu(0)\rangle=\int \frac{d^3k}{(2\pi)^3} G_{\mu\nu}(k)e^{i kx}~,
\fe
is given by the Fourier transform of the momentum space propagator 
\begin{align}\label{propagator_U(1)}
    G_{\mu\nu}(k)=\frac{1}{g^2}\langle a_\mu(k) a_\nu(-k)\rangle=\frac{k^2\delta_{\mu\nu}-k_\mu k_\nu-m\epsilon_{\mu\nu\rho}k^\rho}{k^2\left(k^2+m^2\right)}~,
\end{align}
where $m=g^2\frac{K}{2\pi}$ and $|m|$ is the mass of the photon.

The real part of the Green's function is given by
\ie\label{eq:real_green}
\text{Re}[G_{\mu\nu}(x)]=\int\frac{d^3k}{(2\pi)^3}\frac{k^2\delta_{\mu\nu}-k_\mu k_\nu}{k^2(k^2+m^2)}e^{ikx}~.
\fe
The spacetime symmetry constrains it to be of the form
\ie
\text{Re}[G_{\mu\nu}(x)]=f(r)\delta_{\mu\nu} + g(r) \frac{x_\mu x_\nu}{r^2}~,
\fe
where $r^2=x_\mu x^\mu$.
We can fix $f(r)$ and $g(r)$ by specializing to $(\tau,x,y)=(r,0,0)$. Taking the trace on  \eqref{eq:real_green}, we get
\ie
3f(r)+g(r)&=\int\frac{d^3k}{(2\pi)^3}\frac{2}{k^2+m^2}e^{ikx}
\\
&=\int_0^{\infty}\frac{k^2dk}{2\pi}\int_{0}^\pi\frac{\sin(\theta)d\theta}{2\pi}\frac{2}{k^2+m^2}e^{ik r\cos(\theta)}
\\
&=\frac{e^{-|m| r}}{2\pi r}
~.
\fe
Taking $\mu=\nu=\tau$ in \eqref{eq:real_green}, we get
\ie
f(r)+g(r)&=\int\frac{d^3k}{(2\pi)^3}\frac{k^2-k_\tau^2}{k^2(k^2+m^2)}e^{ikx}
\\
&=\int_0^{\infty}\frac{k^2dk}{2\pi}\int_{0}^\pi\frac{\sin(\theta)d\theta}{2\pi}\frac{\sin^2(\theta)}{k^2+m^2}e^{ik r\cos(\theta)}
\\
&=\frac{1}{2\pi r}\frac{1-(1+|m|r)e^{-|m|r}}{m^2r^2}
\fe
After solving for $f(r)$ and $g(r)$, we obtain
\ie\label{eq:U(1)_green_re}
\text{Re}[G_{\mu\nu}(x)]=\frac{e^{-|m|r}}{6\pi r}\delta_{\mu\nu}+\frac{1}{4\pi r}\left[\frac{1}{m^2r^2}-\left(\frac{1}{3}+\frac{1}{|m|r}+\frac{1}{m^2r^2}\right)e^{-|m|r}\right]Q_{\mu\nu}~,
\fe
where
\ie
Q_{\mu\nu}=\frac{3x_\mu x_\nu-r^2\delta_{\mu\nu}}{r^2}
\fe
is the (traceless) quadrupole tensor.

The imaginary part of the Green's function is given by
\ie
\text{Im}[G_{\mu\nu}(x)]=\int\frac{d^3k}{(2\pi)^3}\frac{i m\epsilon_{\mu\nu\rho}k^\rho}{k^2(k^2+m^2)}e^{ik x}~.
\fe
The spacetime symmetry constrains it to be of the form
\ie
\text{Im}[G_{\mu\nu}(x)]=h(r)\frac{\epsilon_{\mu\nu\rho}x^\rho}{r}~.
\fe
We can fix $h(r)$ by specializing to $(\tau,x,y)=(r,0,0)$. Taking $\mu=x$, $\nu=y$, we get
\ie
h(r)&=\int\frac{d^3k}{(2\pi)^3}\frac{i m k_\tau}{k^2(k^2+m^2)}e^{ik x}
\\
&=\int_0^{\infty}\frac{k^2dk}{2\pi}\int_{0}^\pi\frac{\sin(\theta)d\theta}{2\pi}\frac{i m k\cos(\theta)}{k^2(k^2+m^2)}e^{ik r\cos(\theta)}
\\
&=-\frac{1}{4\pi r}\frac{1-(1+|m|r)e^{-|m|r}}{mr}~.
\fe
This gives
\ie\label{eq:U(1)_green_im}
\text{Im}[G_{\mu\nu}(x)]=-\frac{1}{4\pi r}\frac{1-(1+|m|r)e^{-|m|r}}{mr}\frac{\epsilon_{\mu\nu\rho}x^\rho}{r}~.
\fe

We consider two limits of the Green's function. In the $|m|r\gg1$ limit, the Lagrangian \eqref{U(1)_action} is dominated by the Chern-Simons term and the Green's function reduces to
\ie\label{eq:CS_green}
G_{\mu\nu}(x)=
-\frac{1}{4\pi r}\frac{i}{mr}\frac{\epsilon_{\mu\nu\rho}x^\rho}{r}~.
\fe
The real part is suppressed by a factor of ${1}/{|m|r}$.
In the $|m|r\ll 1$ limit, the Lagrangian \eqref{U(1)_action} is dominated by the Maxwell term and the Green's function reduces to
\ie\label{eq:Maxwell_green}
G_{\mu\nu}(x)= \frac{1}{4\pi r}\frac{x_\mu x_\nu+r^2\delta_{\mu\nu}}{2r^2}~.
\fe
The imaginary part is suppressed by a factor of $|m|r$.

\subsection{Correlation Functions of Wilson Lines}\label{sec:U(1)correlation}
Consider two Wilson lines $W=\exp(i\oint a)$ on closed curves $C_1$ and $C_2$. Their correlation function is given by 
\ie\label{eq:Wilson_line_correlation}
\langle W(C_1) W(C_2)\rangle=\exp\left(-g^2\oint_{C_1} dx_1^\mu\oint_{C_2} dx_2^\nu\, G_{\mu\nu}(x_1-x_2) \right)~.
\fe
We are interested in the behavior of this correlation function at long distance. When $K\neq0$, the long-distance Green's function is given by \eqref{eq:CS_green}.
Substituting it into \eqref{eq:Wilson_line_correlation}, we obtain a topological correlation function at long distance \cite{Polyakov:1988md}:
\ie\label{eq:topological_Wilson}
\langle W(C_1) W(C_2)\rangle= \exp\left(-\frac{2\pi i}{K}\text{link}(C_1,C_2)\right)~.
\fe
The correlation function is topological in the sense that it depends only on the linking number $\text{link}(C_1,C_2)$ between $C_1$ and $C_2$. When $K=0$, the Green's function is given by \eqref{eq:Maxwell_green}. Substituting it into \eqref{eq:Wilson_line_correlation}, we obtain a non-topological real-valued correlation function that decays faster than any perimeter law. It signals the confinement in the Maxwell theory.

\subsection{Electric Potential and Braiding Statistics}

We can couple the theory to electrically charged particles via the coupling $\mathcal{L}\supset ia_\mu j^\mu$ where $j_\mu$ denotes the current of the charged particles. The coupling modifies the equation of motion to
\ie
\frac{i}{g^2}\partial_\nu f^{\mu\nu}-\frac{1}{2\pi}K \epsilon^{\mu\nu\rho} \d_\nu a_\rho=j_\mu~.
\fe
For a static gauge charge, $j_\mu = \delta_{\mu\tau}\delta(\vec{x})$ and the solution to the equation of motion is
\ie\label{eq:guage_source}
&a_\tau =-ig^2\int \frac{d^2\vec{k}}{(2\pi)^2}\frac{e^{i\vec{k}\cdot\vec{x}}}{\vec{k}^2+m^2}= -\frac{ig^2}{2\pi} K_0(|m|r)~,
\\
&a_i =-ig^2\int \frac{d^2\vec{k}}{(2\pi)^2}\frac{m\epsilon_{ij}k^je^{i\vec{k}\cdot\vec{x}}}{\vec{k}^2(\vec{k}^2+m^2)}= \frac{g^2\epsilon_{ij}x^j}{2\pi mr^2}\Big[1-|m|rK_1(|m|r)\Big]~.
\fe
where $K_n(x)$ are the modified Bessel functions of the second kind. 

The electric potential $V(r)$ sourced by a static gauge charge is given by
\ie\label{eq:U(1)_potential}
V(r)=i a_\tau(r)=\frac{g^2}{2\pi} K_0(|m|r)~.
\fe
where $a_\tau$ is given by the solution \eqref{eq:guage_source}.
We can also extract the electric potential from the correlation functions of the Wilson lines. Take the Euclidean time to be a circle with a large periodicity $T$. Consider two Wilson lines wrapping around the Euclidean time cycle separated by a distance $r$. They represent the worldlines of two static charges. The electric potential $V(r)$ is encoded in the correlation function via
\ie\label{eq:potential}
\langle W(r) W(0)\rangle=\exp(-V(r)T)~.
\fe
Substituting \eqref{eq:Wilson_line_correlation}, \eqref{eq:green} and \eqref{propagator_U(1)} into \eqref{eq:potential}, 
we obtain the electric potential \eqref{eq:U(1)_potential}.

For $K=0$, we should take the $m\rightarrow0$ limit of \eqref{eq:U(1)_potential} and it yields the familiar Coulomb potential in 2+1D
\ie
V(r) = -\frac{g^2}{2\pi}\log(r)+\text{const}~.
\fe
Since the potential diverges logarithmically with $r$, the theory is confined.\footnote{If we include monopole operators in the Lagrangian, the electric potential diverges linearly with $r$ \cite{Polyakov:1976fu}.}

For $K\neq0$,  the potential decays exponentially at large $r$
\ie
V(r)\sim \frac{g^2}{2\sqrt{2\pi}}\frac{e^{-|m|r}}{\sqrt{|m|r}}~.
\fe
It implies that the theory is deconfined.

We now study the braiding of gauge charges. Consider the following braiding process: we take a gauge charge around a circle of radius $r$ with another one sitting at the center of the circle. The braiding picks up an Aharonov–Bohm phase 
\ie\label{eq:braiding_U(1)}
\exp(i\phi(r))=\exp\left(i\oint \vec{a}\cdot d\vec{x} \right)= \exp\left[-\frac{2\pi i}{K}\big(1-|m|rK_1(|m|r)\big)\right]
\fe
where $\vec{a}=(a_x,a_y)$ is given by the solution \eqref{eq:guage_source}. The braiding phase in this setup has been computed in \cite{PhysRevX.6.041003}.  

For $K\neq 0$, in the long-distance limit $|m|r\rightarrow\infty$, the braiding phase \eqref{eq:braiding_U(1)} converges to
\ie
\exp(i\phi_\infty)=\exp\left[-\frac{2\pi i}{K}\right]~.
\fe
This reproduced the expected fractional braiding statistics of the gauge charges i.e. anyons. 

For $K=0$, we should take the $K\rightarrow 0$ limit of the braiding phase \eqref{eq:braiding_U(1)}. This gives a vanishing braiding phase as expected.

The braiding phase can also be extracted from the phase of the correlation functions of the Wilson lines. For the above setup, we should consider the correlation function of two Wilson lines: one extends in the time-direction at the spatial origin and the other one wraps around a circle of radius $r$ with its center at the origin. Substituting  \eqref{eq:green} and \eqref{propagator_U(1)} into \eqref{eq:Wilson_line_correlation} and taking the phase, we obtain \eqref{eq:braiding_U(1)}. We can also consider the braiding phase for more general curves. Denote the characteristic scale of the curves by $r$. When $g^2 r\gg 1$ and $K\neq0$, the correlation functions of the Wilson lines \eqref{eq:topological_Wilson} are topological and hence the braiding phase is also topological.

\subsection{Conductivity}
The theory has a $U(1)$ symmetry with a conserved current $J^\mu = \frac{1}{2\pi}\epsilon^{\mu\nu\rho}\partial_\nu a_\rho$. We can couple the current to an external electromagnetic gauge field $A_\mu$ via the coupling $\mathcal{L}\supset iA_\mu J^\mu$. 
The conductivity tensor $\sigma_{ij}(\omega)$ at frequency $\omega$ is defined by the relation,
\ie
J_i(\omega) = \sigma_{ij}(\omega) E_j(\omega)~, \quad i,j=x,y~,
\fe
where $J_i(\omega)$ and $E_i(\omega)$ are the Fourier transform of the current $J_i$ and external electric field $E_i=\partial_0 A_i - \partial_i A_0$.

The conductivity tensor $\sigma_{ij}(\omega)$ can be computed using standard linear response theory. In the Lorentzian signature and the $A_0=0$ gauge, the current is related to the vector potential $A_i$ via
\ie
J_i(t,x,y)=\int d^3x'\,\Pi^{\text{L}}_{ij}(t-t',x-x',y-y')A_j(t',x',y')~,
\fe
where $\Pi^{\text{L}}_{ij}(t,x,y)$ is the retarded response function  
\ie
\Pi^{\text{L}}_{ij}(t,x,y)\equiv-i\Theta(t)\left<\left[J_i(t,\vec{x}),J_j(0,\vec{0})\right]\right>~.
\fe
Define $\Pi^{\text{L}}_{ij}(\omega,k_x,k_y)$ to be the Fourier transform of $\Pi^{\text{L}}_{ij}(t,x,y)$. Because of the relation $E_j(\omega)=i\omega A_j(\omega)$ in the $A_0=0$ gauge, the conductivity tensor is
\begin{align}
    \sigma_{ij}(\omega)=\frac{1}{i\omega}\Pi^{\text{L}}_{ij}(\omega,0,0)~.
\end{align}

We can compute $\Pi^{\text{L}}_{ij}(\omega,k_x,k_y)$ by doing a Wick rotation on the Euclidean momentum space current-current correlator $\Pi^{\text{E}}_{ij}(k_\alpha)\equiv\left\langle J_i(k_\alpha)J_j(-k_\alpha)\right>$:
\ie
\Pi^{\text{L}}_{ij}(\omega,k_x,k_y)=-\Pi^{\text{E}}_{ij}(-i(\omega+i0^+),k_x,k_y)~.
\fe
In our system, we have
\ie
\Pi^{\text{E}}_{ij}(k_\alpha)
=\frac{g^2}{4\pi^2}\epsilon_{i\beta\gamma }\epsilon_{j\rho\lambda}k^\beta k^\rho G^{\gamma\lambda}(k_\alpha)~,
\fe
where $G^{\gamma\lambda}(k_\alpha)$ is the propagator \eqref{propagator_U(1)}.
For our computation of conductivity, the relevant components are
\begin{align}
    \Pi^{\text{E}}_{xx}(k_\tau,0,0)&=\frac{g^2}{4\pi^2}\frac{k_\tau^2}{m^2+k_\tau^2}~, \label{corr_11}\\
    \Pi^{\text{E}}_{xy}(k_\tau,0,0)&=\frac{g^2}{4\pi^2}\frac{mk_\tau}{m^2+k_\tau^2} \label{corr_12}~.
\end{align}
They lead to the conductivity tensor
\ie
    \sigma_{xx}(\omega)&=\sigma_{yy}(\omega)=\frac{g^2}{4\pi^2}\frac{-i\omega}{m^2-\omega^2}~,\\
    \sigma_{xy}(\omega)&=-\sigma_{yx}(\omega)=\frac{g^2}{4\pi^2}\frac{m}{m^2-\omega^2}~. \label{single_12_U(1)}
\fe
Here, more precisely, all the $\omega$ should be $\omega+i0^+$. 

To obtain the DC conductivity, we take the $\omega\rightarrow 0$ limit. For $K\neq0$, we have
\ie\label{eq:DC_CS}
\sigma_{xx}(0) = 0~,\quad \sigma_{xy}(0)=\frac{1}{2\pi K}~.
\fe
The DC longitudinal conductivity $\sigma_{xx}(0)$ vanishes because the theory is gapped. The DC Hall conductivity $\sigma_{xy}(0)$ is a fraction of the quantized Hall conductivity ${1}/{(2\pi)}$ as expected. For $K=0$, we have
\ie\label{eq:DC_Maxwell}
\sigma_{xx}(\omega)=\frac{g^2}{4\pi^2}\frac{i}{\omega}~,\quad \sigma_{xy}(0)=0~.
\fe
The DC longitudinal conductivity $\sigma_{xx}(0)$ diverges because the $U(1)$ symmetry is spontaneously broken. The DC Hall conductivity vanishes because the theory is parity invariant.

\section{Integer Bounded Null Vectors of Infinite-dimensional \texorpdfstring{$K$}{} Matrices}\label{app:null_vector}

In this appendix, we study the integer bounded null vectors of an infinite-dimensional translation invariant $K$ matrix \eqref{Kmatrix}.  The $K$ matrix has a set of bounded eigenvectors $v_I(q)=e^{iq I}$ with eigenvalue $\lambda(q)$ where $q\sim q+2\pi$. Among these bounded vectors, those with vanishing eigenvalues are the bounded null vectors. Their $q$'s are the gapless momenta. We call the bounded null vectors with rational $\frac{q}{2\pi}$, \emph{the commensurate null vectors}, and the vector space they span, \emph{the commensurate subspace}. The incommensurate null vectors and incommensurate subspace are defined similarly. Denote the dimension of the commensurate subspace by $\ell_c$. We will prove that the integer bounded null vectors of the infinite-dimensional $K$ matrix form a $\ell_c$-dimensional lattice within the commensurate subspace.

Note that here we are restricted to integer bounded null vectors. An infinite-dimensional $K$ matrix in principle can have integer unbounded null vectors.  As an example, consider the $K$ matrix associated to the Laurent polynomial $p(u)=u+3+u^{-1}$. The null vectors $w_I$ obey
\ie
w_{I-1}+3w_I+w_{I+1}=0~.
\fe
It can be solved recursively after specifying the initial condition on $w_0$ and $w_1$. Since in the recurrence relation the coefficients of $w_{I-1}$ and $w_{I+1}$ are both 1, the solutions are integral if $w_0$ and $w_1$ are integers. These integral solutions are
\ie
w_I&=\left(\frac{5-3\sqrt{5}}{10}w_0-\frac{1}{\sqrt{5}}w_1\right)\left(\frac{-3-\sqrt{5}}{2}\right)^I+\left(\frac{5+3\sqrt{5}}{10}w_0+\frac{1}{\sqrt{5}}w_1\right)\left(\frac{-3+\sqrt{5}}{2}\right)^I~,
\fe
where $w_0$ and $w_1$ are integers.
They are clearly all unbounded except for the trivial solution that has $w_0=w_1=0$. This example is consistent with the above statement that we would like to prove. It has no null vectors in the form of $v_I(q)=e^{iqI}$ so $\ell_c=0$. This agrees with the fact that there are no integer bounded null vectors.

We now prove the above statement. We will first show that the integer bounded null vectors contain a $\ell_c$-dimensional sub-lattice in the commensurate subspace. Denote the $\ell_c$ commensurate gapless momenta by $q_j$ with $j=1,\ldots,{\ell_c}$. Let $\mathcal{N}$ be the minimum integer such that $q_j\mathcal{N}=0\text{ mod }2\pi$ for all commensurate gapless momenta $q_j$. We can truncate the translation invariant $K$ matrix to a $\mathcal{N}$-dimensional matrix $\mathcal{K}$ with a periodic boundary condition. The matrix $\mathcal{K}$ has $\ell_c$ number of null vectors $\mathcal{V}_I(q_j)=e^{i q_j I}$ where $j=1,\ldots,\ell_c$. As discussed in Section \ref{sec:SNF}, the integer matrix $\mathcal{K}$ has $\ell_c$ number of linearly-independent integer null vectors denoted by $\mathcal{W}_{I,j}$ where $j=1,\ldots,\ell_c$. Repeating these integer null vectors, we can construct $\ell_c$ number of linearly-independent integer bounded null vectors of the infinite-dimensional $K$ matrix:\ $W_{I,j} = \mathcal{W}_{(I\text{ mod }N),j}$.  Since $\mathcal{W}_{I,j}$ are linear combinations of $\mathcal{V}_I(q_j)$ and $v_I(q_j)$ are the repetitions of $\mathcal{V}_I(q_j)$, the infinite-dimensional vectors $W_{I,j}$ are also linear-combinations of $v_I(q_j)$ where $j=1,\ldots,\ell_c$. Hence, the $\ell_c$ integer bounded null vectors $\mathcal{W}_{I,j}$, $j=1,\ldots,\ell_c$, form a $\ell_c$-dimensional lattice in the commensurate subspace and we complete the first part of the proof.

Next, we complete the proof by showing that all the integer bounded null vectors must be in the commensurate subspace. Let $W_I$ be an integer bounded null vector of the infinite-dimensional $K$ matrix \eqref{Kmatrix}. It obeys a $2\xi$-th order recurrence relation
\ie
\sum_{k=-\xi}^\xi c_k W_{I+k}=0~,
\fe
for every integer $I$.
Knowing the values of any $2\xi$ consecutive entries, we can deduce all the other entries of $W_I$ by iterations. Since $W_I$ is bounded, there exists an integer $R$  such that $|W_I|\leq R$ for all $I$. It implies that there are at most $(2R+1)^{2\xi}$ possible combinations for $2\xi$ consecutive entries of $W_I$ so $W_I$ must repeat itself after certain translation, i.e. $W_{I+\mathcal{N}} = W_I$ for some $\mathcal{N}\leq (2R+1)^{2\xi}$. Let $\mathcal{W}_I$ be a $\mathcal{N}$-dimensional trunction of $W_I$. Then, $\mathcal{W}_I$ should be a null vector of the $\mathcal{N}$-dimensional truncation of the infinite-dimensional $K$ matrix with a periodic boundary condition. Hence it must be the linear-combinations of $\mathcal{V}_I(q_j)$, $j=1,\ldots,\ell_c$. Repeating $\mathcal{W}_I$ and $\mathcal{V}_I(q_j)$, $j=1,\ldots,\ell_c$, we recover the infinite-dimensional vectors $W_I$ and $v_I(q_n)$, $j=1,\ldots,\ell_c$. Hence the integer bounded null vector $W_I$ is a vector in the commensurate subspace. This completes the proof.

\section{An Effective Field Theory for Spontaneous Breaking of Translation Symmetry}\label{app:CDW}

In this appendix, we discuss a 1+1d lattice model, which shares many similarities with the iCSM theories. In particular, this lattice model spontaneously breaks the translation symmetry. We will derive an effective continuum field theory description of this system. Similar method is used in Section \ref{sec:effective_theory} to derive the effective field theory of the iCSM theories.

We work with a continuous time and a discrete space. The space is a 1D lattice with sites labeled by $ x$. The Euclidean action is
\ie
S =\int d\tau\sum_{ x} \left[ \frac{1}{2 U}(\partial_\tau \phi_{x})^2+\frac{ K}{2}\left(m \phi_{ x-1} +n \phi_{ x} + m \phi_{ x+1}\right)^2\right]~,
\fe
where $ \phi_{ x}$ is a non-compact real field on the site $ x$.

Let us first consider the spectrum. In Lorentzian signature, there are plane wave states
\ie\label{eq:lattice_plane_wave}
 \phi_{ x}(t) =  C e^{i\omega t + i  k  x}~,
\fe
with momentum $ k\in[-\pi,\pi)$ and the dispersion relation
\ie
\omega( k)^2= U  K\left(n+2m\cos( k)\right)^2~.
\fe
The spectrum is gapless if $|n/m|\leq 2$. The gapless momenta are $ q=\pm \arccos(-n/2m)$. Expanding around the gapless momentum $\pm  q$, we obtain a linear dispersion relation
\ie\label{eq:low_energy_spec}
\omega( k)^2=  v^2( k\mp  q)^2+\cdots~,\quad   v^2=4m^2 U K\sin( q)^2= U K(4m^2-n^2)~.
\fe
If $n/m=\pm 2$, $v$ vanishes and the dispersion relation becomes quadratic in the leading order. 

We will focus on the gapless theories with $|n/m|<2$. These theories spontaneously break the translation symmetry and have linear dispersion relations around the gapless momenta $\pm q$. 

Let us define a set of continuum variables:
\ie
\hat x=  x \mathbf{a}~,\quad  \hat U = U \mathbf{a}~,\quad \hat K= K\mathbf{a}
\fe
where $\mathbf{a}$ is the lattice spacing. In the continuum limit, we send $\mathbf{a}\rightarrow 0$ while holding the continuum coupling $\hat U$ and $\hat K$ fixed. Effectively, this zooms into the low-energy spectrum \eqref{eq:low_energy_spec} near the two gapless momenta $\pm  q$. This motivates us to define a complex slow varying continuum field $\hat \phi_{  q}(\hat x)$, which are related to the lattice field $ \phi( x)$ by
\ie\label{eq:lattice_continuum_field}
 \phi_{ x}(\tau)= e^{i q  x}\hat \phi_{ q}(\tau,x\mathbf{a}  )+e^{-iq  x}\hat \phi_{ q}^\dagger(\tau,x\mathbf{a}  )+\cdots~,
\fe
where we omit the other heavy fields with momenta $k$ away from the gapless momenta $\pm  q$.
In the continuum limit, the action becomes
\ie\label{eq:effective_action_1+1d}
S =\int d\tau d\hat x \left[ \frac{1}{\hat U} |\partial_\tau\hat \phi_{  q}|^2+4m^2\hat K\sin( q)^2|\partial_{\hat x}\hat \phi_{  q}|^2\right]~,
\fe
where we replace $\mathbf{a}\sum_{ x}\rightarrow \int d\hat x$ and use the fact that $\phi_{\pm q}(\hat x)$ are slow varying fields so  they obey the relation
\ie
\sum_{ x} \Big(e^{i q x}\hat \phi_{ q}(x\mathbf{a})\Big)^{n_1}\Big(e^{-i q x}\hat \phi_{ q}^\dagger( x\mathbf{a})\Big)^{n_2}=\delta_{n_1,n_2}\sum_{ x} \hat \phi_{ q}( x\mathbf{a})^{n_1}\hat \phi_{ q}^\dagger(x\mathbf{a})^{n_2}~.
\fe

The continuum field theory \eqref{eq:effective_action_1+1d} has plane wave state
\ie\label{eq:continuum_plane_wave}
\phi_{ q}(t,\hat x) = \hat C e^{i\omega t + i  \hat k  \hat x}~,
\fe
with the dispersion relation
\ie
\omega(\hat k)^2 = \hat v^2\hat k^2~,\quad \hat v^2=4m^2\hat U\hat K\sin( q)^2=\hat U\hat K(4m^2-n^2)~. 
\fe
It reproduces the low-energy spectrum \eqref{eq:low_energy_spec} of the lattice model if we identify $\hat k \mathbf{a}=k\mp q$ in the neighborhood of the gapless momentum $\pm  q$. The number of degrees of freedom also matches. On the lattice, there are two gapless momenta and $ C$ in \eqref{eq:lattice_plane_wave} is a real variable while in the continuum, there is one complex field and $\hat C$ in \eqref{eq:continuum_plane_wave} is a complex variable.

The continuum  field theory \eqref{eq:effective_action_1+1d} also captures the long-distance correlation functions. Since the theory is free, all of its correlation functions can be obtained from the Euclidean two-point function. On the lattice, it is given by
\ie
\langle \phi_{ x}(\tau) \phi_0(0)\rangle&\,=\int^{\infty}_{-\infty} \frac{d\omega}{2\pi} \int_{-\pi}^\pi\frac{d k}{2\pi}\frac{ U e^{i\omega \tau +i k x}}{\omega^2+ U K\left(n+2m\cos( k)\right)^2}
=\frac{ U}{2} \int_{-\pi}^\pi\frac{d k}{2\pi}\frac{e^{-\omega( k) |\tau| +i k x}}{\omega( k)}~,
\fe
where $\omega( k)=\sqrt{ U K}|n+2m\cos( k)|$. For $ \sqrt{UK}\tau\gg1$, the two-point function is dominated by the light modes with small $\omega( k)$ near the gapless momenta $\pm  q$ so it can be approximated by
\ie\label{eq:lattice_2pt}
\langle \phi_{ x}(\tau) \phi_0(0)\rangle&\,= U \int_{-\infty}^\infty\frac{d  k}{2\pi}\frac{e^{- v| k| |\tau| +i k x}}{ v| k|}\cos( q  x)~.
\fe
Using the relation \eqref{eq:lattice_continuum_field}, the long-distance lattice two-point function can be reproduced from the continuum theory \eqref{eq:effective_action_1+1d}:
\ie\label{eq:continuum_2pt}
\langle \phi_{ x}(\tau) \phi_0(0)\rangle&\,=e^{-iq  x}\langle \hat \phi_{ q}^\dagger(\tau,  x\mathbf{a}))\hat \phi_{ q}(0,0)\rangle+e^{i q  x}\langle \hat \phi_{ q}(\tau,x \mathbf{a})\hat \phi_{ q}^\dagger(0,0)\rangle
\\
&\,=\hat U\int_{-\infty}^\infty \frac{d\hat k}{2\pi}\frac{e^{-\hat v|\hat k|\tau +i\hat k x\mathbf{a}}}{\hat v|\hat k|}\cos( q  x)~.
\fe
The continuum expression agrees with \eqref{eq:lattice_2pt} if we rescale $\hat k = k/\mathbf{a}$.

\bibliographystyle{apsrev4-2}
\bibliography{ref}

\begin{thebibliography}{44}%
\makeatletter
\providecommand \@ifxundefined [1]{%
 \@ifx{#1\undefined}
}%
\providecommand \@ifnum [1]{%
 \ifnum #1\expandafter \@firstoftwo
 \else \expandafter \@secondoftwo
 \fi
}%
\providecommand \@ifx [1]{%
 \ifx #1\expandafter \@firstoftwo
 \else \expandafter \@secondoftwo
 \fi
}%
\providecommand \natexlab [1]{#1}%
\providecommand \enquote  [1]{``#1''}%
\providecommand \bibnamefont  [1]{#1}%
\providecommand \bibfnamefont [1]{#1}%
\providecommand \citenamefont [1]{#1}%
\providecommand \href@noop [0]{\@secondoftwo}%
\providecommand \href [0]{\begingroup \@sanitize@url \@href}%
\providecommand \@href[1]{\@@startlink{#1}\@@href}%
\providecommand \@@href[1]{\endgroup#1\@@endlink}%
\providecommand \@sanitize@url [0]{\catcode `\\12\catcode `\$12\catcode
  `\&12\catcode `\#12\catcode `\^12\catcode `\_12\catcode `\%12\relax}%
\providecommand \@@startlink[1]{}%
\providecommand \@@endlink[0]{}%
\providecommand \url  [0]{\begingroup\@sanitize@url \@url }%
\providecommand \@url [1]{\endgroup\@href {#1}{\urlprefix }}%
\providecommand \urlprefix  [0]{URL }%
\providecommand \Eprint [0]{\href }%
\providecommand \doibase [0]{https://doi.org/}%
\providecommand \selectlanguage [0]{\@gobble}%
\providecommand \bibinfo  [0]{\@secondoftwo}%
\providecommand \bibfield  [0]{\@secondoftwo}%
\providecommand \translation [1]{[#1]}%
\providecommand \BibitemOpen [0]{}%
\providecommand \bibitemStop [0]{}%
\providecommand \bibitemNoStop [0]{.\EOS\space}%
\providecommand \EOS [0]{\spacefactor3000\relax}%
\providecommand \BibitemShut  [1]{\csname bibitem#1\endcsname}%
\let\auto@bib@innerbib\@empty
\bibitem [{\citenamefont {Wen}\ and\ \citenamefont
  {Zee}(1992)}]{PhysRevB.46.2290}%
  \BibitemOpen
  \bibfield  {author} {\bibinfo {author} {\bibfnamefont {X.~G.}\ \bibnamefont
  {Wen}}\ and\ \bibinfo {author} {\bibfnamefont {A.}~\bibnamefont {Zee}},\
  }\href {https://doi.org/10.1103/PhysRevB.46.2290} {\bibfield  {journal}
  {\bibinfo  {journal} {Phys. Rev. B}\ }\textbf {\bibinfo {volume} {46}},\
  \bibinfo {pages} {2290} (\bibinfo {year} {1992})}\BibitemShut {NoStop}%
\bibitem [{\citenamefont {Ma}\ \emph {et~al.}(2022)\citenamefont {Ma},
  \citenamefont {Shirley}, \citenamefont {Cheng}, \citenamefont {Levin},
  \citenamefont {McGreevy},\ and\ \citenamefont {Chen}}]{ma2020fractonic}%
  \BibitemOpen
  \bibfield  {author} {\bibinfo {author} {\bibfnamefont {X.}~\bibnamefont
  {Ma}}, \bibinfo {author} {\bibfnamefont {W.}~\bibnamefont {Shirley}},
  \bibinfo {author} {\bibfnamefont {M.}~\bibnamefont {Cheng}}, \bibinfo
  {author} {\bibfnamefont {M.}~\bibnamefont {Levin}}, \bibinfo {author}
  {\bibfnamefont {J.}~\bibnamefont {McGreevy}},\ and\ \bibinfo {author}
  {\bibfnamefont {X.}~\bibnamefont {Chen}},\ }\href
  {https://doi.org/10.1103/PhysRevB.105.195124} {\bibfield  {journal} {\bibinfo
   {journal} {Phys. Rev. B}\ }\textbf {\bibinfo {volume} {105}},\ \bibinfo
  {pages} {195124} (\bibinfo {year} {2022})}\BibitemShut {NoStop}%
\bibitem [{\citenamefont {Arkani-Hamed}\ \emph {et~al.}(2001)\citenamefont
  {Arkani-Hamed}, \citenamefont {Cohen},\ and\ \citenamefont
  {Georgi}}]{Arkani-Hamed:2001kyx}%
  \BibitemOpen
  \bibfield  {author} {\bibinfo {author} {\bibfnamefont {N.}~\bibnamefont
  {Arkani-Hamed}}, \bibinfo {author} {\bibfnamefont {A.~G.}\ \bibnamefont
  {Cohen}},\ and\ \bibinfo {author} {\bibfnamefont {H.}~\bibnamefont
  {Georgi}},\ }\href {https://doi.org/10.1103/PhysRevLett.86.4757} {\bibfield
  {journal} {\bibinfo  {journal} {Phys. Rev. Lett.}\ }\textbf {\bibinfo
  {volume} {86}},\ \bibinfo {pages} {4757} (\bibinfo {year} {2001})},\ \Eprint
  {https://arxiv.org/abs/hep-th/0104005} {arXiv:hep-th/0104005} \BibitemShut
  {NoStop}%
\bibitem [{\citenamefont {Hill}\ \emph {et~al.}(2001)\citenamefont {Hill},
  \citenamefont {Pokorski},\ and\ \citenamefont {Wang}}]{Hill:2000mu}%
  \BibitemOpen
  \bibfield  {author} {\bibinfo {author} {\bibfnamefont {C.~T.}\ \bibnamefont
  {Hill}}, \bibinfo {author} {\bibfnamefont {S.}~\bibnamefont {Pokorski}},\
  and\ \bibinfo {author} {\bibfnamefont {J.}~\bibnamefont {Wang}},\ }\href
  {https://doi.org/10.1103/PhysRevD.64.105005} {\bibfield  {journal} {\bibinfo
  {journal} {Phys. Rev. D}\ }\textbf {\bibinfo {volume} {64}},\ \bibinfo
  {pages} {105005} (\bibinfo {year} {2001})},\ \Eprint
  {https://arxiv.org/abs/hep-th/0104035} {arXiv:hep-th/0104035} \BibitemShut
  {NoStop}%
\bibitem [{\citenamefont {Arkani-Hamed}\ \emph {et~al.}(2003)\citenamefont
  {Arkani-Hamed}, \citenamefont {Cohen}, \citenamefont {Kaplan}, \citenamefont
  {Karch},\ and\ \citenamefont {Motl}}]{Arkani-Hamed:2001wsh}%
  \BibitemOpen
  \bibfield  {author} {\bibinfo {author} {\bibfnamefont {N.}~\bibnamefont
  {Arkani-Hamed}}, \bibinfo {author} {\bibfnamefont {A.~G.}\ \bibnamefont
  {Cohen}}, \bibinfo {author} {\bibfnamefont {D.~B.}\ \bibnamefont {Kaplan}},
  \bibinfo {author} {\bibfnamefont {A.}~\bibnamefont {Karch}},\ and\ \bibinfo
  {author} {\bibfnamefont {L.}~\bibnamefont {Motl}},\ }\href
  {https://doi.org/10.1088/1126-6708/2003/01/083} {\bibfield  {journal}
  {\bibinfo  {journal} {JHEP}\ }\textbf {\bibinfo {volume} {01}},\ \bibinfo
  {pages} {083}},\ \Eprint {https://arxiv.org/abs/hep-th/0110146}
  {arXiv:hep-th/0110146} \BibitemShut {NoStop}%
\bibitem [{\citenamefont {Razamat}(2021)}]{Razamat:2021jkx}%
  \BibitemOpen
  \bibfield  {author} {\bibinfo {author} {\bibfnamefont {S.~S.}\ \bibnamefont
  {Razamat}},\ }\href {https://doi.org/10.1103/PhysRevLett.127.141603}
  {\bibfield  {journal} {\bibinfo  {journal} {Phys. Rev. Lett.}\ }\textbf
  {\bibinfo {volume} {127}},\ \bibinfo {pages} {141603} (\bibinfo {year}
  {2021})},\ \Eprint {https://arxiv.org/abs/2107.06465} {arXiv:2107.06465
  [hep-th]} \BibitemShut {NoStop}%
\bibitem [{\citenamefont {Geng}\ \emph {et~al.}(2021)\citenamefont {Geng},
  \citenamefont {Kachru}, \citenamefont {Karch}, \citenamefont {Nally},\ and\
  \citenamefont {Rayhaun}}]{Geng:2021cmq}%
  \BibitemOpen
  \bibfield  {author} {\bibinfo {author} {\bibfnamefont {H.}~\bibnamefont
  {Geng}}, \bibinfo {author} {\bibfnamefont {S.}~\bibnamefont {Kachru}},
  \bibinfo {author} {\bibfnamefont {A.}~\bibnamefont {Karch}}, \bibinfo
  {author} {\bibfnamefont {R.}~\bibnamefont {Nally}},\ and\ \bibinfo {author}
  {\bibfnamefont {B.~C.}\ \bibnamefont {Rayhaun}},\ }\href
  {https://doi.org/10.1002/prop.202100133} {\bibfield  {journal} {\bibinfo
  {journal} {Fortsch. Phys.}\ }\textbf {\bibinfo {volume} {69}},\ \bibinfo
  {pages} {2100133} (\bibinfo {year} {2021})},\ \Eprint
  {https://arxiv.org/abs/2108.08322} {arXiv:2108.08322 [hep-th]} \BibitemShut
  {NoStop}%
\bibitem [{\citenamefont {Franco}\ and\ \citenamefont
  {Rodriguez-Gomez}(2022)}]{Franco:2022ziy}%
  \BibitemOpen
  \bibfield  {author} {\bibinfo {author} {\bibfnamefont {S.}~\bibnamefont
  {Franco}}\ and\ \bibinfo {author} {\bibfnamefont {D.}~\bibnamefont
  {Rodriguez-Gomez}},\ }\href {https://doi.org/10.1103/PhysRevLett.128.241603}
  {\bibfield  {journal} {\bibinfo  {journal} {Phys. Rev. Lett.}\ }\textbf
  {\bibinfo {volume} {128}},\ \bibinfo {pages} {241603} (\bibinfo {year}
  {2022})},\ \Eprint {https://arxiv.org/abs/2203.01335} {arXiv:2203.01335
  [hep-th]} \BibitemShut {NoStop}%
\bibitem [{\citenamefont {Qiu}\ \emph {et~al.}(1989)\citenamefont {Qiu},
  \citenamefont {Joynt},\ and\ \citenamefont {MacDonald}}]{PhysRevB.40.11943}%
  \BibitemOpen
  \bibfield  {author} {\bibinfo {author} {\bibfnamefont {X.}~\bibnamefont
  {Qiu}}, \bibinfo {author} {\bibfnamefont {R.}~\bibnamefont {Joynt}},\ and\
  \bibinfo {author} {\bibfnamefont {A.~H.}\ \bibnamefont {MacDonald}},\ }\href
  {https://doi.org/10.1103/PhysRevB.40.11943} {\bibfield  {journal} {\bibinfo
  {journal} {Phys. Rev. B}\ }\textbf {\bibinfo {volume} {40}},\ \bibinfo
  {pages} {11943} (\bibinfo {year} {1989})}\BibitemShut {NoStop}%
\bibitem [{\citenamefont {Qiu}\ \emph {et~al.}(1990)\citenamefont {Qiu},
  \citenamefont {Joynt},\ and\ \citenamefont {MacDonald}}]{PhysRevB.42.1339}%
  \BibitemOpen
  \bibfield  {author} {\bibinfo {author} {\bibfnamefont {X.}~\bibnamefont
  {Qiu}}, \bibinfo {author} {\bibfnamefont {R.}~\bibnamefont {Joynt}},\ and\
  \bibinfo {author} {\bibfnamefont {A.~H.}\ \bibnamefont {MacDonald}},\ }\href
  {https://doi.org/10.1103/PhysRevB.42.1339} {\bibfield  {journal} {\bibinfo
  {journal} {Phys. Rev. B}\ }\textbf {\bibinfo {volume} {42}},\ \bibinfo
  {pages} {1339} (\bibinfo {year} {1990})}\BibitemShut {NoStop}%
\bibitem [{\citenamefont {Naud}\ \emph {et~al.}(2000)\citenamefont {Naud},
  \citenamefont {Pryadko},\ and\ \citenamefont {Sondhi}}]{Naud_2000}%
  \BibitemOpen
  \bibfield  {author} {\bibinfo {author} {\bibfnamefont {J.~D.}\ \bibnamefont
  {Naud}}, \bibinfo {author} {\bibfnamefont {L.~P.}\ \bibnamefont {Pryadko}},\
  and\ \bibinfo {author} {\bibfnamefont {S.~L.}\ \bibnamefont {Sondhi}},\
  }\href {https://doi.org/10.1103/physrevlett.85.5408} {\bibfield  {journal}
  {\bibinfo  {journal} {Physical Review Letters}\ }\textbf {\bibinfo {volume}
  {85}},\ \bibinfo {pages} {5408} (\bibinfo {year} {2000})}\BibitemShut
  {NoStop}%
\bibitem [{\citenamefont {Naud}\ \emph {et~al.}(2001)\citenamefont {Naud},
  \citenamefont {Pryadko},\ and\ \citenamefont {Sondhi}}]{Naud:2000xa}%
  \BibitemOpen
  \bibfield  {author} {\bibinfo {author} {\bibfnamefont {J.~D.}\ \bibnamefont
  {Naud}}, \bibinfo {author} {\bibfnamefont {L.~P.}\ \bibnamefont {Pryadko}},\
  and\ \bibinfo {author} {\bibfnamefont {S.~L.}\ \bibnamefont {Sondhi}},\
  }\href {https://doi.org/10.1016/S0550-3213(00)00679-9} {\bibfield  {journal}
  {\bibinfo  {journal} {Nucl. Phys. B}\ }\textbf {\bibinfo {volume} {594}},\
  \bibinfo {pages} {713} (\bibinfo {year} {2001})},\ \Eprint
  {https://arxiv.org/abs/cond-mat/0006506} {arXiv:cond-mat/0006506}
  \BibitemShut {NoStop}%
\bibitem [{\citenamefont {Chamon}(2005)}]{Chamon:2004lew}%
  \BibitemOpen
  \bibfield  {author} {\bibinfo {author} {\bibfnamefont {C.}~\bibnamefont
  {Chamon}},\ }\href {https://doi.org/10.1103/physrevlett.94.040402} {\bibfield
   {journal} {\bibinfo  {journal} {Phys. Rev. Lett.}\ }\textbf {\bibinfo
  {volume} {94}},\ \bibinfo {pages} {040402} (\bibinfo {year} {2005})},\
  \Eprint {https://arxiv.org/abs/cond-mat/0404182} {arXiv:cond-mat/0404182}
  \BibitemShut {NoStop}%
\bibitem [{\citenamefont {Haah}(2011)}]{Haah:2011drr}%
  \BibitemOpen
  \bibfield  {author} {\bibinfo {author} {\bibfnamefont {J.}~\bibnamefont
  {Haah}},\ }\href {https://doi.org/10.1103/physreva.83.042330} {\bibfield
  {journal} {\bibinfo  {journal} {Phys. Rev. A}\ }\textbf {\bibinfo {volume}
  {83}},\ \bibinfo {pages} {042330} (\bibinfo {year} {2011})},\ \Eprint
  {https://arxiv.org/abs/1101.1962} {arXiv:1101.1962 [quant-ph]} \BibitemShut
  {NoStop}%
\bibitem [{\citenamefont {Vijay}\ \emph {et~al.}(2016)\citenamefont {Vijay},
  \citenamefont {Haah},\ and\ \citenamefont {Fu}}]{Vijay:2016phm}%
  \BibitemOpen
  \bibfield  {author} {\bibinfo {author} {\bibfnamefont {S.}~\bibnamefont
  {Vijay}}, \bibinfo {author} {\bibfnamefont {J.}~\bibnamefont {Haah}},\ and\
  \bibinfo {author} {\bibfnamefont {L.}~\bibnamefont {Fu}},\ }\href
  {https://doi.org/10.1103/PhysRevB.94.235157} {\bibfield  {journal} {\bibinfo
  {journal} {Phys. Rev. B}\ }\textbf {\bibinfo {volume} {94}},\ \bibinfo
  {pages} {235157} (\bibinfo {year} {2016})},\ \Eprint
  {https://arxiv.org/abs/1603.04442} {arXiv:1603.04442 [cond-mat.str-el]}
  \BibitemShut {NoStop}%
\bibitem [{\citenamefont {Seiberg}(2020)}]{Seiberg:2019vrp}%
  \BibitemOpen
  \bibfield  {author} {\bibinfo {author} {\bibfnamefont {N.}~\bibnamefont
  {Seiberg}},\ }\href {https://doi.org/10.21468/SciPostPhys.8.4.050} {\bibfield
   {journal} {\bibinfo  {journal} {SciPost Phys.}\ }\textbf {\bibinfo {volume}
  {8}},\ \bibinfo {pages} {050} (\bibinfo {year} {2020})},\ \Eprint
  {https://arxiv.org/abs/1909.10544} {arXiv:1909.10544 [cond-mat.str-el]}
  \BibitemShut {NoStop}%
\bibitem [{\citenamefont {Seiberg}\ and\ \citenamefont
  {Shao}(2021{\natexlab{a}})}]{Seiberg:2020bhn}%
  \BibitemOpen
  \bibfield  {author} {\bibinfo {author} {\bibfnamefont {N.}~\bibnamefont
  {Seiberg}}\ and\ \bibinfo {author} {\bibfnamefont {S.-H.}\ \bibnamefont
  {Shao}},\ }\href {https://doi.org/10.21468/SciPostPhys.10.2.027} {\bibfield
  {journal} {\bibinfo  {journal} {SciPost Phys.}\ }\textbf {\bibinfo {volume}
  {10}},\ \bibinfo {pages} {027} (\bibinfo {year} {2021}{\natexlab{a}})},\
  \Eprint {https://arxiv.org/abs/2003.10466} {arXiv:2003.10466
  [cond-mat.str-el]} \BibitemShut {NoStop}%
\bibitem [{\citenamefont {Seiberg}\ and\ \citenamefont
  {Shao}(2020)}]{Seiberg:2020wsg}%
  \BibitemOpen
  \bibfield  {author} {\bibinfo {author} {\bibfnamefont {N.}~\bibnamefont
  {Seiberg}}\ and\ \bibinfo {author} {\bibfnamefont {S.-H.}\ \bibnamefont
  {Shao}},\ }\href {https://doi.org/10.21468/SciPostPhys.9.4.046} {\bibfield
  {journal} {\bibinfo  {journal} {SciPost Phys.}\ }\textbf {\bibinfo {volume}
  {9}},\ \bibinfo {pages} {046} (\bibinfo {year} {2020})},\ \Eprint
  {https://arxiv.org/abs/2004.00015} {arXiv:2004.00015 [cond-mat.str-el]}
  \BibitemShut {NoStop}%
\bibitem [{\citenamefont {Seiberg}\ and\ \citenamefont
  {Shao}(2021{\natexlab{b}})}]{Seiberg:2020cxy}%
  \BibitemOpen
  \bibfield  {author} {\bibinfo {author} {\bibfnamefont {N.}~\bibnamefont
  {Seiberg}}\ and\ \bibinfo {author} {\bibfnamefont {S.-H.}\ \bibnamefont
  {Shao}},\ }\href {https://doi.org/10.21468/SciPostPhys.10.1.003} {\bibfield
  {journal} {\bibinfo  {journal} {SciPost Phys.}\ }\textbf {\bibinfo {volume}
  {10}},\ \bibinfo {pages} {003} (\bibinfo {year} {2021}{\natexlab{b}})},\
  \Eprint {https://arxiv.org/abs/2004.06115} {arXiv:2004.06115
  [cond-mat.str-el]} \BibitemShut {NoStop}%
\bibitem [{\citenamefont {Gorantla}\ \emph
  {et~al.}(2022{\natexlab{a}})\citenamefont {Gorantla}, \citenamefont {Lam},
  \citenamefont {Seiberg},\ and\ \citenamefont {Shao}}]{Gorantla:2022eem}%
  \BibitemOpen
  \bibfield  {author} {\bibinfo {author} {\bibfnamefont {P.}~\bibnamefont
  {Gorantla}}, \bibinfo {author} {\bibfnamefont {H.~T.}\ \bibnamefont {Lam}},
  \bibinfo {author} {\bibfnamefont {N.}~\bibnamefont {Seiberg}},\ and\ \bibinfo
  {author} {\bibfnamefont {S.-H.}\ \bibnamefont {Shao}},\ }\href
  {https://doi.org/10.1103/PhysRevB.106.045112} {\bibfield  {journal} {\bibinfo
   {journal} {Phys. Rev. B}\ }\textbf {\bibinfo {volume} {106}},\ \bibinfo
  {pages} {045112} (\bibinfo {year} {2022}{\natexlab{a}})},\ \Eprint
  {https://arxiv.org/abs/2201.10589} {arXiv:2201.10589 [cond-mat.str-el]}
  \BibitemShut {NoStop}%
\bibitem [{\citenamefont {Nandkishore}\ and\ \citenamefont
  {Hermele}(2019)}]{Nandkishore:2018sel}%
  \BibitemOpen
  \bibfield  {author} {\bibinfo {author} {\bibfnamefont {R.~M.}\ \bibnamefont
  {Nandkishore}}\ and\ \bibinfo {author} {\bibfnamefont {M.}~\bibnamefont
  {Hermele}},\ }\href
  {https://doi.org/10.1146/annurev-conmatphys-031218-013604} {\bibfield
  {journal} {\bibinfo  {journal} {Ann. Rev. Condensed Matter Phys.}\ }\textbf
  {\bibinfo {volume} {10}},\ \bibinfo {pages} {295} (\bibinfo {year} {2019})},\
  \Eprint {https://arxiv.org/abs/1803.11196} {arXiv:1803.11196
  [cond-mat.str-el]} \BibitemShut {NoStop}%
\bibitem [{\citenamefont {Pretko}\ \emph {et~al.}(2020)\citenamefont {Pretko},
  \citenamefont {Chen},\ and\ \citenamefont {You}}]{Pretko:2020cko}%
  \BibitemOpen
  \bibfield  {author} {\bibinfo {author} {\bibfnamefont {M.}~\bibnamefont
  {Pretko}}, \bibinfo {author} {\bibfnamefont {X.}~\bibnamefont {Chen}},\ and\
  \bibinfo {author} {\bibfnamefont {Y.}~\bibnamefont {You}},\ }\href
  {https://doi.org/10.1142/S0217751X20300033} {\bibfield  {journal} {\bibinfo
  {journal} {Int. J. Mod. Phys. A}\ }\textbf {\bibinfo {volume} {35}},\
  \bibinfo {pages} {2030003} (\bibinfo {year} {2020})},\ \Eprint
  {https://arxiv.org/abs/2001.01722} {arXiv:2001.01722 [cond-mat.str-el]}
  \BibitemShut {NoStop}%
\bibitem [{\citenamefont {Shirley}\ \emph {et~al.}(2018)\citenamefont
  {Shirley}, \citenamefont {Slagle}, \citenamefont {Wang},\ and\ \citenamefont
  {Chen}}]{Shirley:2017suz}%
  \BibitemOpen
  \bibfield  {author} {\bibinfo {author} {\bibfnamefont {W.}~\bibnamefont
  {Shirley}}, \bibinfo {author} {\bibfnamefont {K.}~\bibnamefont {Slagle}},
  \bibinfo {author} {\bibfnamefont {Z.}~\bibnamefont {Wang}},\ and\ \bibinfo
  {author} {\bibfnamefont {X.}~\bibnamefont {Chen}},\ }\href
  {https://doi.org/10.1103/PhysRevX.8.031051} {\bibfield  {journal} {\bibinfo
  {journal} {Phys. Rev. X}\ }\textbf {\bibinfo {volume} {8}},\ \bibinfo {pages}
  {031051} (\bibinfo {year} {2018})},\ \Eprint
  {https://arxiv.org/abs/1712.05892} {arXiv:1712.05892 [cond-mat.str-el]}
  \BibitemShut {NoStop}%
\bibitem [{\citenamefont {Shirley}\ \emph
  {et~al.}(2019{\natexlab{a}})\citenamefont {Shirley}, \citenamefont {Slagle},\
  and\ \citenamefont {Chen}}]{Shirley:2018nhn}%
  \BibitemOpen
  \bibfield  {author} {\bibinfo {author} {\bibfnamefont {W.}~\bibnamefont
  {Shirley}}, \bibinfo {author} {\bibfnamefont {K.}~\bibnamefont {Slagle}},\
  and\ \bibinfo {author} {\bibfnamefont {X.}~\bibnamefont {Chen}},\ }\href
  {https://doi.org/10.1016/j.aop.2019.167922} {\bibfield  {journal} {\bibinfo
  {journal} {Annals Phys.}\ }\textbf {\bibinfo {volume} {410}},\ \bibinfo
  {pages} {167922} (\bibinfo {year} {2019}{\natexlab{a}})},\ \Eprint
  {https://arxiv.org/abs/1806.08625} {arXiv:1806.08625 [cond-mat.str-el]}
  \BibitemShut {NoStop}%
\bibitem [{\citenamefont {Shirley}\ \emph
  {et~al.}(2019{\natexlab{b}})\citenamefont {Shirley}, \citenamefont {Slagle},\
  and\ \citenamefont {Chen}}]{Shirley:2018hkm}%
  \BibitemOpen
  \bibfield  {author} {\bibinfo {author} {\bibfnamefont {W.}~\bibnamefont
  {Shirley}}, \bibinfo {author} {\bibfnamefont {K.}~\bibnamefont {Slagle}},\
  and\ \bibinfo {author} {\bibfnamefont {X.}~\bibnamefont {Chen}},\ }\href
  {https://doi.org/10.1103/PhysRevB.99.115123} {\bibfield  {journal} {\bibinfo
  {journal} {Phys. Rev. B}\ }\textbf {\bibinfo {volume} {99}},\ \bibinfo
  {pages} {115123} (\bibinfo {year} {2019}{\natexlab{b}})},\ \Eprint
  {https://arxiv.org/abs/1806.08633} {arXiv:1806.08633 [cond-mat.str-el]}
  \BibitemShut {NoStop}%
\bibitem [{\citenamefont {Shirley}\ \emph
  {et~al.}(2019{\natexlab{c}})\citenamefont {Shirley}, \citenamefont {Slagle},\
  and\ \citenamefont {Chen}}]{Shirley:2018vtc}%
  \BibitemOpen
  \bibfield  {author} {\bibinfo {author} {\bibfnamefont {W.}~\bibnamefont
  {Shirley}}, \bibinfo {author} {\bibfnamefont {K.}~\bibnamefont {Slagle}},\
  and\ \bibinfo {author} {\bibfnamefont {X.}~\bibnamefont {Chen}},\ }\href
  {https://doi.org/10.21468/SciPostPhys.6.4.041} {\bibfield  {journal}
  {\bibinfo  {journal} {SciPost Phys.}\ }\textbf {\bibinfo {volume} {6}},\
  \bibinfo {pages} {041} (\bibinfo {year} {2019}{\natexlab{c}})},\ \Eprint
  {https://arxiv.org/abs/1806.08679} {arXiv:1806.08679 [cond-mat.str-el]}
  \BibitemShut {NoStop}%
\bibitem [{\citenamefont {Shirley}\ \emph {et~al.}(2020)\citenamefont
  {Shirley}, \citenamefont {Slagle},\ and\ \citenamefont
  {Chen}}]{Shirley:2019uou}%
  \BibitemOpen
  \bibfield  {author} {\bibinfo {author} {\bibfnamefont {W.}~\bibnamefont
  {Shirley}}, \bibinfo {author} {\bibfnamefont {K.}~\bibnamefont {Slagle}},\
  and\ \bibinfo {author} {\bibfnamefont {X.}~\bibnamefont {Chen}},\ }\href
  {https://doi.org/10.1103/PhysRevB.102.115103} {\bibfield  {journal} {\bibinfo
   {journal} {Phys. Rev. B}\ }\textbf {\bibinfo {volume} {102}},\ \bibinfo
  {pages} {115103} (\bibinfo {year} {2020})},\ \Eprint
  {https://arxiv.org/abs/1907.09048} {arXiv:1907.09048 [cond-mat.str-el]}
  \BibitemShut {NoStop}%
\bibitem [{\citenamefont {Sullivan}\ \emph {et~al.}(2021)\citenamefont
  {Sullivan}, \citenamefont {Dua},\ and\ \citenamefont
  {Cheng}}]{Sullivan:2021rbk}%
  \BibitemOpen
  \bibfield  {author} {\bibinfo {author} {\bibfnamefont {J.}~\bibnamefont
  {Sullivan}}, \bibinfo {author} {\bibfnamefont {A.}~\bibnamefont {Dua}},\ and\
  \bibinfo {author} {\bibfnamefont {M.}~\bibnamefont {Cheng}},\ }\href@noop {}
  {\  (\bibinfo {year} {2021})},\ \Eprint {https://arxiv.org/abs/2109.13267}
  {arXiv:2109.13267 [cond-mat.str-el]} \BibitemShut {NoStop}%
\bibitem [{\citenamefont {Gaiotto}\ \emph {et~al.}(2015)\citenamefont
  {Gaiotto}, \citenamefont {Kapustin}, \citenamefont {Seiberg},\ and\
  \citenamefont {Willett}}]{Gaiotto_2015}%
  \BibitemOpen
  \bibfield  {author} {\bibinfo {author} {\bibfnamefont {D.}~\bibnamefont
  {Gaiotto}}, \bibinfo {author} {\bibfnamefont {A.}~\bibnamefont {Kapustin}},
  \bibinfo {author} {\bibfnamefont {N.}~\bibnamefont {Seiberg}},\ and\ \bibinfo
  {author} {\bibfnamefont {B.}~\bibnamefont {Willett}},\ }\bibfield  {journal}
  {\bibinfo  {journal} {Journal of High Energy Physics}\ }\textbf {\bibinfo
  {volume} {2015}},\ \href {https://doi.org/10.1007/jhep02(2015)172}
  {10.1007/jhep02(2015)172} (\bibinfo {year} {2015})\BibitemShut {NoStop}%
\bibitem [{\citenamefont {Chen}\ \emph {et~al.}()\citenamefont {Chen},
  \citenamefont {Lam},\ and\ \citenamefont {Ma}}]{appear}%
  \BibitemOpen
  \bibfield  {author} {\bibinfo {author} {\bibfnamefont {X.}~\bibnamefont
  {Chen}}, \bibinfo {author} {\bibfnamefont {H.~T.}\ \bibnamefont {Lam}},\ and\
  \bibinfo {author} {\bibfnamefont {X.}~\bibnamefont {Ma}},\ }\href@noop {}
  {\bibinfo  {journal} {work in progress}\ }\BibitemShut {NoStop}%
\bibitem [{\citenamefont {Niven}(1956)}]{Niven}%
  \BibitemOpen
\bibfield  {journal} {  }\bibfield  {author} {\bibinfo {author} {\bibfnamefont
  {I.}~\bibnamefont {Niven}},\ }\href@noop {} {\emph {\bibinfo {title}
  {Irrational Numbers}}}\ (\bibinfo  {publisher} {Mathematical Association of
  America},\ \bibinfo {year} {1956})\BibitemShut {NoStop}%
\bibitem [{\citenamefont {Borokhov}\ \emph {et~al.}(2002)\citenamefont
  {Borokhov}, \citenamefont {Kapustin},\ and\ \citenamefont
  {Wu}}]{Borokhov:2002ib}%
  \BibitemOpen
  \bibfield  {author} {\bibinfo {author} {\bibfnamefont {V.}~\bibnamefont
  {Borokhov}}, \bibinfo {author} {\bibfnamefont {A.}~\bibnamefont {Kapustin}},\
  and\ \bibinfo {author} {\bibfnamefont {X.-k.}\ \bibnamefont {Wu}},\ }\href
  {https://doi.org/10.1088/1126-6708/2002/11/049} {\bibfield  {journal}
  {\bibinfo  {journal} {JHEP}\ }\textbf {\bibinfo {volume} {11}},\ \bibinfo
  {pages} {049}},\ \Eprint {https://arxiv.org/abs/hep-th/0206054}
  {arXiv:hep-th/0206054} \BibitemShut {NoStop}%
\bibitem [{\citenamefont {Callan}\ and\ \citenamefont
  {Harvey}(1985)}]{Callan:1984sa}%
  \BibitemOpen
  \bibfield  {author} {\bibinfo {author} {\bibfnamefont {C.~G.}\ \bibnamefont
  {Callan}, \bibfnamefont {Jr.}}\ and\ \bibinfo {author} {\bibfnamefont
  {J.~A.}\ \bibnamefont {Harvey}},\ }\href
  {https://doi.org/10.1016/0550-3213(85)90489-4} {\bibfield  {journal}
  {\bibinfo  {journal} {Nucl. Phys. B}\ }\textbf {\bibinfo {volume} {250}},\
  \bibinfo {pages} {427} (\bibinfo {year} {1985})}\BibitemShut {NoStop}%
\bibitem [{\citenamefont {Polyakov}(1977)}]{Polyakov:1976fu}%
  \BibitemOpen
  \bibfield  {author} {\bibinfo {author} {\bibfnamefont {A.~M.}\ \bibnamefont
  {Polyakov}},\ }\href {https://doi.org/10.1016/0550-3213(77)90086-4}
  {\bibfield  {journal} {\bibinfo  {journal} {Nucl. Phys. B}\ }\textbf
  {\bibinfo {volume} {120}},\ \bibinfo {pages} {429} (\bibinfo {year}
  {1977})}\BibitemShut {NoStop}%
\bibitem [{\citenamefont {Affleck}\ \emph {et~al.}(1989)\citenamefont
  {Affleck}, \citenamefont {Harvey}, \citenamefont {Palla},\ and\ \citenamefont
  {Semenoff}}]{Affleck:1989qf}%
  \BibitemOpen
  \bibfield  {author} {\bibinfo {author} {\bibfnamefont {I.}~\bibnamefont
  {Affleck}}, \bibinfo {author} {\bibfnamefont {J.~A.}\ \bibnamefont {Harvey}},
  \bibinfo {author} {\bibfnamefont {L.}~\bibnamefont {Palla}},\ and\ \bibinfo
  {author} {\bibfnamefont {G.~W.}\ \bibnamefont {Semenoff}},\ }\href
  {https://doi.org/10.1016/0550-3213(89)90220-4} {\bibfield  {journal}
  {\bibinfo  {journal} {Nucl. Phys. B}\ }\textbf {\bibinfo {volume} {328}},\
  \bibinfo {pages} {575} (\bibinfo {year} {1989})}\BibitemShut {NoStop}%
\bibitem [{\citenamefont {Pace}\ and\ \citenamefont
  {Wen}(2022)}]{Pace:2022cnh}%
  \BibitemOpen
  \bibfield  {author} {\bibinfo {author} {\bibfnamefont {S.~D.}\ \bibnamefont
  {Pace}}\ and\ \bibinfo {author} {\bibfnamefont {X.-G.}\ \bibnamefont {Wen}},\
  }\href@noop {} {\  (\bibinfo {year} {2022})},\ \Eprint
  {https://arxiv.org/abs/2207.03544} {arXiv:2207.03544 [cond-mat.str-el]}
  \BibitemShut {NoStop}%
\bibitem [{\citenamefont {Gorantla}\ \emph
  {et~al.}(2022{\natexlab{b}})\citenamefont {Gorantla}, \citenamefont {Lam},\
  and\ \citenamefont {Shao}}]{Gorantla:2022mrp}%
  \BibitemOpen
  \bibfield  {author} {\bibinfo {author} {\bibfnamefont {P.}~\bibnamefont
  {Gorantla}}, \bibinfo {author} {\bibfnamefont {H.~T.}\ \bibnamefont {Lam}},\
  and\ \bibinfo {author} {\bibfnamefont {S.-H.}\ \bibnamefont {Shao}},\
  }\href@noop {} {\  (\bibinfo {year} {2022}{\natexlab{b}})},\ \Eprint
  {https://arxiv.org/abs/2207.08585} {arXiv:2207.08585 [cond-mat.str-el]}
  \BibitemShut {NoStop}%
\bibitem [{\citenamefont {Gorantla}\ \emph
  {et~al.}(2022{\natexlab{c}})\citenamefont {Gorantla}, \citenamefont {Lam},
  \citenamefont {Seiberg},\ and\ \citenamefont {Shao}}]{Gorantla:2022pii}%
  \BibitemOpen
  \bibfield  {author} {\bibinfo {author} {\bibfnamefont {P.}~\bibnamefont
  {Gorantla}}, \bibinfo {author} {\bibfnamefont {H.~T.}\ \bibnamefont {Lam}},
  \bibinfo {author} {\bibfnamefont {N.}~\bibnamefont {Seiberg}},\ and\ \bibinfo
  {author} {\bibfnamefont {S.-H.}\ \bibnamefont {Shao}},\ }\href@noop {} {\
  (\bibinfo {year} {2022}{\natexlab{c}})},\ \Eprint
  {https://arxiv.org/abs/2210.03727} {arXiv:2210.03727 [cond-mat.str-el]}
  \BibitemShut {NoStop}%
\bibitem [{\citenamefont {Wen}\ and\ \citenamefont
  {Zee}(1993)}]{PhysRevB.47.2265}%
  \BibitemOpen
  \bibfield  {author} {\bibinfo {author} {\bibfnamefont {X.~G.}\ \bibnamefont
  {Wen}}\ and\ \bibinfo {author} {\bibfnamefont {A.}~\bibnamefont {Zee}},\
  }\href {https://doi.org/10.1103/PhysRevB.47.2265} {\bibfield  {journal}
  {\bibinfo  {journal} {Phys. Rev. B}\ }\textbf {\bibinfo {volume} {47}},\
  \bibinfo {pages} {2265} (\bibinfo {year} {1993})}\BibitemShut {NoStop}%
\bibitem [{\citenamefont {Burnell}\ \emph {et~al.}(2022)\citenamefont
  {Burnell}, \citenamefont {Devakul}, \citenamefont {Gorantla}, \citenamefont
  {Lam},\ and\ \citenamefont {Shao}}]{Burnell:2021reh}%
  \BibitemOpen
  \bibfield  {author} {\bibinfo {author} {\bibfnamefont {F.~J.}\ \bibnamefont
  {Burnell}}, \bibinfo {author} {\bibfnamefont {T.}~\bibnamefont {Devakul}},
  \bibinfo {author} {\bibfnamefont {P.}~\bibnamefont {Gorantla}}, \bibinfo
  {author} {\bibfnamefont {H.~T.}\ \bibnamefont {Lam}},\ and\ \bibinfo {author}
  {\bibfnamefont {S.-H.}\ \bibnamefont {Shao}},\ }\href
  {https://doi.org/10.1103/PhysRevB.106.085113} {\bibfield  {journal} {\bibinfo
   {journal} {Phys. Rev. B}\ }\textbf {\bibinfo {volume} {106}},\ \bibinfo
  {pages} {085113} (\bibinfo {year} {2022})},\ \Eprint
  {https://arxiv.org/abs/2110.09529} {arXiv:2110.09529 [cond-mat.str-el]}
  \BibitemShut {NoStop}%
\bibitem [{\citenamefont {Lake}(2018)}]{Lake:2018dqm}%
  \BibitemOpen
  \bibfield  {author} {\bibinfo {author} {\bibfnamefont {E.}~\bibnamefont
  {Lake}},\ }\href@noop {} {\  (\bibinfo {year} {2018})},\ \Eprint
  {https://arxiv.org/abs/1802.07747} {arXiv:1802.07747 [hep-th]} \BibitemShut
  {NoStop}%
\bibitem [{\citenamefont {MacDonald}(1988)}]{PhysRevB.37.4792}%
  \BibitemOpen
  \bibfield  {author} {\bibinfo {author} {\bibfnamefont {A.~H.}\ \bibnamefont
  {MacDonald}},\ }\href {https://doi.org/10.1103/PhysRevB.37.4792} {\bibfield
  {journal} {\bibinfo  {journal} {Phys. Rev. B}\ }\textbf {\bibinfo {volume}
  {37}},\ \bibinfo {pages} {4792} (\bibinfo {year} {1988})}\BibitemShut
  {NoStop}%
\bibitem [{\citenamefont {Polyakov}(1988)}]{Polyakov:1988md}%
  \BibitemOpen
  \bibfield  {author} {\bibinfo {author} {\bibfnamefont {A.~M.}\ \bibnamefont
  {Polyakov}},\ }\href {https://doi.org/10.1142/S0217732388000398} {\bibfield
  {journal} {\bibinfo  {journal} {Mod. Phys. Lett. A}\ }\textbf {\bibinfo
  {volume} {3}},\ \bibinfo {pages} {325} (\bibinfo {year} {1988})}\BibitemShut
  {NoStop}%
\bibitem [{\citenamefont {Knapp}\ \emph {et~al.}(2016)\citenamefont {Knapp},
  \citenamefont {Zaletel}, \citenamefont {Liu}, \citenamefont {Cheng},
  \citenamefont {Bonderson},\ and\ \citenamefont {Nayak}}]{PhysRevX.6.041003}%
  \BibitemOpen
  \bibfield  {author} {\bibinfo {author} {\bibfnamefont {C.}~\bibnamefont
  {Knapp}}, \bibinfo {author} {\bibfnamefont {M.}~\bibnamefont {Zaletel}},
  \bibinfo {author} {\bibfnamefont {D.~E.}\ \bibnamefont {Liu}}, \bibinfo
  {author} {\bibfnamefont {M.}~\bibnamefont {Cheng}}, \bibinfo {author}
  {\bibfnamefont {P.}~\bibnamefont {Bonderson}},\ and\ \bibinfo {author}
  {\bibfnamefont {C.}~\bibnamefont {Nayak}},\ }\href
  {https://doi.org/10.1103/PhysRevX.6.041003} {\bibfield  {journal} {\bibinfo
  {journal} {Phys. Rev. X}\ }\textbf {\bibinfo {volume} {6}},\ \bibinfo {pages}
  {041003} (\bibinfo {year} {2016})}\BibitemShut {NoStop}%
\end{thebibliography}%

\end{document}